\documentclass[preprint,12pt,authoryear]{elsarticle}

\usepackage{amssymb}
\usepackage{amsmath}
\usepackage{lineno}

\modulolinenumbers[1]
\usepackage[colorlinks]{hyperref}

\journal{Journal of the Mechanics and Physics of Solids}

\usepackage[margin=2cm]{geometry}
\usepackage{bm}
\usepackage{accents}
\usepackage{color}
\usepackage{xcolor}
\usepackage{soul}
\usepackage{multirow}
\usepackage{mathtools}
\usepackage{graphicx,subcaption}
\usepackage{float}
\usepackage{gensymb}
\usepackage{tikz}

\newcommand{\ubar}[1]{\underaccent{\bar}{\bm{#1}}}
\newcommand{\tend}[1]{\hbox{\oalign{$\bm{#1}$\crcr\hidewidth$\scriptscriptstyle\bm{\sim}$\hidewidth}}}
\newcommand{\tene}[1]{\hbox{\oalign{$\bm{#1}$\crcr\hidewidth$\scriptscriptstyle\bm{\simeq}$\hidewidth}}}
\newcommand{\tenq}[1]{\hbox{\oalign{$\bm{#1}$\crcr\hidewidth$\scriptscriptstyle\bm{\approx}$\hidewidth}}}
\newcommand{\pvec}[2]{\hbox{\oalign{$\accentset{\times}{\underaccent{\bar}{\bm{#1}}}^{\mathrm{#2}}$\crcr\hidewidth}}}
\newcommand{\pvecdot}[2]{\hbox{\oalign{$\accentset{\times}{\underaccent{\bar}{\dot{\bm{#1}}}}^{\mathrm{#2}}$\crcr\hidewidth}}}

\DeclareMathOperator{\axi}{axi}

\usepackage[english]{babel}
\addto\captionsenglish{}

\graphicspath{{./}}
\newcommand{\MB}[1]{#1}

\begin{document}

\begin{frontmatter}



\title{A multi-physics model for dislocation driven spontaneous grain nucleation and microstructure evolution in polycrystals} %% Article title

\author[fzj]{I.T. Tandogan}

\author[fzj]{M. Budnitzki\texorpdfstring{\corref{mycorrespondingauthor}}{}}
\cortext[mycorrespondingauthor]{Corresponding author}
\ead{m.budnitzki@fz-juelich.de}

\author[fzj,rwth]{S. Sandfeld}

\affiliation[fzj]{organization={Institute for Advanced Simulations – Materials Data Science and Informatics (IAS‑9), Forschungszentrum Jülich GmbH},
	addressline={},
	city={Jülich},
	postcode={52425},
	state={},
	country={Germany}}

\affiliation[rwth]{organization={Chair of Materials Data Science and Materials Informatics, Faculty 5 – Georesources and Materials Engineering, RWTH Aachen University},
	addressline={},
	city={Aachen},
	postcode={52056},
	state={},
	country={Germany}}

%% Abstract
\begin{abstract}
The granular microstructure of metals evolves significantly during thermomechanical processing through viscoplastic deformation and recrystallization. Microstructural features such as grain boundaries, sub-grains, localized deformation bands, and non-uniform dislocation distributions critically influence grain nucleation and growth during recrystallization. Traditionally, modeling this coupled evolution involves separate, specialized frameworks for mechanical deformation and microstructural kinetics, typically used in a staggered manner. Nucleation is often introduced ad hoc, with nuclei seeded at predefined sites based on criteria like critical dislocation density, stress, or strain. This is a consequence of the inherent limitations of the staggered approach, where newly formed grain boundaries or grains have to be incorporated with additional processing.

In this work, we propose a unified, thermodynamically consistent field theory that enables spontaneous nucleation driven by stored dislocations at grain boundaries. The model integrates Cosserat crystal plasticity with the Henry–Mellenthin–Plapp orientation phase field approach, allowing the simulation of key microstructural defects, as well as curvature- and stored energy-driven grain boundary migration. The unified approach enables seamless identification of grain boundaries that emerge from deformation and nucleation. Nucleation is activated through a coupling function that links dislocation-related free energy contributions to the phase field. Dislocation recovery occurs both at newly formed nuclei and behind migrating grain boundaries.

The model's capabilities are demonstrated using periodic bicrystal and polycrystal simulations, where mechanisms such as strain-induced boundary migration, sub-grain growth, and coalescence are captured. The proposed spontaneous nucleation mechanism offers a novel addition to the capabilities of phase field models for recrystallization simulation.
\end{abstract}


\begin{highlights}
\item A unified multi-physics model for the evolution of grain microstructure is presented
\item A dislocation driven spontaneous nucleation mechanism at grain boundaries is proposed
\item Nucleation is affected by misorientation, dislocation distribution and GB velocity
\item Model is capable of slip/kink band formation and sub-grain nucleation by deformation
\item Recrystallization behavior strongly depends on deformed state of the microstructure
\end{highlights}

%% Keywords

\end{frontmatter}

%\linenumbers


\section{Introduction}

The granular microstructure of crystalline metals significantly transforms during mechanical processing and heat treatment. Viscoplastic deformation generates a large number of dislocations, often non-uniformly and concentrated at defects such as grain boundaries and second phase particles \citep{ashby1970deformation}. Within the grains, localized regions of intense deformation (slip bands), and reorientation (kink bands) can form \citep{asaro1977strain,stinville2023insights}, or the grains may fragment into smaller subgrains \MB{separated by} low-angle grain boundaries \citep{sedlavcek2002subgrain}. These locations of high stored energy \MB{act as} nucleation sites during recrystallization through mechanisms such as strain induced boundary migration (SIBM) and nucleation \MB{as well as} subgrain growth and coalescence, where dislocation-free nuclei expand into other grains reducing total energy. High-angle grain boundaries and triple junctions with large lattice orientation \MB{jumps} are also preferential sites for nucleation \citep{beck1950strain,cahn1950new,rios2005nucleation,raabe2014recovery,alaneme2019recrystallization,ferdinand2022nucleation}. After recrystallization, grain coarsening increases the overall size of the grains reducing the total area of the grain boundaries; \MB{in this regime} the migration of grain boundaries is driven by curvature as well as stored energy \citep{gottstein2009grain}. A non-uniform distribution of the latter can cause abnormal growth of some grains \citep{rollett2017recrystallization}.

The kinetics of recrystallization and grain boundary motion have been studied with a variety of methods such as vertex techniques (cf. \cite{gill1996variational,mcelfresh2023initial}), cellular automata (e.g. \cite{hesselbarth1991simulation,raabe2002cellular,chen2021multiscale,liu2024polycrystal}), Monte Carlo Potts \citep{anderson1984computer,mason2015kinetics,tutcuoglu2019stochastic,yu2021analysis} and level-set methods \citep{chen1995novel,sarrazola2020full,sarrazola2020new,bernacki2024kinetic} \MB{as well as} phase field models \MB{(cf. \cite{tourret2022phase} for a review)}, where the last two can implicitly keep track of moving interfaces. Phase field approaches that treat grain boundaries (GB) as continuous diffuse interfaces, are divided mainly into \MB{two families}. Multi-phase field (MPF) models (see \cite{steinbach1996phase,steinbach1999generalized,fan1997computer}) where each grain is \MB{represented by an individual order parameter}, and orientation phase field models (cf. \cite{kobayashi2000continuum,warren2003extending,henry2012orientation}), where a single order parameter and lattice orientation field describe the whole structure. While the phase field methods originally use curvature as the driving force, by construction, their free energy can be modified to include other physical effects such as stored energy \citep{steinbach2006multi,abrivard2012aphase}. Still, on their own, they do not account for microstructural changes caused by mechanical deformation. The coupled evolution problem is usually tackled by employing a model for GB migration sequentially with a \MB{mechanics model, such as} crystal plasticity (CP), \MB{in order to capture} deformation, where information is passed back and forth between them. The nucleation is incorporated in an intermediate step, \MB{requiring a third model} governed by a probabilistic method or trigger criteria such as critical stress, strain or dislocation density, where a nucleus is planted \MB{in an ad-hoc manner} at a possible nucleation site if requirements are met. Such an approach to modeling static and dynamic recrystallization \MB{was taken by} \cite{takaki2010static,li2020effect,chatterjee2024crystal}, where MPF was combined with CP frameworks. Three dimensional recrystallization and twinning in polycrystals were efficiently simulated by \cite{chen2015integrated}, \cite{zhao2016integrated} and \citep{hu2021spectral} using spectral Fast-Fourier-Transform (FFT) based MPF and CP solvers. \cite{takaki2007phase} coupled KWC type orientation phase field to finite element CP, whereas \cite{abrivard2012aphase,abrivard2012bphase} and \cite{luan2020combining} additionally enhanced the KWC phase field with a stored dislocation energy term.

Along with staggered schemes, some researchers have pursued a unified, thermodynamically consistent framework that strongly couples the evolution of mechanical and kinetic variables. In this regard, orientation phase field models
are well suited for a strong coupling with crystal plasticity as the lattice orientation is \MB{an independent degree of freedom}. \cite{admal2018unified} and \cite{he2021polycrystal} have proposed a model that couples strain gradient CP with Kobayashi-Warren-Carter (KWC) type orientation phase field, by \MB{identifying} the lattice orientation gradient in the KWC \MB{model's} free energy as the geometrically necessary dislocation density tensor \citep{nye1953some}. Hence, it is capable of predicting both shear-induced and curvature driven GB motion, where the GB motion is accommodated by plastic slip processes. At the same time, \cite{ask2018cosserat,ask2018bcosserat,ask2019cosserat,ask2020microstructure} have utilized Cosserat type CP with independent micro-rotation degrees of freedom \citep{forest1997cosserat,forest2000cosserat}, recognizing it as a natural framework for the coupling. Lattice orientation can evolve simultaneously due to viscoplastic deformation and GB migration, and the latter is driven by both \MB{curvature} and accumulated dislocations. The coupling of Cosserat continuum with orientation phase field allows heterogeneous reorientation and subgrain \MB{formation} in the bulk. Recently, \cite{ghiglione2024cosserat} and \cite{doghman2025finite} showed through the torsion of a single crystal rod that the model can also trigger spontaneous grain nucleation. \MB{New grains are formed as the result of an instability caused by the torsion-induced strong gradient in lattice orientation}. Another unified framework, coupling Henry-Mellenthin-Plapp (HMP) type orientation phase field \citep{henry2012orientation,staublin2022phase} with Cosserat CP (CCP) was proposed by \cite{tandogan2025multi}, where the HMP model offered some improvements compared to the KWC \MB{orientation phase field}.

In this paper, we present an improved version of the HMP-CCP model \citep{tandogan2025multi}, enhanced with \MB{model-free} spontaneous dislocation driven grain nucleation in polycrystals \MB{with preferred nucleation sites} at the grain boundaries. The microstructure can evolve by viscoplastic deformation and GB migration, where the latter is driven by curvature and stored energy gradients. The coupled model with the proposed nucleation mechanism reproduce SIBM, subgrain coarsening and coalescence mechanisms seamlessly in a strongly coupled setting. This is achieved by using a modified form of the stored energy contribution of dislocations to the free energy \MB{in the grain boundaries}. By adjusting the \MB{model parameters}, it is possible to strengthen or weaken the nucleation mechanism. Dislocations are recovered at the nuclei and in the wake of GBs using a modified Kocks-Mecking-Teodosiu law. Furthermore, similar to the KWC-CCP model, localized deformation and resulting formation of slip, kink bands, as well as fragmentation of grains into subgrains \MB{are captured}, which significantly affect the nucleation behavior. To the best of authors knowledge, no such a coupled treatment of nucleation in a unified and thermodynamically consistent network has been \MB{proposed to date}.

The paper is structured as follows: Section \ref{sec:model} summarizes the coupled Cosserat crystal plasticity and HMP orientation phase field framework, presenting the governing equations, constitutive model, and the proposed nucleation mechanism. In Section \ref{sec:numex}, the nucleation mechanism and its dependency on misorientation, dislocation distribution and GB velocity is explored. The capabilities of the model are \MB{demonstrated based on} periodic bicrystal and polycrystal examples. Finally, the paper is concluded with a summary and outlook in Section \ref{sec:conc}.

\section{Diffuse interface Cosserat crystal plasticity-HMP orientation phase field framework}\label{sec:model}

In this section, a summary of the modeling framework is presented in a small deformation setting. The details of the model, which couples Cosserat crystal plasticity with the HMP type orientation phase field \citep{henry2012orientation,staublin2022phase}, were introduced previously in \cite{tandogan2025multi}. For the detailed derivations the reader is referred to this publication. In the present work, a fully coupled, dislocation density based \MB{mechanism for} spontaneous grain nucleation is proposed. Sections \ref{ssec:balance} and \ref{ssec:consev} give an overview over the balance equations and the constitutive equations, respectively. Finally, Section \ref{ssec:grainnuc} presents the modifications proposed to enable the grain nucleation mechanism.  

\subsection*{Notation}

In the following, vectors $a_i$ are denoted by $\ubar{a}$, 2nd order tensors $A_{ij}$ by $\tend{A}$, 3rd order Levi-Civita permutation tensor $\epsilon_{ijk}$ by $\tene{\epsilon}$, 4th order tensors $C_{ijkl}$ by $\tenq{C}$. \MB{The Kronecker delta is denoted by $\delta^{ij}$.} Gradient is denoted by $\nabla(.)$, divergence by $\nabla\cdot(.)$, trace by $tr(.)$, transpose by $(.)^\text{T}$, dot product by $(.)\cdot(.)$, double contraction by $(.):(.)$, and tensor product by $(.)\otimes(.)$. \MB{The partial derivative of $(.)$ with respect to $\eta$ is denoted by $(.)_{,\eta}$.} The transformation between pseudo-vector $\pvec{a}{}$ and skew-symmetric tensor $\tend{A}^{\text{skew}}$ is given by

\begin{equation}
	\pvec{a}{}\,\MB{=\axi(\tend{A}^{\text{skew}})}\coloneqq-\frac{1}{2}\tene{\epsilon}:\tend{A}^{\text{skew}} \quad\text{and}\quad \tend{A}^{\text{skew}}=-\tene{\epsilon}\cdot\pvec{a}{}.
\end{equation}
%\MB{and the function $\text{pvec}(\cdot)$ is defined as $\pvec{a}{}=pvec(\tend{A}^{\text{skew}})$}.

\subsection{Balance laws}\label{ssec:balance}

At each material point, the Cosserat continuum enhances the classical displacement degrees of freedom $\ubar{u}$ with 
the additional independent microrotation degrees of freedom represented by the pseudo-vector $\ubar{\Theta}$. In the small deformation setting, the microrotation tensor is given by

\begin{equation}
	\tend{R}=\tend{I}-\tene{\epsilon}\cdot\ubar{\Theta},
\end{equation}
where $\tend{I}$ is the identity tensor. The objective deformation measures are, the deformation tensor $\tend{e}$ and the curvature tensor \tend{\kappa} where

\begin{equation}\label{eqn:ekappa}
	\tend{e}=\nabla\ubar{u}+\tene{\epsilon}\cdot\ubar{\Theta}, \quad\quad \tend{\kappa}=\nabla\ubar{\Theta},
\end{equation}
respectively \citep{eringen1976polar,forest1997cosserat}. The former is additively decomposed into elastic and plastic parts

\begin{equation}\label{eqn:eadddec}
	\tend{e} = \tend{e}^{\mathrm{e}} + \tend{e}^{\mathrm{p}},
\end{equation}
while a plastic curvature is not considered for simplicity, though \MB{its inclusion is} possible [see \cite{forest1997cosserat}].

In the coupled model, the Cosserat theory is enhanced with the orientation phase field model introducing a course-grained measure of the crystalline order, $\eta\in [0,1]$. It is equal to 1 in the bulk of the grains and <1 in the diffuse grain boundaries. Moreover, the lattice orientation is represented by the Cosserat microrotation $\ubar{\Theta}$. By manipulating equations \eqref{eqn:ekappa} and \eqref{eqn:eadddec}, defining the spin tensor $\tend{\omega} = \left[\nabla\dot{\ubar{u}}-\left(\nabla\dot{\ubar{u}}\right)^\text{T}\right]/2$, and its \MB{additive} elastic-plastic decomposition $\pvec{\omega}{\text{e}}:=\pvec{\omega}{}-\pvec{\omega}{\text{p}}$ and $\pvec{\omega}{\text{p}}:=\pvecdot{e}{\text{p}}$, it is found that

\begin{equation}\label{eqn:rotlink}
	\pvecdot{e}{\text{e}} = \pvec{\omega}{\text{e}}-\dot{\ubar{\Theta}}.
\end{equation}
According to equation \eqref{eqn:rotlink}, the rate of change of the lattice orientation and the Cosserat microrotation will be equal if the constraint $\pvec{e}{\text{e}}\equiv 0$ is \MB{fulfilled}. It can be enforced on the constitutive level with a penalty parameter \citep{forest2000cosserat}, or by a more sophisticated and efficient duality-based formulation proposed recently by \cite{baek2022duality}. We follow the former approach for simplicity.

The balance equations and the boundary conditions are derived by applying the principle of virtual power. \MB{The phase field portion of the resulting equations corresponds to Gurtin's microforce balance \citep{gurtin1996generalized,gurtin1999sharp}. The boundary value problem is given by the following equations (cf. \cite{ask2018bcosserat} for a detailed derivation)} 
%where the contribution from phase field variables is included by using the microforce formalism as described by Gurtin \cite{gurtin1996generalized,gurtin1999sharp}. Here the resulting equations are given directly, and the detailed derivation was presented in \cite{ask2018bcosserat}. Using the virtual rates, $\mathcal{V} = \left\{\dot{\eta}^\star,\nabla\dot{\eta}^\star,\dot{\ubar{u}}^\star,\nabla\dot{\ubar{u}}^\star,\dot{\ubar{\Theta}}^\star,\nabla\dot{\ubar{\Theta}}^\star\right\}$, we get,

\begin{alignat}{3}
	&\nabla\cdot\ubar{\xi}_\eta + \pi_\eta +\pi_\eta^{\text{ext}}=0 \quad\quad\quad &&\text{in}\;\;\Omega,&&\label{eqn:balgen}\\
	&\nabla\cdot\tend{\sigma}+\ubar{f}^{\text{ext}}=\ubar{0} \quad\quad\quad &&\text{in}\;\;\Omega,&&\label{eqn:ballin}\\
	&\nabla\cdot\tend{m}+2\pvec{\sigma}{}+\ubar{c}^{\text{ext}}=\ubar{0} \quad\quad\quad &&\text{in}\;\;\Omega,&&\label{eqn:balang}\\
	&\ubar{\xi}_\eta\cdot\ubar{n}=\pi_\eta^\text{c} \quad\quad\quad &&\text{on}\;\;\partial\Omega,&&\\
	&\tend{\sigma}\cdot\ubar{n}=\ubar{f}^\text{c} \quad\quad\quad &&\text{on}\;\;\partial\Omega,&&\\
	&\tend{m}\cdot\ubar{n}=\ubar{c}^\text{c} \quad\quad\quad &&\text{on}\;\;\partial\Omega,&&
\end{alignat}
\MB{where $\ubar{f}$ and $\ubar{c}$ are forces and couples, respectively}. The superscript $(.)^\text{ext}$ denotes external forces and couples, while $(.)^\text{c}$ denotes contact forces and couples. The vector $\ubar{n}$ is the outward normal to the surface $\partial\Omega$ of \MB{material body} $\Omega$. $\pi_\eta$ and $\ubar{\xi}_\eta$ are the generalized microforce and stress, respectively. $\tend{\sigma}$ is the (generally unsymmetrical) stress tensor and $\tend{m}$ is the couple-stress. The work conjugate pairs are given by $\left\{\eta:\pi_\eta, \nabla\eta:\ubar{\xi}_\eta, \tend{e}:\tend{\sigma}, \tend{\kappa}:\tend{m}\right\}$.

\subsection{Constitutive equations}\label{ssec:consev}

We define the Helmholtz free energy \MB{density} as $\psi\coloneqq\psi\left(\eta,\nabla\eta,\tend{e}^{\mathrm{e}},\tend{\kappa},r_\alpha\right)$ where $r_\alpha$ are plasticity related internal variables. Applying the Clausius-Duhem inequality results in

\begin{equation}
	-\left[\pi_\eta+\dfrac{\partial\psi}{\partial\eta}\right]\dot{\eta} + \left[\ubar{\xi}_\eta-\dfrac{\partial\psi}{\partial\nabla\eta}\right]\cdot\nabla\dot{\eta} + \left[\tend{\sigma}-\dfrac{\partial\psi}{\partial\tend{e}^{\mathrm{e}}}\right]:\dot{\tend{e}}^{\mathrm{e}} + \left[\tend{m}-\dfrac{\partial\psi}{\partial\tend{\kappa}}\right]:\dot{\tend{\kappa}} +\tend{\sigma}:\dot{\tend{e}}^{\mathrm{p}}-\sum_\alpha\dfrac{\partial\psi}{\partial r_\alpha}\dot{r}_\alpha \ge 0 .
\end{equation}

The relaxation dynamics of the phase field is recovered by decomposing \MB{the scalar microforce} $\pi_\eta=\pi_\eta^\text{eq} + \pi_\eta^\text{dis}$ into energetic $\pi_\eta^\text{eq}$ and dissipative $\pi_\eta^\text{dis}$ parts. Then, the constitutive relations are given by,

\begin{equation}\label{eqn:const}
	\pi_\eta^\text{eq}=-\dfrac{\partial\psi}{\partial\eta},\qquad\ubar{\xi}_\eta=\dfrac{\partial\psi}{\partial\nabla\eta},\qquad\tend{\sigma}=\dfrac{\partial\psi}{\partial\tend{e}^{\mathrm{e}}},\qquad\tend{m}=\dfrac{\partial\psi}{\partial\tend{\kappa}}.
\end{equation}

Introducing thermodynamic forces related to $r_\alpha$ as $R_\alpha=\partial\psi/\partial r_\alpha$, and the dissipation potential $\Omega=\Omega^{\mathrm{p}}(\tend{\sigma},\MB{R_{\alpha};\eta})+\Omega^\eta(\pi_\eta^\text{dis})$, the evolution of dissipative processes are obtained as

\begin{equation}\label{eqn:evodot}
	\dot{\tend{e}}^{\mathrm{p}}=\dfrac{\partial\Omega^{\mathrm{p}}}{\partial\tend{\sigma}},\qquad \dot{r}_\alpha=-\dfrac{\partial\Omega^\mathrm{p}}{\partial R_\alpha},\qquad \dot{\eta}=-\dfrac{\partial\Omega^\eta}{\partial\pi_\eta^{dis}}.
\end{equation}

Now \MB{specific} forms of the free energy \MB{density} $\psi$ and dissipation potential $\Omega$ are \MB{provided for closure}. Similar to \cite{tandogan2025multi}, the problem is restricted to two dimensions and isotropic grain boundary energies for simplicity. Therefore, it follows that $\ubar{\Theta}=\left[0\;0\;\theta\right]^\text{T}$, $\pvec{\omega}{}=\left[0\;0\;\MB{\accentset{\times}{\omega}}\right]^\text{T}$, $\pvec{e}{}=\left[0\;0\;\accentset{\times}{e}\right]^\text{T}$ and $\ubar{m}_\theta=[m_{31}\;m_{32}\;m_{33}]^\text{T}$. 

\MB{\subsubsection{Free energy density}}

With \MB{the above} considerations, the following free energy \MB{density} is proposed

\begin{align}\label{eqn:freeen2d}
	\begin{split}
		\psi\left(\eta,\nabla\eta,\tend{e}^{\mathrm{e}},\theta,r_\alpha\right) &= f_0\left[\alpha V(\eta)+\frac{\nu^2}{2}|\nabla\eta|^2+\mu^2g(\eta)|\nabla\theta|^2\right]\\
		&+\dfrac{1}{2}\tend{\varepsilon}^{\mathrm{e}}:\tenq{E}^\mathrm{s}:\tend{\varepsilon}^{\mathrm{e}}+2\mu_\mathrm{c}\,\bigl(\accentset{\times}{e}^{\mathrm{e}}\bigr)^2+\phi(\eta)\sum_{\alpha=1}^{N}\frac{\lambda}{2}\mu^{\mathrm{e}}r_{\alpha}^2,
	\end{split}
\end{align}
where $f_0$ is a normalization coefficient. The terms in the first line are inherited from the Henry-Mellenthin-Plapp \citep{henry2012orientation} orientation phase field model. The coefficients $\alpha$, $\nu$ and $\mu$ can be used to tune the equilibrium profiles of diffuse grain boundaries, i.e., the order parameter $\eta$ and lattice orientation $\theta$. The potential $V(\eta)$ is a single-well function penalizing the existence of grain boundaries, i.e. $\eta<1$, while the second and third terms penalize gradients in $\eta$ and $\theta$. It is easy to see that \MB{all} terms are minimized for a single grain solution. The coupling function $g(\eta)$ is constructed such that it is singular for $\eta=1$, which makes localized grain boundary solutions possible \citep{henry2012orientation}. 

The second line of \eqref{eqn:freeen2d} contains the elastic energy contributions, where $\tenq{E}^\mathrm{s}$ is the 4th order elasticity tensor \MB{with minor and major symmetry} and $\mu_\mathrm{c}$ is the Cosserat couple modulus. As shown in \cite{ask2018cosserat} and \cite{tandogan2025multi}, a large $\mu_\mathrm{c}$ penalizes the skew-symmetric part of the \MB{elastic} deformation, which enforces the constraint $\pvec{e}{\text{e}}\equiv 0$; then the Cosserat microrotation follows the lattice orientation. The last term is the energy contribution due to accumulated dislocations where $N$ is the number of slip systems, $\lambda$ is a parameter close to 0.3 \citep{hirth1983theory} and $\mu^{\mathrm{e}}$ is the shear modulus. $\phi(\eta)$ is a coupling function, which results in a driving force for migration of grain boundaries in the presence of stored dislocations. Previously, in \cite{tandogan2025multi}, it was shown that the form of this function directly affects the equilibrium of the order parameter $\eta$ and the dynamics of the grain boundary migration. In this work, we show that the same function can be used to implement dislocation driven grain nucleation.

By inserting \eqref{eqn:freeen2d} into \eqref{eqn:const}, the microforces, couple-stress and stress are derived as
\begin{align}
	\pi_\eta^\text{eq}=& -f_0\left[\alpha V_{,\eta}+\mu^2g_{,\eta}|\nabla\theta|^2\right]-\phi_{,\eta}\sum_{\alpha=1}^{N}\frac{\lambda}{2}\mu^{\mathrm{e}}r_{\alpha}^2,\label{eqn:pieq}\\
	\ubar{\xi}_\eta=& f_0\left[\nu^2\nabla\eta\right],\label{eqn:xieta}\\
	\ubar{m}_\theta=& f_0\left[2\mu^2g(\eta)\nabla\theta\right],\label{eqn:mtheta}\\
	\tend{\sigma}=& \tenq{E}^\mathrm{s}:\tend{\varepsilon}^{\mathrm{e}} - 2\mu_\mathrm{c}\,\tene{\epsilon}\cdot\pvec{e}{\text{e}}.\label{eqn:sigall}
\end{align}
The last term in \eqref{eqn:sigall} is the skew part of the stress tensor and it is equivalent to $\accentset{\times}{\sigma}=2\mu_\mathrm{c}\,\accentset{\times}{e}^{\mathrm{e}}$ in 2D.

\MB{\subsubsection{Dissipation potential}}

The dissipation potential is chosen as

\begin{equation}\label{eqn:plaspot}
	\Omega^{\mathrm{p}}(\tend{\sigma},\MB{R_{\alpha};\eta})=\sum_{\alpha=1}^N\dfrac{K_\mathrm{v}}{n+1}\left<\dfrac{|\tau^\alpha|-R_\alpha/\phi(\eta)}{K_\mathrm{v}}\right>^{n+1}+\dfrac{1}{2}\tau_*^{-1}(\eta)\pvec{\sigma}{}\cdot\pvec{\sigma}{},
\end{equation}
where $<\cdot>$ are Macaulay brackets \MB{defined by $\langle\cdot\rangle\coloneqq\max(0,\cdot)$}, $K_\mathrm{v}$ and $n$ are viscosity parameters. $\tau^\alpha$ is the resolved shear stress given by $\tau^\alpha=\ubar{l}^\alpha\cdot\tend{\sigma}\cdot\ubar{n}^\alpha$ for the slip direction $\ubar{l}^\alpha$ and the slip plane normal $\ubar{n}^\alpha$. Note that in the Cosserat framework the skew part of $\tend{\sigma}$ results in a contribution to $\tau^\alpha$ acting as a size dependent kinematic hardening \citep{forest2008some,forest2023size}. The term $R_\alpha/\phi(\eta)$ is identified as the critically resolved shear stress of slip system $\alpha$. From its definition,

\begin{equation}
	R_\alpha=\dfrac{\partial\psi}{\partial r_\alpha}=\lambda\phi(\eta)\mu^{\mathrm{e}}r_\alpha \quad\text{with}\quad r_\alpha=b\sqrt{\sum_{\beta=1}^{N}h^{\alpha\beta}\rho^\beta},
\end{equation}
where $b$ is the norm of the Burgers vector, $h^{\alpha\beta}$ is the slip system interaction matrix and $\rho^\alpha$ are statically stored dislocation (SSD) densities. Inserting \eqref{eqn:plaspot} in \eqref{eqn:evodot}, we get the plastic deformation rate

\begin{equation}\label{eqn:epdot}
	\dot{\tend{e}}^{\mathrm{p}}=\sum_{\alpha=1}^N\dot{\gamma}^\alpha\ubar{l}^\alpha\otimes\ubar{n}^\alpha+\dfrac{1}{2}\tau_*^{-1}(\eta)\,\text{skew}(\tend{\sigma}) \quad\text{with}\quad \dot{\gamma}^\alpha=\left<\dfrac{|\tau^\alpha|-R_\alpha/\phi(\eta)}{K_\mathrm{v}}\right>^n \text{sign}\,\tau^\alpha
\end{equation}
as the slip rate according to the viscoplastic flow rule from \cite{cailletaud1992micromechanical}. The unconventional purely skew symmetric term in \eqref{eqn:epdot} represents the atomic reshuffling and the accompanying reorientation process during migration of grain boundaries \citep{ask2018bcosserat}. It is restricted to the grain boundaries by choosing an appropriate form of inverse mobility \MB{function} $\tau_*(\eta)$, i.e.,

\begin{equation}\label{eqn:taustarg}
	\tau_*=\hat{\tau}_*g(\eta),
\end{equation}
where the singular function $g(\eta)$ is large in the bulk and small in the grain boundary. Let $\pvec{e}{\text{p}}=\pvec{e}{\text{slip}}+\pvec{e}{*}$, where $\pvec{e}{*}$ is an eigenrotation. Then,

\begin{alignat}{3}
	\pvecdot{e}{\text{slip}}&=\pvec{\omega}{\text{p}}-\pvec{\omega}{*},\qquad\qquad \tend{\omega}^{\mathrm{p}}-&&\tend{\omega}^* =&&\,\text{skew}\left(\sum_{\alpha=1}^N\dot{\gamma}^\alpha\ubar{l}^\alpha\otimes\ubar{n}^\alpha\right),\\
	\pvecdot{e}{*}&=\pvec{\omega}{*}, &&\pvec{\omega}{*} =&&\tau_*^{-1}\pvec{\sigma}{}.\label{eqn:wstar}
\end{alignat}
From \eqref{eqn:ekappa} and \eqref{eqn:eadddec} we have

\begin{equation}\label{eqn:eeskew}
	\pvec{e}{\text{e}}=\axi(\text{skew}[\nabla\ubar{u}])-\pvec{e}{\text{p}}-\ubar{\Theta}=\axi(\text{skew}[\nabla\ubar{u}])-\pvec{e}{\text{slip}}-\pvec{e}{*}-\ubar{\Theta},
\end{equation} 
where the undeformed state is not stress free, due to non-zero $\ubar{\Theta}$, unless the eigenrotation $\pvec{e}{*}$ is initialized as [\cite{ask2018bcosserat,tandogan2025multi}]

\begin{equation}
	\pvec{e}{*}(t=0)=\pvec{e}{\text{p}}(t=0)=-\ubar{\Theta}(t=0).
\end{equation}
According to \eqref{eqn:eeskew}, a migrating grain boundary generates $\pvec{e}{\text{e}}$ and $\pvec{\sigma}{}$ in the region of the moving front, which are relaxed by the evolution of eigenrotation $\pvec{e}{*}$ in \eqref{eqn:wstar}.

A SSD density based hardening is used, whose evolution is governed by a modified Kocks-Mecking-Teodosiu law [\cite{abrivard2012aphase,ask2018cosserat,ask2018bcosserat}]
\begin{equation}\label{eqn:rhodot}
	\dot{\rho}^\alpha=
	\begin{cases}
		\dfrac{1}{b}\left(K\sqrt{\sum _\beta a^{\alpha\beta}\rho^\beta}-2d\rho^\alpha\right)|\dot{\gamma}| -\rho^\alpha C_\mathrm{D}A(\eta)\dot{\eta} \quad&\text{if}\quad \dot{\eta}>0,\vspace{1mm}\\
		\dfrac{1}{b}\left(K\sqrt{\sum _\beta a^{\alpha\beta}\rho^\beta}-2d\rho^\alpha\right)|\dot{\gamma}| &\text{if}\quad \dot{\eta}\le 0,
	\end{cases}
\end{equation}
where $K$ is the mobility constant, $d$ is the critical annihilation distance between dislocations of opposite sign, and $a^{\alpha\beta}$ is an interaction matrix for cross-slip. The extra term \MB{$-\rho^\alpha C_\mathrm{D}A(\eta)\dot{\eta}$} is active only when the order parameter $\eta$ increases, for example in the wake of a migrating grain boundary or in a nucleating grain. It represents the static recovery of dislocations \citep{abrivard2012aphase,ask2018cosserat,bailey1962recrystallization}. \MB{This is a simplistic representation of the combined absorption and annihilation recovery mechanisms that occur during the migration of a grain boundary.} For sufficiently large $C_\mathrm{D}$, full recovery can be achieved. The function $A(\eta)$ localizes this recovery to the grain boundary region. In this work we use the form

\begin{equation}\label{eqn:recfung}
	A(\eta)=\dfrac{7\eta^3-6\eta^4}{(1-\eta)}.
\end{equation}
In the previous works by \cite{ask2018cosserat,ask2018bcosserat,ask2019cosserat,ask2020microstructure} and in \cite{tandogan2025multi}, a hyperbolic tangent form based on $|\nabla\theta|$ was employed. However, we observed that such a form does not allow full recovery when two migrating grain boundaries merge, and that equation \eqref{eqn:recfung} performs better. \MB{This is, firstly, because $|\nabla\theta|>0$ is a more strict representation of the GB which is active in a narrower region compared to the $\eta<1$ condition (cf. Fig. \ref{fig:nucCD}). Secondly, towards the end of a merge the grain in between two GBs rotates to decrease misorientation to zero, and at small misorientations $\tanh(|\nabla\theta|^2)$ is negligible while $1-\eta$ is not.}

The quadratic dissipation potential

\begin{equation}
    \Omega^\eta = \frac{1}{2}\tau_\eta^{-1}\pi_\eta^{\text{dis}\;2}
\end{equation}
is chosen for the evolution of the order parameter $\eta$ \citep{gurtin1996generalized,abrivard2012aphase}. From \eqref{eqn:evodot} we get

\begin{equation}
	\pi_\eta^\text{dis}=-\tau_\eta\dot{\eta}.\label{eqn:pineq}
\end{equation}
Inserting \eqref{eqn:pieq}-\eqref{eqn:mtheta} and \eqref{eqn:pineq} into the balance laws \eqref{eqn:balgen} and \eqref{eqn:balang}, while assuming the absence body forces and couple forces, gives the evolution equations for the order parameter $\eta$ and lattice orientation $\theta$

\begin{align}
	\tau_\eta\dot{\eta}&=f_0\nu^2\nabla^2\eta-f_0\left[\alpha V_{,\eta}+\mu^2g_{,\eta}|\nabla\theta|^2\right]-\phi_{,\eta}\sum_{\alpha=1}^{N}\frac{\lambda}{2}\mu^{\mathrm{e}}r_{\alpha}^2,\label{eqn:hmpetadot}\\
	0&=f_0\nabla\cdot\left[\mu^2g(\eta)\nabla\theta\right]+2\mu_\mathrm{c}\,\accentset{\times}{e}^{\mathrm{e}}.\label{eqn:hmpthetadot}
\end{align}
Eq. \eqref{eqn:hmpthetadot} can be rewritten using \eqref{eqn:wstar} as

\begin{equation}\label{eqn:tdot_hmpccp}
	-\tau_*\accentset{\times}{\dot{e}}^*=f_0\nabla\cdot\left[\mu^2g(\eta)\nabla\theta\right].
\end{equation}
As before \citep{tandogan2025multi}, the potential $V(\eta)$ and the singular coupling function $g(\eta)$ are chosen as

\begin{equation}
	V(\eta)=\frac{1}{2}(1-\eta)^2 \quad\text{and}\quad g(\eta)=\dfrac{7\eta^3-6\eta^4}{(1-\eta)^3}+c\ln(1-\eta)+C_0,\label{eqn:fung}
\end{equation}
where
\begin{equation}
	C_0=\text{min}\left(\dfrac{7\eta_*^3-6\eta_*^4}{(1-\eta_*)^3}+c\ln(1-\eta_*)\right)+0.01 \quad\text{with}\quad \eta_*=[0,1].
\end{equation}
For conditions \MB{restricting the choice} of these functions see \cite{staublin2022phase} and \cite{tandogan2025multi}. The logarithmic term in $g(\eta)$ is used to obtain a Read-Shockley type grain boundary energy.

\subsection{Grain nucleation mechanism}\label{ssec:grainnuc} 

The orientation phase field models by \cite{kobayashi2000continuum} and \cite{henry2012orientation} are constructed such that the free energy is minimized for a localized grain boundary solution. Recently, \cite{ghiglione2024cosserat} showed that the Kobayashi-Warren-Carter (KWC) type phase field is capable of nucleating new grains starting with the (unstable) initial condition of a uniform orientation gradient. Utilizing this mechanism with the KWC-CCP coupled model, these authors were able to reproduce grain nucleation in a single crystal cylindrical rod which was plastically deformed by twisting. 
%We have observed that the HMP type phase field is similarly capable of such a mechanism since the minimization principle is the same. 
However, a prerequisite for this nucleation mechanism is the uniform orientation gradient inside a single crystal, which can be observed in a twisted rod, but is unlikely to occur in a deformed \MB{oligocrystal}. Grain nucleation is observed to occur in deformed polycrystals at the grain boundaries, where energy content is increased due to stored dislocations \citep{rollett2017recrystallization}. In this work, we show that nucleation at the grain boundaries driven by stored dislocations is captured with the HMP-CCP coupled model by using a new form of SSD energy coupling function $\phi(\eta)$.

\begin{figure}[ht]
	\centering
	\includegraphics[width=1\textwidth]{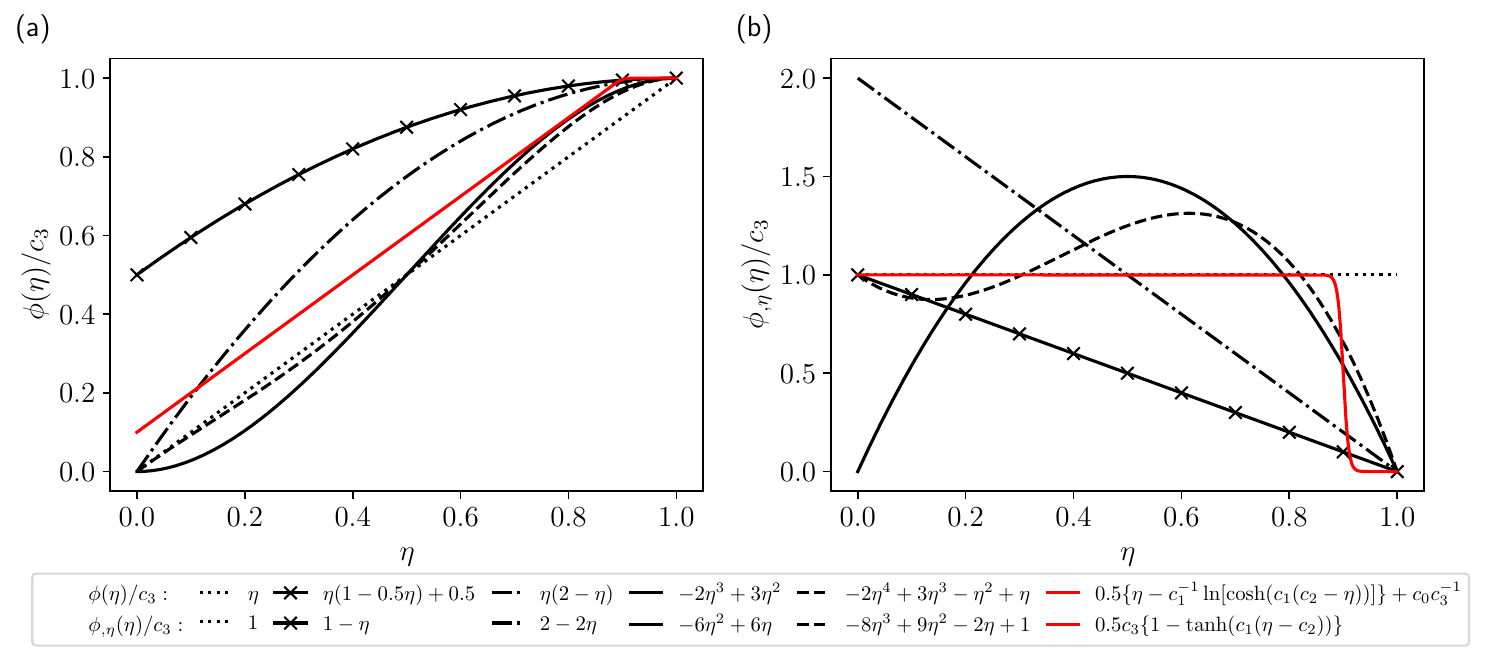}\vspace*{-4mm}
	\caption{Comparison of the SSD energy multiplier function $\phi(\eta)$ forms (left) and their derivatives $\phi'(\eta)$ (right). For the new form, $c_1=100$ and $c_2=0.9$ is used. \MB{$c_0=c_3(1-\phi|_{\eta=1})$ for visualization purposes.}}
	\label{fig:gbmigphi}
\end{figure}

Previously, we have shown that the form of the coupling function $\phi(\eta)$ \MB{and its derivative $\phi_{,\eta}$} in the SSD energy term in \eqref{eqn:hmpetadot} are important for the equilibrium \MB{profile of the order parameter} and the grain boundary dynamics \citep{tandogan2025multi}. When \MB{$\phi(\eta)=\eta$ and $\phi_{,\eta}=1$}, the equilibrium value of the order parameter $\eta$ becomes <1 in the presence of stored dislocations \citep{ask2018cosserat}, which is incompatible with the HMP model since it breaks the singularity of $g(\eta)$ \MB{in the bulk}. Therefore, we have proposed some polynomial forms shown in Fig. \ref{fig:gbmigphi} with $\phi_{,\eta}(\eta=1)=0$, which do not affect the equilibrium in the bulk \citep{tandogan2025multi}. Building upon that, \MB{in this work we use},

\begin{equation}\label{eqn:philog}
	\phi(\eta)=\dfrac{1}{2}c_3\left\{\eta-c_1^{-1}\ln\left[\cosh(c_1(c_2-\eta))\right]\right\} + c_0 ,
\end{equation}
with

\begin{equation}\label{eqn:phiderlog}
	\phi_{,\eta}(\eta)=\dfrac{c_3}{2}-\dfrac{c_3\tanh(c_1(\eta-c_2))}{2},
\end{equation}
where the coefficients $c_1$, $c_2$ and $c_3$ are \MB{non-negative} scalars.

Fig. \ref{fig:gbmigphi} shows the comparison of the \MB{coupling} function \MB{used in this work} with the previous polynomial forms. Since $\eta$ is a measure of order, the physical interpretation of $\phi(\eta)$ is that the energetic contribution of SSDs is more significant in the ordered, \MB{crystalline} state. The new form sets a saturation value, of this dependence on the order parameter, such that above a given $\eta=\overline{\eta}$, the \MB{material} is assumed to be \MB{'crystalline enough'}, in terms of SSD energy contribution. This assumption makes the diffuse GB representation, which is some orders wider than a real GB, somewhat more physical, since the outer region of the diffuse GB is \MB{treated as bulk material}. From the modeling perspective, the main feature of the new function is the step like form of its derivative, where the steepness and the position of the step are controlled by $c_1$ and $c_2$, respectively. As seen in Eq. \eqref{eqn:hmpetadot}, $\phi_{,\eta}$ creates a driving force on $\eta$ in the presence of SSDs. However, for $\eta>c_2$, $\phi_{,\eta}/c_3$ asymptotically approaches 0, while for $\eta<c_2$ it approaches 1. Hence, the equilibrium of $\eta$ is not affected in the bulk of the grain, while it is affected inside the grain boundary with the magnitude \MB{of the effect} depending on the value of the SSD density. Note that if $c_2>>1$, the function reduces to the form proposed by \cite{abrivard2012aphase}, where $\phi_{,\eta}=1$, and if $c_2=0$, the driving force \MB{on the order parameter} is effectively deactivated. When the SSD density is too large, the minimum value of $\eta$ can become less than zero, which is incompatible with the model. To prevent this, $c_3$ is used as a scaling coefficient to keep $\eta$ inside the limits. The consequence of the new form is a dislocation driven grain nucleation and growth, which is demonstrated in the next section using numerical examples.

\section{Numerical Examples}\label{sec:numex}

In this section, the capabilities of the model are demonstrated with 2D finite element simulations. In Section \ref{ssec:modparam} the chosen model parameters are given, where the grain boundary energy for misorientations up to 30$^{\degree}$ is fitted to data from literature for pure Copper (Cu). Section \ref{ssec:disnucmig} focuses on the proposed grain nucleation mechanism, and presents plastic deformation driven grain boundary migration and grain nucleation with a periodic bicrystal example. Finally, in Section \ref{ssec:polycrys} the potential of the coupled HMP-CCP model is \MB{explored} with periodic polycrystal examples containing 6 and 32 grains.

The model has been implemented in the FEniCS 2019 open-source finite element library \citep{alnaes2015fenics} used together with the MFront code generator for the material models \citep{helfer_introducing_2015}. FEniCS handles the finite element framework and the global Newton \MB{procedure}, while MFront is responsible for the constitutive \MB{portion of the} model and iteration at the material point \MB{level}. The communication between them is handled through the MGIS:fenics library \citep{helfer2020mfrontgenericinterfacesupport}, which has been modified and extended for our purposes. The system of equations is solved with a monolithic approach where, in two dimensions, each node has 4 degrees of freedom: the order parameter $\eta$, the Cosserat microrotation $\theta$ \MB{as well as two} displacements $(u_1,u_2)$. A semi-implicit time discretization is used, and the resulting nonlinear system of equations is solved with the Newton-Raphson algorithm. The continuous and evolving lattice orientation $\theta$ is used to rotate between the global and the local coordinate systems.

\subsection{Model parameters}\label{ssec:modparam}

In the simulations, the model parameters presented in Tab. \ref{tab:param} are used, unless otherwise stated. The material considered is pure copper. For the mechanical part of the model, the parameters for elasticity and plasticity are adopted from the literature \citep{gerard2009hardening,cheong2004discrete}. The Cosserat coupling modulus is chosen high enough to penalize Eq. \eqref{eqn:rotlink}. The recovery parameter $C_\mathrm{D}$ in \eqref{eqn:rhodot} is chosen to allow full recovery of dislocations. 

\begin{figure}[ht]
	\centering
	\includegraphics[width=0.6\textwidth]{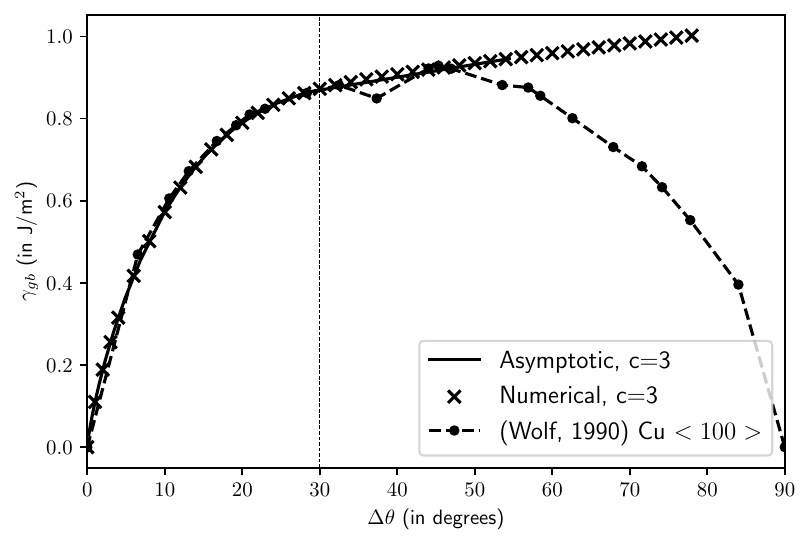}\vspace*{-4mm}
	\caption{Grain boundary energies at increasing misorientations calibrated to atomistic simulations of $<100>$ tilt grain boundaries \citep{wolf1990structure}. Only the Read-Shockley part of the curve up to 30$^{\degree}$ is calibrated since the model does not capture the cusps at larger misorientations. Singular function $g(\eta)$ has the form $(7\eta^3-6\eta^4)/(1-\eta)^3+c\ln(1-\eta)+C_0$ to generate Read-Shockley type grain boundary energy, with $c=3$.}
	\label{fig:gben}
\end{figure}

The phase field parameters $\nu$, $\alpha$ and $\mu$ determine the equilibrium profile of the order parameter and the orientation. They should be chosen by considering the length scale of the problem, \MB{which determines the admissible} grain boundary thickness. The asymptotic analysis of the HMP model, which is also valid for the undeformed state in the HMP-CCP model, can be used to find equilibrium profiles and the grain boundary energy dependence on misorientation \citep{staublin2022phase,tandogan2025multi}. As an alternative, it is straightforward to calculate them from 1D grain boundary simulations. Fig. \ref{fig:gben} shows the calibration of the grain boundary energy to the values from atomistic simulations of copper for $<100>$ tilt grain grain boundaries. Fitting is achieved by adjusting $c$ in \eqref{eqn:fung} and $f_0$ in  \eqref{eqn:freeen2d}. Calibration is limited to 30$^{\degree}$ \MB{misorientation}, since the model does not account for the symmetry and cusps at larger angles. 

\begin{table}[ht]
	\caption{Parameter set used for the coupled Cosserat Crystal Plasticity - HMP orientation phase field model.}\label{tab:param}
	\centering
    \setlength{\tabcolsep}{3pt}
	\begin{tabular}{l|ccccccc}
		\multicolumn{1}{c|}{\multirow{2}{*}{Phase field}} & $f_0$ (in kPa) & $\tau_\eta$ (in J\,s\,m$^{-3}$) & $\hat{\tau}_*$ (in J\,s\,m$^{-3}$) & $\nu$ (in \textmu m) & $\alpha$  & $\mu$ (in \textmu m) & $c$   \\ \cline{2-8} 
		\multicolumn{1}{c|}{}                             & \rule{0pt}{2.5ex}     371  & $10^2f_0t_0$; $10^4f_0t_0$  & $10^2f_0t_0$; $10^1f_0t_0$ & 1  & 20 & 2.5/$\pi$ & 3 \\ \hline\hline
		\multirow{4}{*}{Mechanics} & C11 (in GPa) & C12 (in GPa) & C44 (in GPa) & $\mu_\mathrm{c}$ (in GPa) & $\lambda$ & $b$ (in nm) &  \\ \cline{2-8} 
		& \rule{0pt}{2.5ex} 160 & 110 & 75 & 75 & 0.3 & 0.2556 &   \\ \cline{2-8}
		& \rule{0pt}{2.5ex} $K_\mathrm{v}$ (in MPa$\,$s$^{1/n}$) & $n$ & $K$ & $d$ (in nm) & $h^{\alpha\beta}$ & $a^{\alpha\beta}$ & $C_\mathrm{D}$ \\ \cline{2-8} 
		& \rule{0pt}{2.5ex} 10 & 10 & 1 & 10 & $\delta^{\alpha\beta}$ & $\delta^{\alpha\beta}$ & 100  \\ 
	\end{tabular}
\end{table}

The \MB{motion} of grain boundaries in the coupled model has two main sources, curvature as well as stored dislocation energy difference \MB{between neighboring grains}. Currently, there is no available asymptotic procedure to include both effects, but it is possible to estimate them with simple numerical simulations. In this work, we focus on qualitative examples to show grain nucleation and grain boundary migration mechanisms, instead of fitting to experimental mobility data. If desired, the mobility data in \cite{vandermeer1997grain} for pure copper can be used to \MB{adjust the mobility to the desired corresponding temperature}. The inverse mobility $\tau_\eta$ is chosen to obtain a reasonable time scale for recrystallization, while $\tau_*$ is limited by the choice of $\tau_\eta$ \citep{tandogan2025multi}. During the mechanical loading phase, $\tau_\eta$ is chosen as $10^2f_0t_0$, whereas in the recrystallization phase it is $10^4f_0t_0$ with $t_0=1$, \MB{in order to emulate a temperature dependency.} Accordingly, $\hat{\tau}_*$ is $10^2f_0t_0$ and $10^1f_0t_0$ during loading and recrystallization, respectively. 

\MB{The reasoning for the two different values of $\tau_\eta$ is that, as a simple constant inverse mobility coefficient, it is not able to cover the separate physics and different time scales of both mechanical loading and heat treatment phases at the same time. Lattice orientation $\theta$ evolves in both phases, where during deformation it changes very rapidly in several seconds, while during heat treatment it changes slowly in the span of several hours. The $\tau_\eta$ that is inherited from the HMP orientation phase field is physically meaningful for the heat treatment phase, where it corresponds to a temperature, and we choose the value $10^4f_0t_0$ according to that. If the same value is used during the deformation phase, the evolution of the order parameter $\eta$ cannot keep up with the rapid changes in orientation $\theta$, and does not represent the deformed microstructure correctly. Therefore, it is necessary make $\tau_\eta$ process dependent. We assume}

\begin{equation}
    \tau_\eta\coloneqq\tau_\eta(\overline{\dot{\theta}}),
\end{equation}
\MB{such that in the presence of rapid reorientation, i.e. during loading, $\tau_\eta$ takes the value $10^2f_0t_0$, where $\overline{(\cdot)}$ is the domain average.}

\subsection{Dislocation driven grain nucleation and grain boundary migration}\label{ssec:disnucmig}

\begin{figure}[ht]
	\centering
	\includegraphics[width=0.8\textwidth]{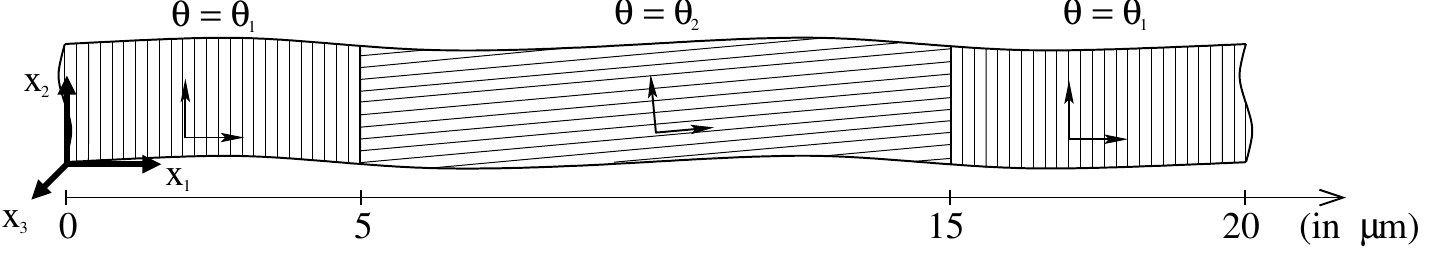}\vspace*{-3mm}
	\caption{Periodic bicrystal structure with variation in $x_1$ direction.}
	\label{fig:bicrystal}
\end{figure}

In the following simulations, the 2D periodic bicrystal structure shown in Fig. \ref{fig:bicrystal} is considered. The grains are 10\,\textmu m wide and have alternating orientations of $\theta_1$ and $\theta_2$. \MB{Periodic boundary conditions are applied on the solution variables at opposite points of the surface.} The displacement $\ubar{u}$ is decomposed into a mean displacement field $\tend{B}\cdot\ubar{x}$ and periodic fluctuations $\ubar{v}$ as,

\begin{equation}
	\underline{\bm{u}}=\tend{B}\cdot\underline{\bm{x}}+\underline{\bm{v}}, \qquad\text{with}\qquad \tend{B}=
	\begin{bmatrix}
		0 & B_{12} & 0\;\;\\
		B_{21} & 0 & 0\\
		0 & 0 & 0
	\end{bmatrix},\label{eqn:Bx}
\end{equation}
where $\ubar{x}$ is the position measured relative to lower left corner. The fluctuations $\ubar{v}$ are set to zero at the corners, and $B_{12}$ or $B_{21}$ is used to apply shear loading. Due to boundary conditions, the solution field is constant along $x_2$; \MB{hence, the extension of the domain} along $x_2$ is only for visualization purposes. The domain is discretized into 400 \MB{blocks in $x_1$ direction, each with two} second order triangular elements with reduced integration.

In order to use the coupled model, first it is necessary to initialize the phase field. A detailed investigation of this process was presented by \cite{tandogan2025multi}, showing the effects of the model parameters such as the mobilities $\tau_\eta$ and $\tau_*$, and the Cosserat coupling modulus $\mu_\mathrm{c}$. Here, the focus is on plastic deformation and the new nucleation mechanism. The order parameter is initialized with $\eta_0=0.99$. The eigenrotation is set to $\pvec{e}{*}\hspace{-1mm}_0=-\ubar{\Theta}_0$ with $\ubar{\Theta}_0=\left[0\;0\;\theta_0\right]^\text{T}$, so that stresses in the undeformed \MB{state} are zero. Then, the fields are relaxed until equilibrium is reached, where $\pvec{e}{*}$ evolves according to \eqref{eqn:wstar}, following the changes of $\theta$ inside the grain boundary.

\begin{figure}[ht!]
	\centering\vspace*{-4mm}
	\includegraphics[width=0.9\textwidth]{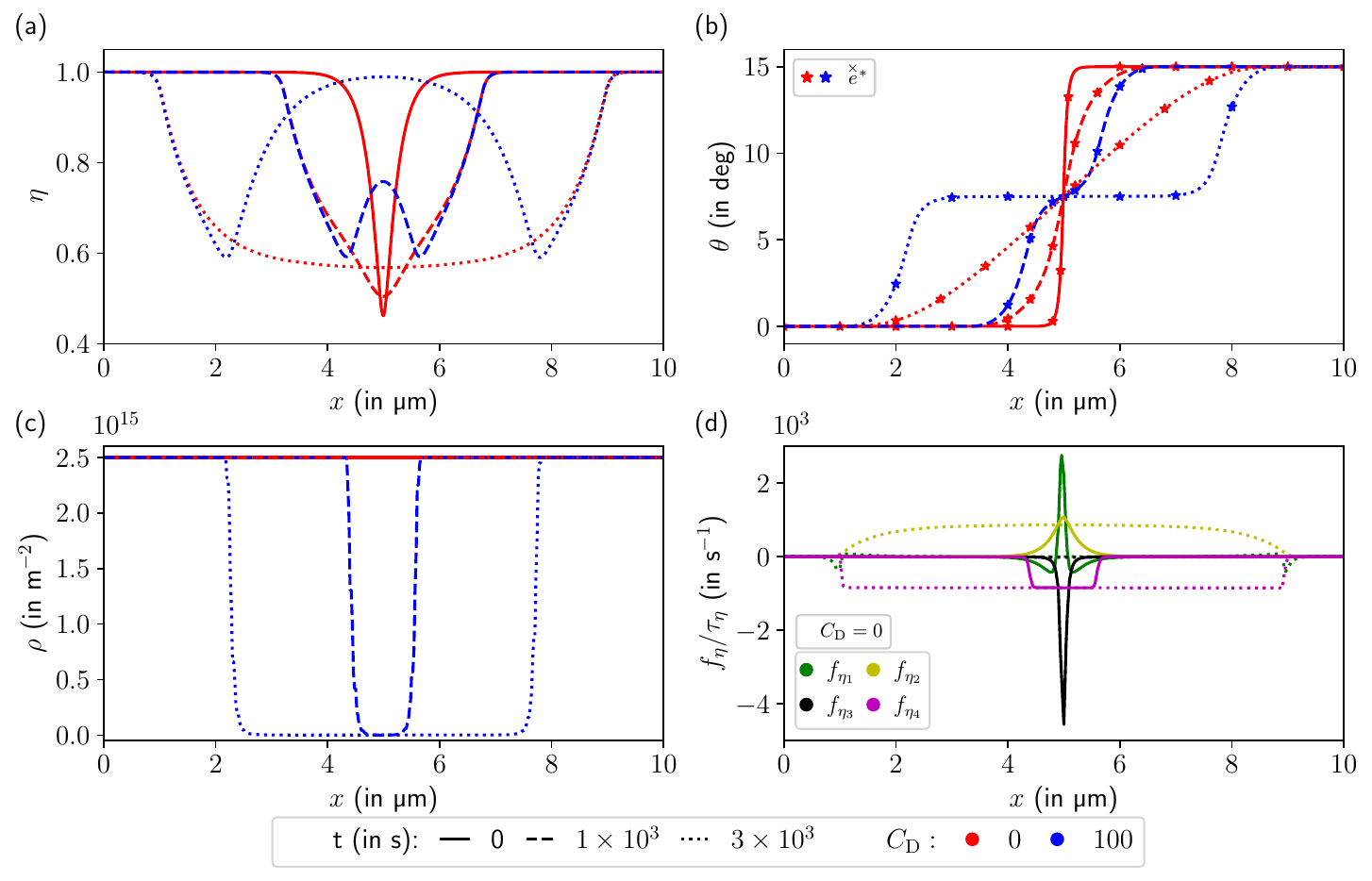}\vspace*{-4mm}
	\caption{Order parameter $\eta$ (a), orientation $\theta$ (b) and SSD density $\rho$ (c) with SSD recovery activated ($C_\mathrm{D}=100$) or deactivated ($C_\mathrm{D}=0$) at times 0, $1\times 10^3$ and $3\times 10^3$\;s. \MB{The profiles are the same at $t=0$\;s.} The driving terms for $\dot{\eta}$ on the right hand side of equation \eqref{eqn:hmpetadot} (d) for $C_\mathrm{D}=0$. Parameters of $\phi(\eta)$ are $c_1=100$, $c_2=0.95$ and $c_3=1.7$.}
	\label{fig:nucCD}
\end{figure}

In order to demonstrate the grain nucleation mechanism we focus on the GB at $x_1=5$\,\textmu m in Fig. \ref{fig:bicrystal}. The initial profiles of $\eta$ and $\theta$ at $t=0$\;s are shown in Fig. \ref{fig:nucCD}a/b. Fig. \ref{fig:nucCD}d shows the \MB{forces} in \eqref{eqn:hmpetadot} that drive the evolution of the order parameter $\eta$, where

\begin{equation}
	f_{\eta_1}=f_0\nu^2\nabla^2\eta, \quad f_{\eta_2}=-f_0\left[\alpha V_{,\eta}\right], \quad f_{\eta_3}=-f_0\left[\mu^2g_{,\eta}|\nabla\theta|^2\right], \quad f_{\eta_4}=-\phi_{,\eta}\frac{\lambda}{2}\mu^{\mathrm{e}}b^2\rho.\label{eqn:etadotterms}
\end{equation}
At $t=0$\;s, we introduce a constant dislocation density of $\rho_0=2.5\times 10^{15}$\;m$^{-2}$ in both grains, assuming that we have a single slip system. Before the introduction of $\rho_0$, we have $f_{\eta_4}=0$, and the forces $f_{\eta_1}$, $f_{\eta_2}$ and $f_{\eta_3}$ are in equilibrium. A non-zero dislocation density creates a \MB{net} driving force on $\eta$ [see \eqref{eqn:etadotterms} and Fig. \ref{fig:nucCD}d] and \MB{changes} the equilibrium profile. $f_{\eta_4}$ is active only inside the grain boundary and not in the bulk, because of the form of $\phi_{,\eta}$ shown in Fig. \ref{fig:gbmigphi}. Its magnitude is controlled by the coefficient $c_3$ and the dislocation density $\rho$. The sharpness and width of the $f_{\eta_4}$ profile in Fig. \ref{fig:nucCD}d at $t=0$\;s is controlled by the coefficients $c_1$ and $c_2$ in equation \eqref{eqn:phiderlog}, respectively. When the magnitude of $f_{\eta_4}$ is small, i.e., the dislocation density $\rho$ is small, then the change in the $\eta$ profile is minimal since $f_{\eta_4}$ can be compensated by the other terms. However, if $\rho$ is \MB{sufficiently large}, the grain boundary starts to evolve. Two distinct evolution \MB{patterns} are observed, depending on whether or not the dislocation recovery mechanism in \eqref{eqn:rhodot} is active. 

\subsubsection{\MB{No dislocation recovery}}

\MB{In this case} $C_\mathrm{D}=0$ in equation \eqref{eqn:rhodot}. After $\rho$ is introduced, due to \MB{the driving force} $f_{\eta_4}$, \MB{the value of} $\eta$ decreases in the outer region of the diffuse grain boundary, as can be predicted from Fig. \ref{fig:nucCD}d, resulting in the widening of the $\eta$ profile at the GB (Fig \ref{fig:nucCD}a). The orientation $\theta$ follows this change and its gradient $|\nabla\theta|$ decreases (Fig \ref{fig:nucCD}b) resulting in a \MB{further diffused profile}. As the grain boundary expands, at $t=3\times 10^3$\;s, only the terms $f_{\eta_2}$ and $f_{\eta_4}$ in Fig \ref{fig:nucCD}d remain relevant, since the gradients are so small ($f_{\eta_1}, f_{\eta_3}\approx 0$). Hence, the equilibrium value of $\eta$ inside the GB at this state can be found from,

\begin{equation}\label{eqn:etaeq}
	\eta^\text{eq}=1-\dfrac{\phi_{,\eta}(\eta^\text{eq})\frac{\lambda}{2}\mu^{\mathrm{e}}b^2\rho}{f_0\alpha},
\end{equation}
where $\phi_{,\eta}(\eta^\text{eq})=c_3$ according to \eqref{eqn:phiderlog}.

This state of expanding grain boundary constitutes the first part of the grain nucleation mechanism, where we obtain a non-localized uniform orientation gradient similar to \cite{ghiglione2024cosserat}. The non-localized orientation gradient is an unstable state for the KWC and HMP orientation phase field models, since their energy is minimized for a localized GB solution. The new formulation of $\phi(\eta)$ creates such an unstable state in the presence of SSDs at GBs, whereas \cite{ghiglione2024cosserat} created it by \MB{external torsional loading}. 

\subsubsection{\MB{With dislocation recovery}}

The final piece of the puzzle for nucleation is provided by the dislocation recovery mechanism in equation \eqref{eqn:rhodot}, which is activated when $\dot{\eta}>0$ and $C_\mathrm{D}$ is non-zero. This recovery mechanism was originally proposed by \cite{abrivard2012aphase} to model recovery of dislocations in the wake of a moving grain boundary, hence the $\dot{\eta}>0$ condition. Combining the expansion of the grain boundary due to dislocations and the recovery mechanism results in the dislocation driven spontaneous nucleation of a new grain. For $C_\mathrm{D}=100$, Fig. \ref{fig:nucCD}a/b shows the nucleation of a new grain, where its orientation is the average \MB{of the orientations} of the two neighboring grains. As the new grain nucleates, dislocations are recovered at the nucleation site; \MB{further recovery takes place during grain growth} in the wake of the \MB{moving} boundaries, as seen in Fig. \ref{fig:nucCD}c. In the end, we obtain a \MB{dislocation-free} new grain. From a physical perspective, this means that the stored energy of dislocations is \MB{expended} to nucleate and grow the new grain. 
%In the model what happens is that the value of $\eta_\text{eq}$ (see \eqref{eqn:etaeq}) in the state of expanding grain boundary with $C_\mathrm{D}=0$ is higher than the initial $\eta_{min}$ at the center of the GB at t=0 before dislocations are introduced (see Fig. \ref{fig:nucCD}a). Therefore, for the case with $C_\mathrm{D}=100$, since the order parameter $\eta$ at the center of the GB increases (Fig \ref{fig:nucCD}a), the recovery mechanism is activated. 
\MB{Dislocation recovery is activated, since the equilibrium value for $\eta$ given by Eq. \eqref{eqn:etaeq} inside the GB in the presence of dislocations is higher compared to the initial equilibrium value at $t=0$\;s, i.e., the introduction of a dislocation density $\rho>0$ inside the GB leads to $\dot{\eta}>0$, which activates the recovery term in Eq. \eqref{eqn:rhodot}}.
This reduces $\rho$ (Fig \ref{fig:nucCD}c) and thus the magnitude of $f_{\eta_4}$, which in turn causes $\eta$ to increase even more \MB{resulting in a self-perpetuating process}. This continues until $\rho$ reduces to 0 and $\eta$ increases to 1. As a result, two separate grain boundaries are created from the single expanding boundary, which move away from each other, resembling nucleation by SIBM at grain boundaries. As $\eta$ goes to 1 in the nucleus, $|\nabla\theta|$ reduces to 0 (Fig \ref{fig:nucCD}b), and a stable bulk \MB{crystal is formed}. Throughout this process the eigenrotation $\accentset{\times}{e}^*$ evolves together with $\theta$ (see Fig. \ref{fig:nucCD}b), keeping the new grain stress free. This \MB{completes} the basic mechanism of nucleation in the model. 

\subsubsection{\MB{The effects of misorientation, dislocation distribution and GB velocity}}

\begin{figure}[ht!]
	\centering\vspace*{-4mm}
	\includegraphics[width=0.9\textwidth]{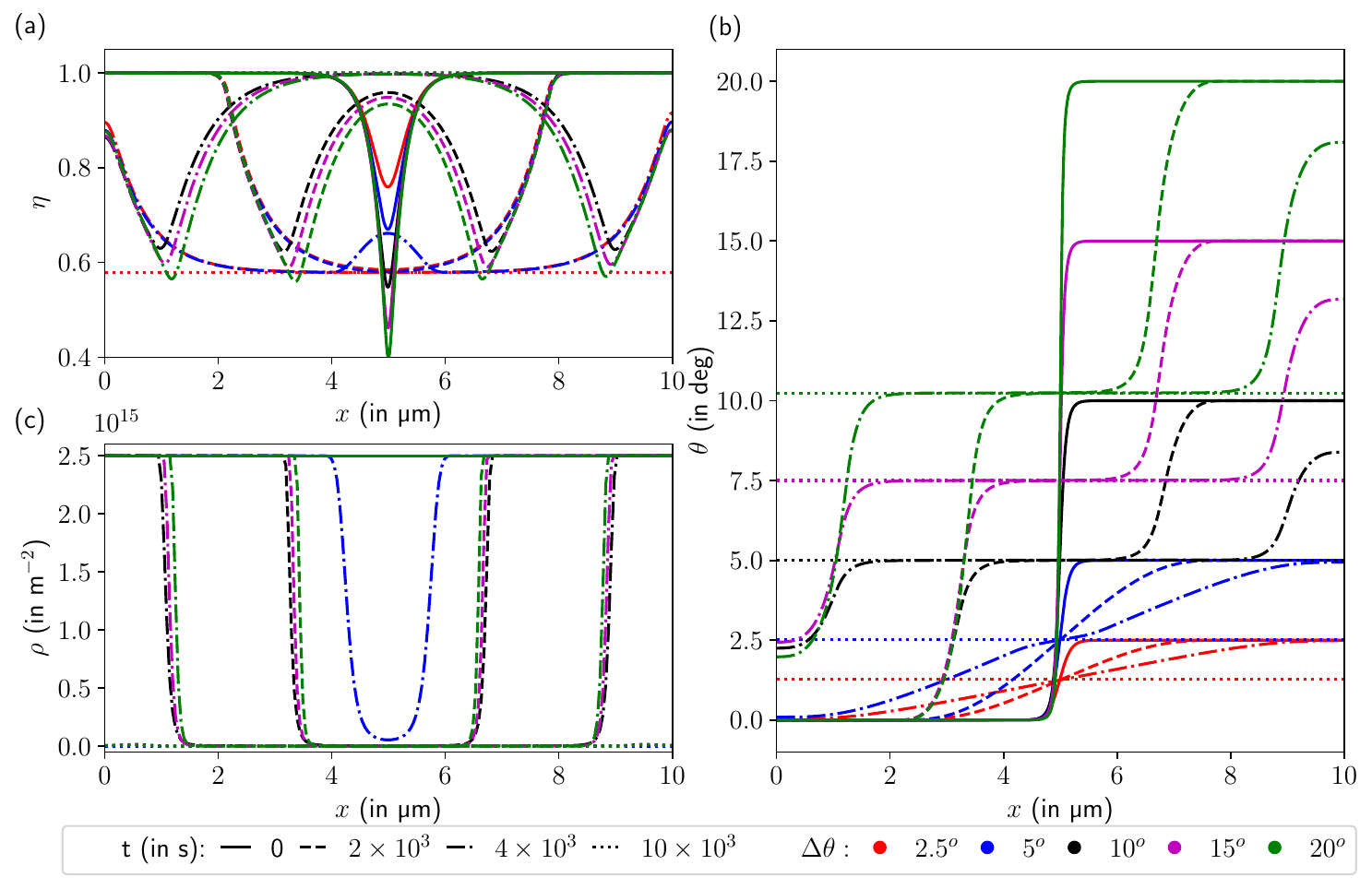}\vspace*{-4mm}
	\caption{Order parameter $\eta$ (a), orientation $\theta$ (b) and SSD density $\rho$ (c) with $C_\mathrm{D}=100$ plotted at different times for varying misorientations. Parameters of $\phi(\eta)$ are $c_1=100$, $c_2=0.95$ and $c_3=1.7$.}
	\label{fig:nucdtheta}
\end{figure}

In Fig. \ref{fig:nucdtheta}, the effect of misorientation $\Delta\theta$ on the nucleation behavior is shown, where $\Delta\theta$ varies between $2.5^{\degree}-20^{\degree}$. In the model, the degree of misorientation \MB{between neighboring grains} determines the minimum value of the order parameter at the grain boundaries. As misorientation increases, we obtain a deeper well in the equilibrium profile of $\eta$ as seen in Fig. \ref{fig:nucdtheta}a for $t=0$\;s. This fact directly affects the nucleation behavior, because, as discussed previously, the nucleation and the recovery of $\rho$ are triggered if $\eta$ initially increases inside the GB, where its equilibrium value $\eta_\text{eq}$ is determined by \eqref{eqn:etaeq} (cf. the dotted red line in Fig \ref{fig:nucdtheta}a). This is true for the cases with $\Delta\theta=$ 10$^{\degree}$, 15$^{\degree}$, 20$^{\degree}$ (Fig \ref{fig:nucdtheta}a). However, for the cases with $\Delta\theta=$ 2.5$^{\degree}$, 5$^{\degree}$, the initial \MB{value of} $\eta$ at the center of the GB is already higher than $\eta_\text{eq}$, \MB{causing it to} decrease instead and nucleation is not triggered immediately. For $\Delta\theta=5^{\degree}$, nucleation is still triggered at $t=4000$\;s after the grain boundary expands significantly, \MB{as seen by the recovery at the nucleation site ($x=5$\,\textmu m) in Fig. \ref{fig:nucdtheta}c. Nucleation can also be recognized by the bulge of order parameter $\eta$ forming at the same position in Fig. \ref{fig:nucdtheta}a, and the corresponding crystalline bulk, i.e., $\nabla\theta=0$, region forming in Fig. \ref{fig:nucdtheta}b}. For $\Delta\theta=2.5^{\degree}$ 
\MB{dislocation recovery is never triggered. Due to the presence of dislocations and the resulting $f_{\eta_4}>0$, the initially localized gradient in lattice orientation $\theta$ becomes more and more diffuse until it disappears (see Fig. \ref{fig:nucdtheta}b). Dislocation density $\rho$ keeps its initial value since it is not recovered, and order parameter $\eta$ becomes a constant $\eta_\text{eq}=0.579$ (see Fig \ref{fig:nucdtheta}a), which is determined by Eq. \eqref{eqn:etaeq}.}
%it never happens, and instead the GB keeps expanding until it meets the other side, after which $\theta$ becomes constant where its value is again the average of the neighboring grains (Fig \ref{fig:nucdtheta}b). 
%$\eta$ becomes a constant $\eta_\text{eq}$ everywhere with a value less than 1 (Fig \ref{fig:nucdtheta}b), which is determined by equation \eqref{eqn:etaeq}, while $\rho$ keeps its initial value (Fig \ref{fig:nucdtheta}c). 
Consequently, we can state that there is a threshold value \MB{$5^{\degree}<\Delta\theta_\mathrm{T}<10^{\degree}$} above which the nucleation is \MB{easily} triggered, and below which \MB{it may trigger after the GB widens significantly, ($\Delta\theta=$5$^{\degree}$), or does not trigger at all ($\Delta\theta=$2.5$^{\degree}$}). \MB{This threshold can be controlled by adjusting the values of $\alpha$ and $c_3$.}
%The \MB{cases where the nucleation is not triggered properly} can be prevented by \MB{increasing} the value of $\eta_\text{eq}$ in Eq. \eqref{eqn:etaeq}, for example by increasing $\alpha$ or decreasing $c_3$, \MB{making sure that for the given $\Delta\theta$ the corresponding initial $\eta$ at the GB increases and activates dislocation recovery}. 
%The value of $\rho$ has the same effect as $c_3$, and this issue becomes irrelevant as $\rho$ decreases. Here we consider the extreme case where $\rho$ has reached its saturation value according to Kocks-Mecking-Teodosiu law in \eqref{eqn:rhodot}, which corresponds to $2.5\times 10^{15}$ for the parameters in Table \ref{tab:param}. 
The parameter $c_3$ can also be used to scale the magnitude of $\rho$ that triggers the nucleation mechanism. We have chosen the value of $c_3=1.7$ in these examples by considering the saturation value of $\rho$ according to Kocks-Mecking-Teodosiu law in Eq. \eqref{eqn:rhodot}, which corresponds to $2.5\times 10^{15}$\;m$^{-2}$ for the parameters in Table \ref{tab:param}.

\begin{figure}[ht!]
	\centering\vspace*{-4mm}
	\includegraphics[width=0.9\textwidth]{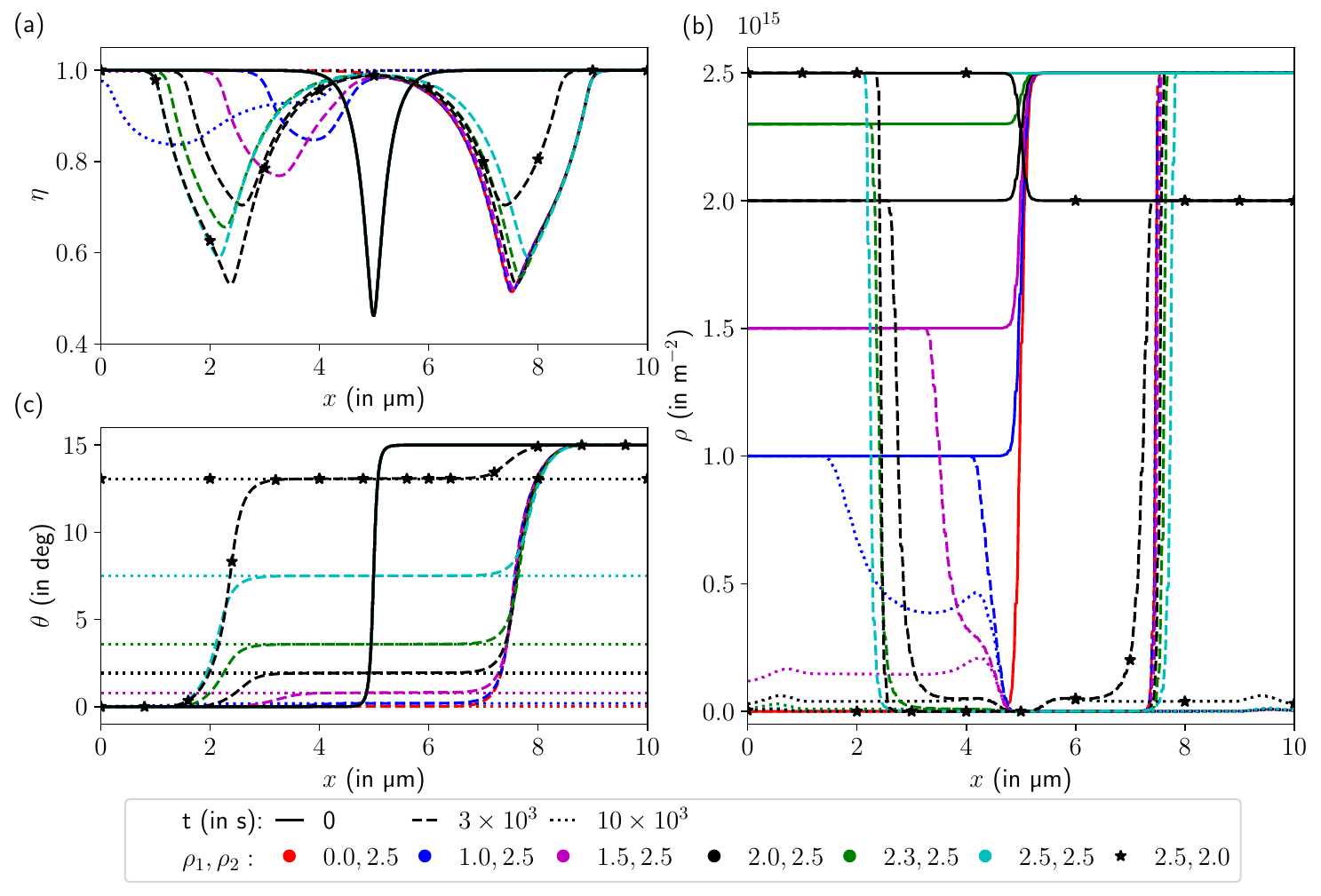}\vspace*{-4mm}
	\caption{Order parameter $\eta$ (a), SSD density $\rho$ (b) and orientation $\theta$ (c) with $C_\mathrm{D}=100$ plotted at different times for varying SSD distributions around grain boundary. Parameters of $\phi(\eta)$ are $c_1=100$, $c_2=0.95$ and $c_3=1.7$.}
	\label{fig:nucssd}
\end{figure}

So far, we have only considered a constant distribution of dislocation density \MB{$\rho=2.5\times 10^{15}$\;m$^{-2}$ throughout the domain}. In Fig. \ref{fig:nucssd}, we explore cases, where $\rho$ \MB{takes different values in different grains}  (Fig. \ref{fig:nucssd}b). 
\MB{We investigate cases ranging from $\rho_1=2.5\times 10^{15}$\;m$^{-2}$; $\rho_2=2.5\times 10^{15}$\;m$^{-2}$, i.e.,the constant distribution, to $\rho_1=0$; $\rho_2=2.5\times 10^{15}$\;m$^{-2}$, i.e., one grain is dislocation-free and the other has maximum value. In the latter case, nucleation is not possible since the GB is under a driving force to only one side, hence it represents a pure SSD driven GB migration.}
%The extreme cases are $\rho_1=0$; $\rho_2=2.5\times 10^{15}$ and $\rho_1=2.5\times 10^{15}$; $\rho_2=2.5\times 10^{15}$, where the latter is the same as the previous examples. The former corresponds to a pure dislocation driven grain boundary migration example, where the GB moves to reduce the SSD energy similar to the examples in \cite{tandogan2025multi}, however the profile of $\eta$ is different due to the new function $\phi(\eta)$ (Fig. \ref{fig:nucssd}a). 
For the cases in-between \MB{nucleation is possible, and} we have an interesting result. If the nucleation is triggered, the orientation of the nucleus is determined by the ratio of the \MB{values of the} dislocation \MB{density on both sides of} the grain boundary (Fig. \ref{fig:nucssd}c); it is not always the average orientation of the neighboring grains. Let us define the smaller dislocation density as $\rho_\mathrm{s}$ and the larger one as $\rho_\mathrm{l}$. The corresponding orientations of the grains are $\theta_\mathrm{s}$ and $\theta_\mathrm{l}$, respectively, and that of the nucleus is $\theta_{\text{nuc}}$. Then, we \MB{observe that $\theta_{\text{nuc}}$ is determined by}

\begin{equation}
	\label{eqn:nucrules}
	\begin{cases}
		\theta_{\text{nuc}}=\frac{\theta_\mathrm{s}+\theta_\mathrm{l}}{2} &\text{if } \rho_\mathrm{s}=\rho_\mathrm{l},\\
		\theta_\mathrm{s}<\theta_{\text{nuc}}<\frac{\theta_\mathrm{s}+\theta_\mathrm{l}}{2} &\text{if } \rho_\mathrm{s}<\rho_\mathrm{l},\\
		\theta_\mathrm{s}<\theta_{\text{nuc}}<<\frac{\theta_\mathrm{s}+\theta_\mathrm{l}}{2} &\text{if } \rho_\mathrm{s}<<\rho_\mathrm{l}.
	\end{cases}
\end{equation}
The extreme case is $\theta_{\text{nuc}}=\theta_\mathrm{s}$ for $\rho_\mathrm{s}=0$, which is effectively migration of the GB \MB{into the grain with the larger dislocation density}. Otherwise, the nucleus prefers an orientation that is closer to the orientation of the grain with \MB{lower stored} energy. 
%and the actual value depends on the ratio $\rho_\mathrm{s}/\rho_\mathrm{l}$. 
This is \MB{illustrated} when we compare the cases $\rho_1=2\times 10^{15}$\;m$^{-2}$; $\rho_2=2.5\times 10^{15}$\;m$^{-2}$ and $\rho_1=2.5\times 10^{15}$\;m$^{-2}$; $\rho_2=2\times 10^{15}$\;m$^{-2}$ in Fig. \ref{fig:nucssd}c, where $\theta_{\text{nuc}}$ is mirrored about the average value. Another interesting case is $\rho_1=1\times 10^{15}$\;m$^{-2}$; $\rho_2=2.5\times 10^{15}$\;m$^{-2}$, where the new grain nucleates with a very small misorientation compared to $\theta_\mathrm{s}$. At $t=10000$\;s, the nucleated grains fully expand into their neighbors (Fig. \ref{fig:nucssd}c), and the dislocation density is mostly recovered. 
%which gets closer to full recovery if initial $\rho_0$ is higher (Fig. \ref{fig:nucssd}b). 
The amount of recovery can be controlled by adjusting the value of $C_\mathrm{D}=100$.

\begin{figure}[ht!]
	\centering\vspace*{-4mm}
	\includegraphics[width=1\textwidth]{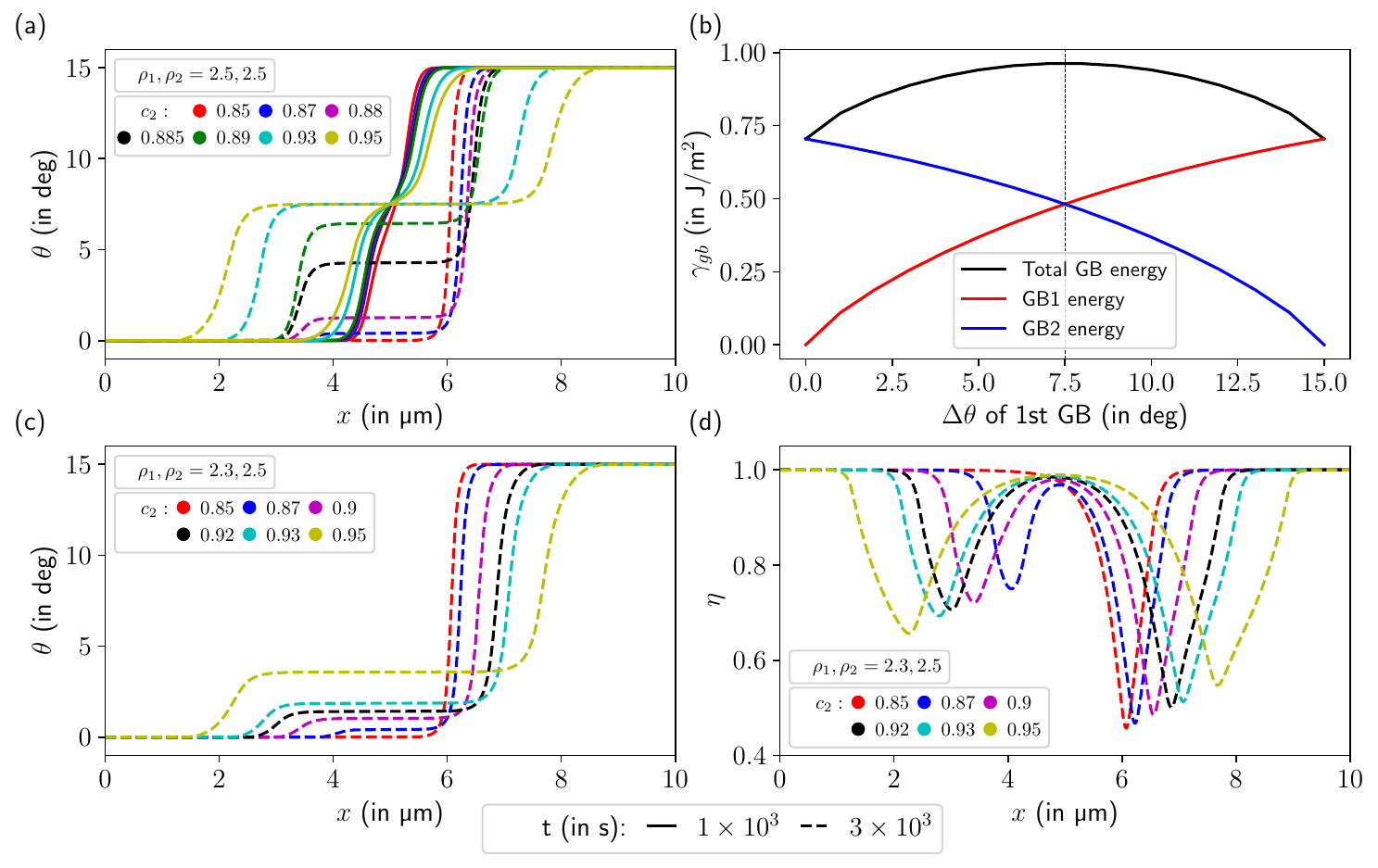}\vspace*{-4mm}
	\caption{Orientation $\theta$ for $\rho_1/\rho_2=1$ distribution (a) and energy distribution when a 15$^{\degree}$ grain boundary divides into two grain boundaries (b). Orientation $\theta$ for $\rho_1/\rho_2=2.3/2.5$ distribution (c) and order parameter $\eta$ (d). Parameters of $\phi(\eta)$ are $c_1=100$ and $c_3=1.7$.}
	\label{fig:nucc2}
\end{figure}

While these observations physically make sense, it is not intuitive, why a single grain boundary would \MB{split} into two, since it is energetically unfavorable, \MB{if only} the grain boundary energy \MB{is considered}. Indeed, as shown in Fig. \ref{fig:nucc2}b, when we \MB{split} a GB with 15$^{\degree}$ misorientation into two GBs with $\Delta\theta_1$, $\Delta\theta_2$ and $\Delta\theta_1+\Delta\theta_2=15^{\degree}$, the total grain boundary energy increases (cf. Fig. \ref{fig:gben}). 
%So, how is this possible? The answer is obtained by examining Fig. \ref{fig:nucc2}, where the cases with different $c_2$ values of function $\phi_{,\eta}$ are compared. 
\MB{In order to understand, why new grains are formed under these circumstances, it is helpful to consider the effect of different values of the parameter $c_2$ [cf. Eq. \eqref{eqn:phiderlog}] as shown in Fig. \ref{fig:nucc2}.}
In Fig. \ref{fig:gbmigphi} and Fig. \ref{fig:nucCD} \MB{we showed} that $c_2$ controls the width of \MB{the distribution of the} driving force $f_{\eta_4}$ due to dislocations. As a consequence, the higher the value of $c_2$, the higher the GB velocity due to stored dislocations (see Fig. \ref{fig:nucc2}a/c). Fig. \ref{fig:nucc2}d shows that, in addition to velocity, the profile of $\eta$ is significantly affected by the value of $c_2$. In Fig. \ref{fig:nucc2}a with $\rho_1/\rho_2=1$ at $t=1\times 10^3$\;s, we see that in all of these cases nuclei with the same average $\theta_{\text{nuc}}$ \MB{are formed}. \MB{Later} the orientation of the new grain $\theta_\text{new}$ may stabilize at a value between $\theta_\mathrm{s}=0$ and $\theta_{\text{nuc}}$, with the final value depending on $c_2$. If $c_2$ is not \MB{sufficiently large}, the nuclei rotate towards $\theta_\mathrm{s}$ and disappear as in the case with $c_2=0.85$. Fig. \ref{fig:nucc2}c shows similar behavior for $\rho_1=2.3\times 10^{15}$\;m$^{-2}$; $\rho_2=2.5\times 10^{15}$\;m$^{-2}$, where the nucleus initially has an orientation $\theta_{\text{nuc}}$ closer to $\theta_\mathrm{s}$. Based on these observations, we expand the set of rules \eqref{eqn:nucrules}:

\begin{enumerate}
	\item Nucleation is triggered due to stored dislocation energy at a GB, \MB{that splits} into two GBs. While this increases the GB energy, the total energy is decreased since dislocation density is recovered at the nucleation site. The grain is nucleated with $\theta_{\text{nuc}}$ according to \eqref{eqn:nucrules}, \MB{and subsequently expands} to reduce $\rho$ \MB{and the associated stored energy}. 
	\item If the nucleus \MB{grows sufficiently} fast, $\theta_\text{new}$ stabilizes at $\theta_{\text{nuc}}$, since the order parameter $\eta$ reaches 1 in the \MB{newly formed grain}, and \MB{lattice} rotation is prevented by the singular function $g(\eta)$. \MB{The lack of} long range interactions \MB{between} GBs \MB{in the HMP model helps to} quickly stabilize the new grain.
	\item If the GBs do not move fast enough, then the nucleus starts to rotate \MB{in order to} to minimize the energy \MB{$\gamma_\text{gb}$} following Fig. \ref{fig:nucc2}b. Depending on the velocity, it may stabilize at a value $\theta_\mathrm{s}<\theta_\text{new}<\theta_{\text{nuc}}$, or fully rotate towards $\theta_\mathrm{s}$, \MB{causing} the nucleus to disappear before expanding.
\end{enumerate}

\begin{figure}[ht!]
	\centering
	\vspace*{-4mm}
	\includegraphics[width=1\textwidth]{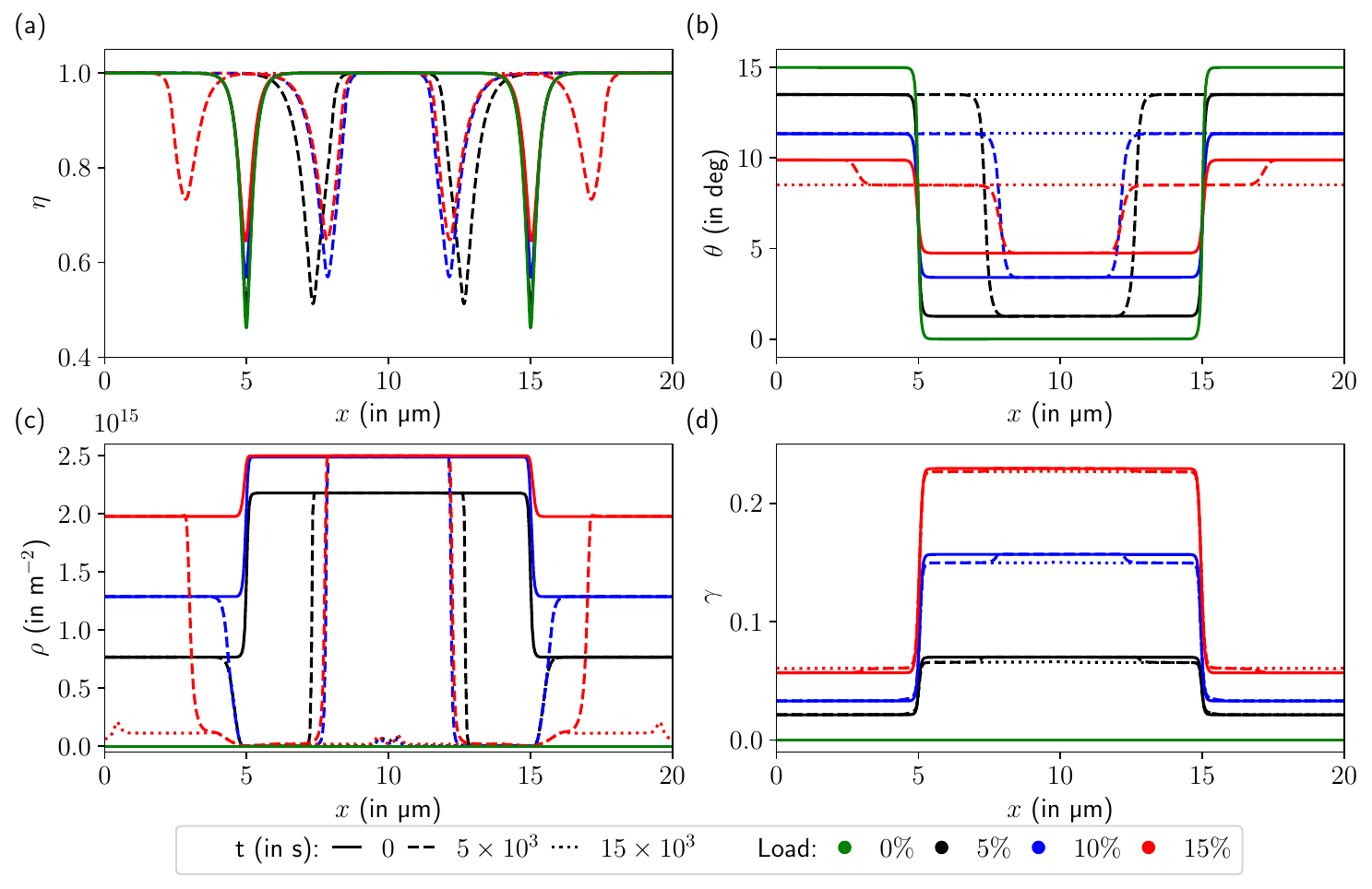}\vspace*{-4mm}
	\caption{Order parameter $\eta$ (a), orientation $\theta$ (b) SSD density $\rho$ (c) and viscoplastic slip $\gamma$ (d). Specimen is deformed by different amounts before applying heat treatment. Parameters of $\phi(\eta)$ are $c_1=100$, $c_2=0.9$ and $c_3=1.7$.}
	\label{fig:nucload}
\end{figure}

\MB{\subsection{Plastic deformation and microstructure evolution}}

The above rules summarize the \MB{observed} grain nucleation mechanism \MB{in our model}. In the next examples we explore the full potential of the HMP-CCP coupling model \MB{by plastically deforming} periodic bicrystal and polycrystal specimens in order to \MB{predict} deformation driven grain boundary migration and grain nucleation. We define four possible slip systems rotated by 0$^{\degree}$, 90$^{\degree}$, 45$^{\degree}$ and 135$^{\degree}$ with respect to \MB{the crystal frame}. Corresponding slip $(\ubar{l}^\alpha)$ and normal $(\ubar{n}^\alpha)$ directions are given by
\begin{alignat*}{4}
	&\ubar{l}^1=(1,0),\quad &&\ubar{l}^2=(0,1),\quad &&\ubar{l}^3=(1/\sqrt{2},1/\sqrt{2}),\quad &&\ubar{l}^4=(-1/\sqrt{2},1/\sqrt{2}), \\
	&\ubar{n}^1=(0,1),\quad &&\ubar{n}^2=(-1,0),\quad &&\ubar{n}^3=(-1/\sqrt{2},1/\sqrt{2}),\quad &&\ubar{n}^4=(-1/\sqrt{2},-1/\sqrt{2}).
\end{alignat*}

\MB{\subsubsection{Bi-crystal}}

For the bicrystal example in Fig. \ref{fig:bicrystal}, we assume that only slip system 1 is active, and it is initialized with $\rho=10^{11}$\,m$^{-2}$. Orientations of the grains are $\theta_1=15^{\degree}$ and $\theta_2=0^{\degree}$ initially. After the initialization of the phase field, the bicrystal is loaded with $B_{12}=0.05$, $B_{12}=0.10$ or $B_{12}=0.15$ \MB{[cf. Eq. \eqref{eqn:Bx}]} with a rate of 0.01 per second. The last case is \MB{technically} out of the scope of a small deformation setting, but it converges without issues for this simple example, and we use it for the sake of demonstration. During the loading phase, the dislocation recovery and grain nucleation mechanisms are disabled by setting $C_\mathrm{D}=0$ and $c_2=0$, and a dimensionless inverse mobility of $\overline{\tau}_\eta=10^2$ is used, which allows $\eta$ to conform to the reorientation due to deformation. After the loading phase, a heat treatment phase is applied for $15\times 10^{3}$\;s with $\overline{\tau}_\eta=10^4$, where the microstructure evolves by migration of the grain boundary and nucleation of grains. During the heat treatment phase the displacements $\ubar{u}$ are held constant. The recovery and migration mechanisms are activated by setting $C_\mathrm{D}=100$ and $c_2=0.9$.

\MB{The coupled orientation phase field Cosserat crystal plasticity model has the potential to simulate the coupled physics of mechanical deformation and GB structure evolution at the same time. In this work, we are making some simplifications in order to distinguish and focus on distinct mechanisms. For example, it is possible to simulate dynamic recrystallization if the nucleation and migration mechanisms are kept active during the loading phase; however, we are focusing on the bulk and localized evolution of the lattice orientation $\theta$ caused by the applied deformation, which is possible due to the strong coupling. Similarly, during the heat treatment phase, the migration of the GBs and the corresponding changes in the lattice orientation can induce mechanical deformation in the domain. However, this is omitted by holding displacements $\ubar{u}$ constant in order to clearly show the nucleation and the growth mechanisms. A demonstration comparing the evolution during heat treatment when the structure is free to deform is included for the 32 grain polycrystal in Sec. \ref{ssec:polycrys}. However, the detailed investigation of the complex interaction between mechanisms is left to future work.}

Fig. \ref{fig:nucload}b at $t=0$\;s shows the reorientation of the grains due to plastic deformation at different strain levels, resulting in the decrease of the initial $15^{\degree}$\hspace{-1mm}-misorientation. This is straightforward to simulate with the coupled model, since $\theta$ is a degree of freedom in the Cosserat continuum. Similarly, Fig. \ref{fig:nucload}a at $t=0$\;s shows the decrease in the depth of the well in the order parameter $\eta$ profile due to the decrease in the misorientation. Due to the orientation of the grains relative to the loading direction, the outer grain deforms less, as seen in Fig. \ref{fig:nucload}c and Fig. \ref{fig:nucload}d. In the heat treatment phase, we have two distinct behaviors: For the 5\% and 10\% \MB{pre-strain}, the outer grain expands towards the inner grain (Fig. \ref{fig:nucload}a/b) since the inner grain has a larger energy\MB{-density} due to stored dislocations. Recovery takes place in the wake of the moving grain boundaries (Fig. \ref{fig:nucload}c), and in the end the inner grain is fully absorbed (Fig. \ref{fig:nucload}b at $15\times 10^{3}$\;s). In the case of 10\% pre-strain, the GB has slightly higher velocity since $\rho$ is higher compared to the 5\% case. For 15\% \MB{pre-strain}, the deformation generates enough dislocations to nucleate a new grain (Fig. \ref{fig:nucload}a/b), which fully expands into the neighboring grains. When the GBs move, the change in $\theta$ creates a skew-symmetric stress interpreted as a result of the reshuffling of atoms, which slightly evolves the viscoplastic slip $\gamma$ as seen in Fig. \ref{fig:nucload}d.
\\

\subsubsection{Polycrystalline examples}\label{ssec:polycrys}

In this section, the model is tested with periodic polycrystal specimens containing 6 and 32 grains. In addition to grain nucleation, we can observe other interesting phenomena during deformation, such as subgrain and kink band formation. We use the microstructures presented in \cite{ask2020microstructure} that were generated using Voronoi tessellation, where orientations of the grains are between $0^{\degree}$ and $35^{\degree}$. The specimen with 6 grains has a domain size of 20$\times$20\,\textmu m, while the size of 32 grain specimen is 50$\times$50\,\textmu m. Both are discretized using second order triangular elements with reduced integration. The former has 48496 nodes, while the latter has 195128 nodes with 4 degrees of freedom per node. \MB{Periodic boundary conditions are applied on the opposing points of the surface. The periodic fluctuations $\ubar{v}$ are additionally fixed at the corners.} Shear deformation is applied \MB{with mean deformation gradient} $B_{12}$ while $B_{21}=0$ \MB{in Eq. \eqref{eqn:Bx}}. Dislocation densities are initialized with $\rho^\alpha=2\times 10^{11}$ m$^{-2}$; during deformation $C_\mathrm{D}=0$ and $c_2=0$.

\begin{figure}[ht!]
	\centering
	\begin{subfigure}[t]{0.32\textwidth}
		\centering\includegraphics[width=0.86\textwidth,trim={0 1.2cm 0 1.2cm},clip]{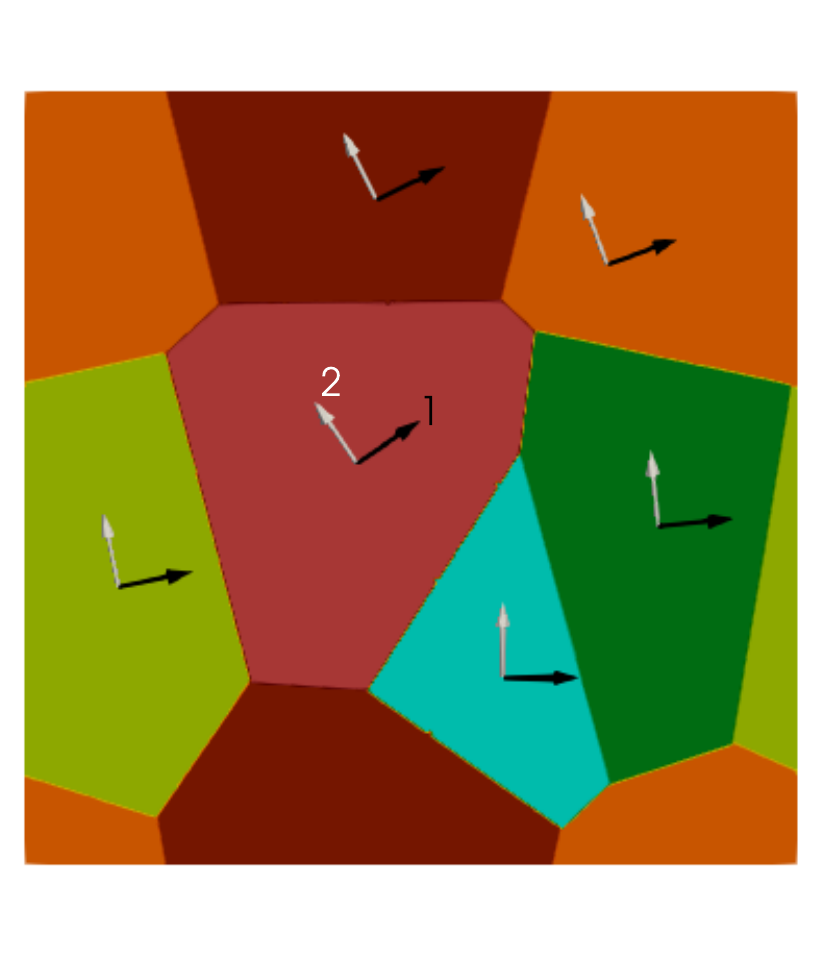}
		\caption{}
	\end{subfigure}
	\begin{subfigure}[t]{0.32\textwidth}
        \centering
        \begin{tikzpicture}
	       \draw(0,0)node[inner sep=0]{\includegraphics[width=0.99\textwidth]{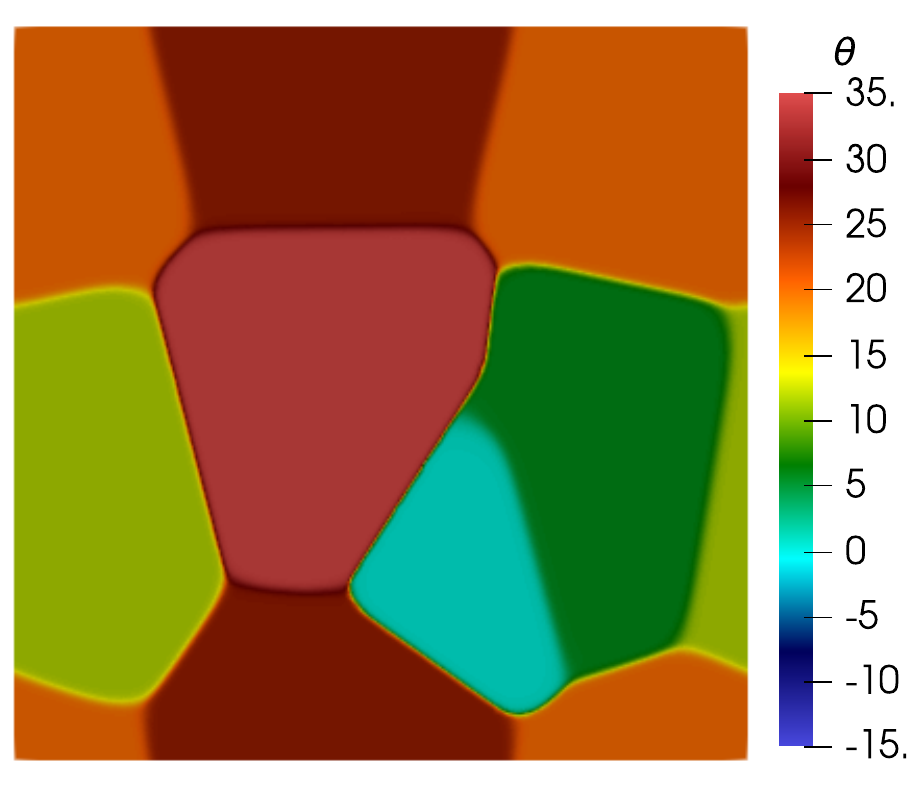}};
          \draw(-2.25,-0.5)node{A}; \draw(1.,0)node{B}; \draw(-0.75,0)node{C}; \draw(0.,-1.)node{D}; \draw(1.,1.5)node{E};
        \end{tikzpicture}
		\caption{}
	\end{subfigure}
	\begin{subfigure}[t]{0.32\textwidth}
		\centering\includegraphics[width=1\textwidth]{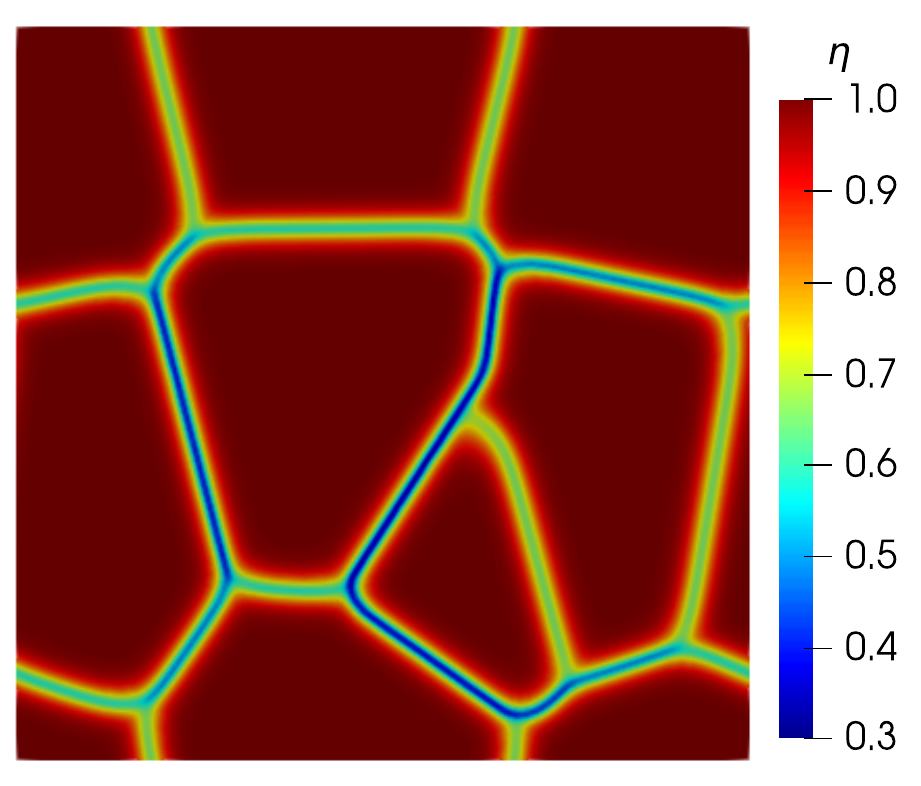}
		\caption{}
	\end{subfigure}
	\vspace*{-3mm}
	\caption{Periodic polycrystal structure with 6 grains and their orientations, where arrows show slip direction of slip system 1 and 2 (a). Fields after the phase field is initialized: orientation $\theta$ (b) and order parameter $\eta$ (c).}
	\label{fig:6grainsinit}
\end{figure}
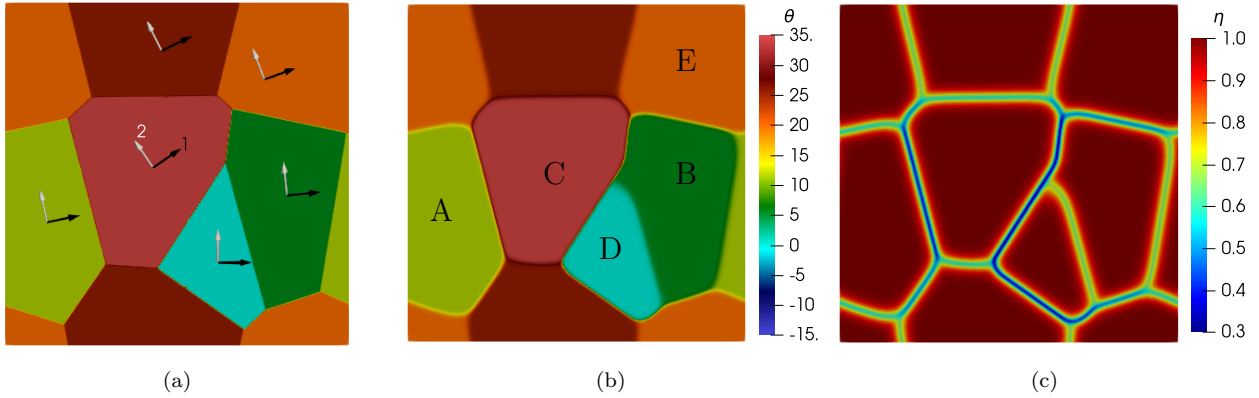

\begin{figure}[ht!]
	\centering
	\vspace*{-2mm}\hspace*{-2mm}
    \begin{tikzpicture}
	\draw(0,0)node[inner sep=0]{\includegraphics[width=0.29\textwidth,trim={0 0 0 0},clip]{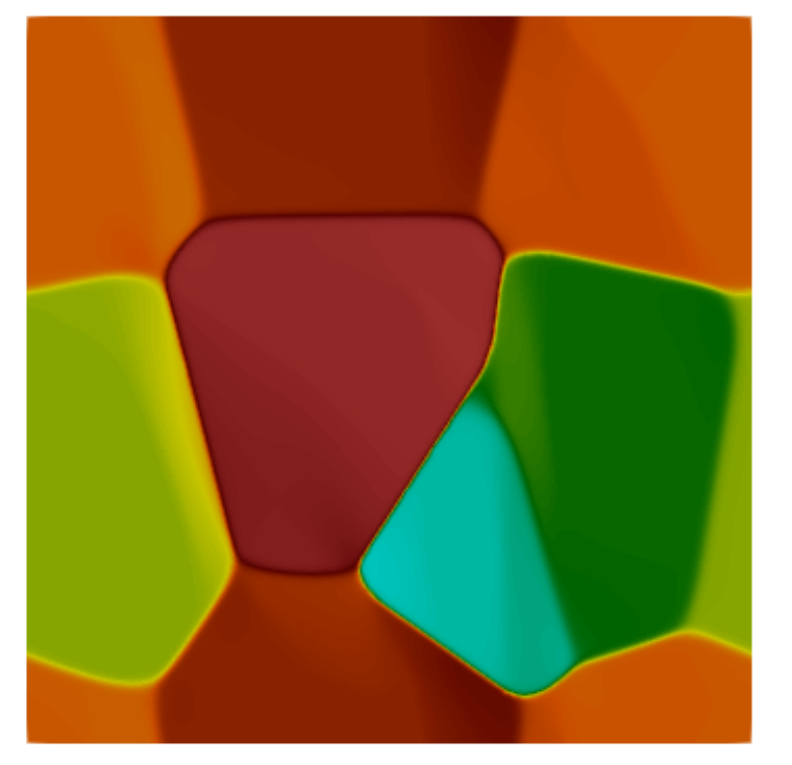}};
    \draw(1.25,-0.5)node{\textcolor{white}{$\Leftarrow$}};
    \end{tikzpicture}
    \begin{tikzpicture}
	\draw(0,0)node[inner sep=0]{\includegraphics[width=0.29\textwidth]{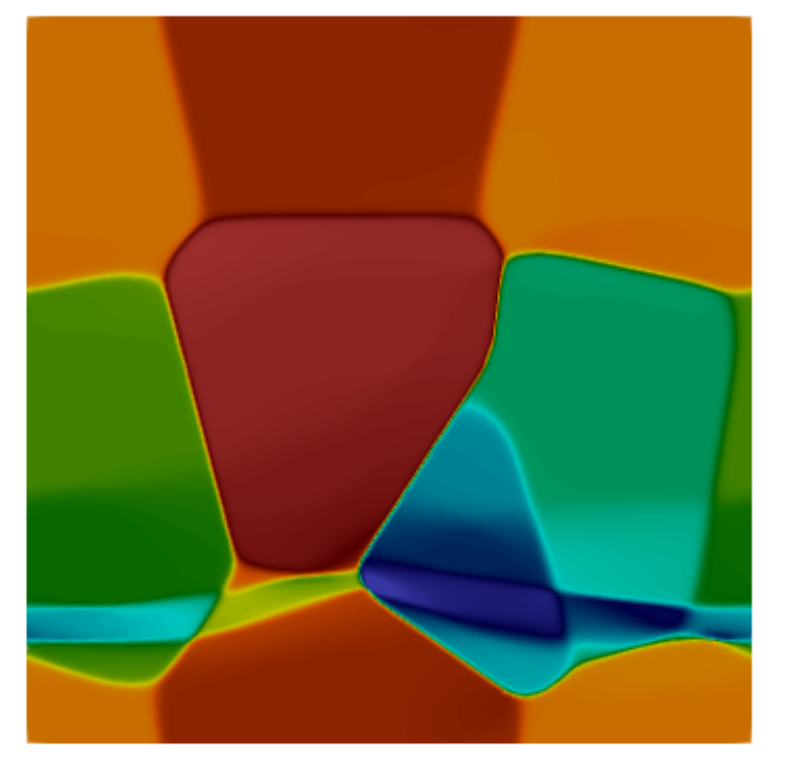}};
    \draw(-2,-0.)node{A$_1$}; \draw(-2,-1.25)node{A$_2$}; \draw(1.5,0)node{B$_1$}; \draw(1.5,-1.25)node{B$_2$};
    \draw(0.5,-1.75)node{\textcolor{white}{$\Uparrow$}}; \draw(-1.25,-1.85)node{\textcolor{white}{$\Uparrow$}};
    \end{tikzpicture}
	\includegraphics[width=0.33\textwidth]{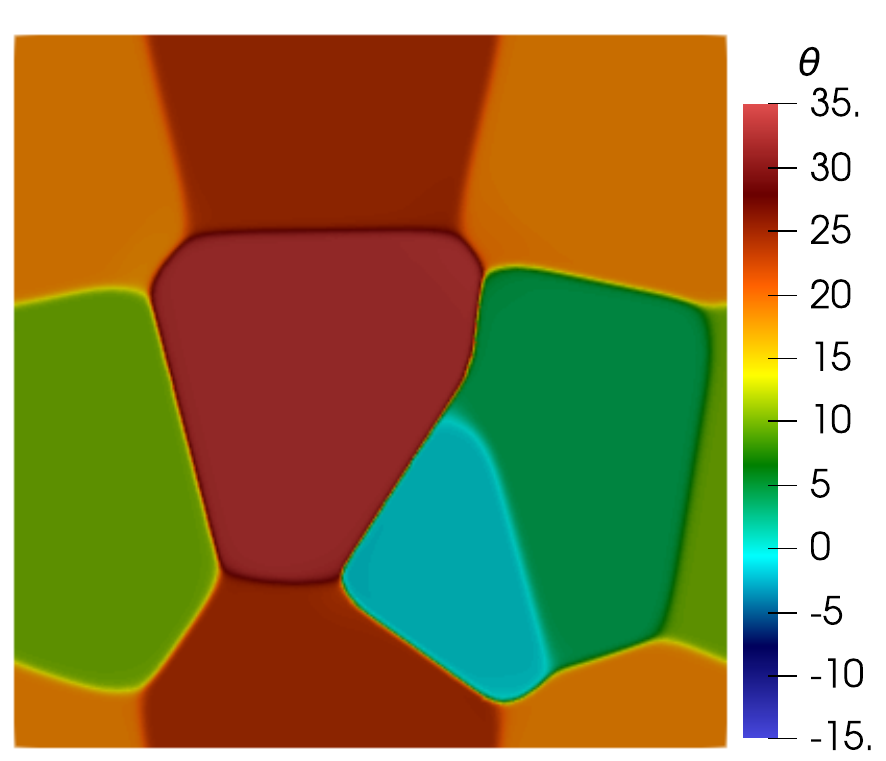}
	\vspace*{0mm}\hspace*{-0mm}
	\includegraphics[width=0.29\textwidth]{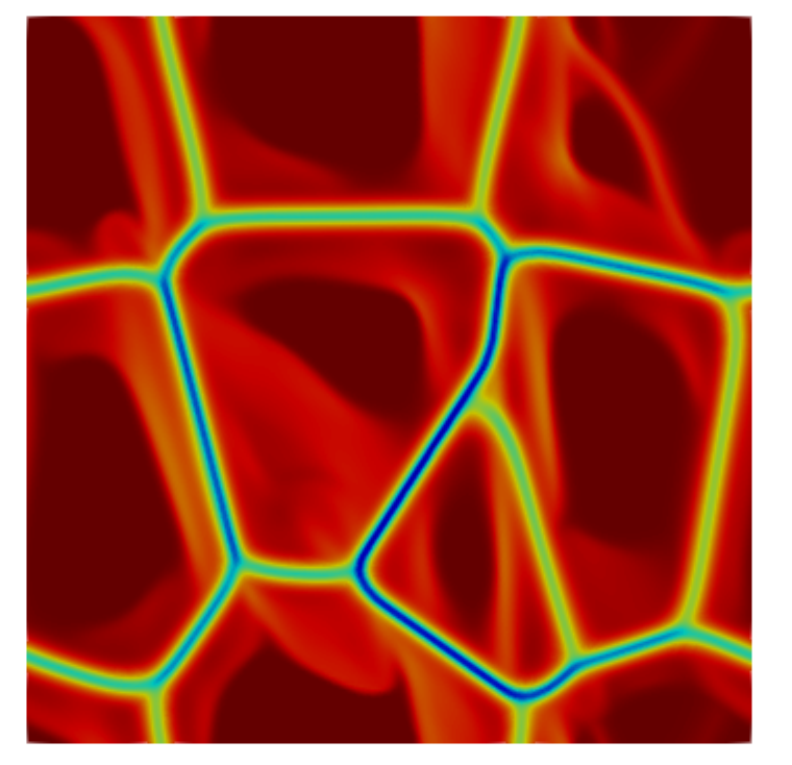}
	\includegraphics[width=0.29\textwidth]{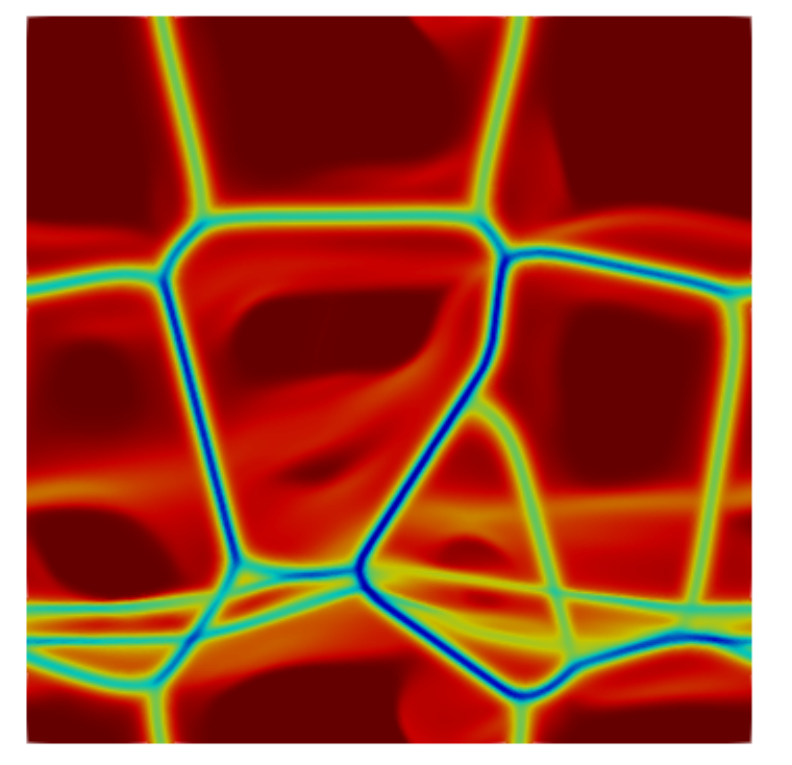}
	\includegraphics[width=0.35\textwidth]{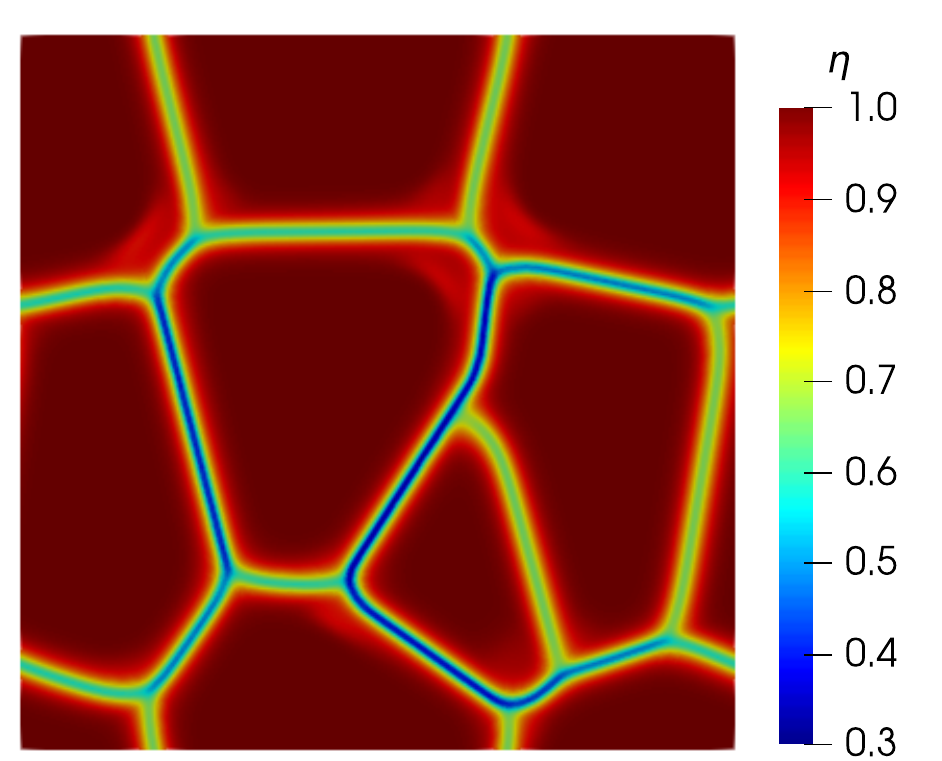}
	\vspace*{-4mm}\hspace*{4mm}
	\includegraphics[width=0.29\textwidth]{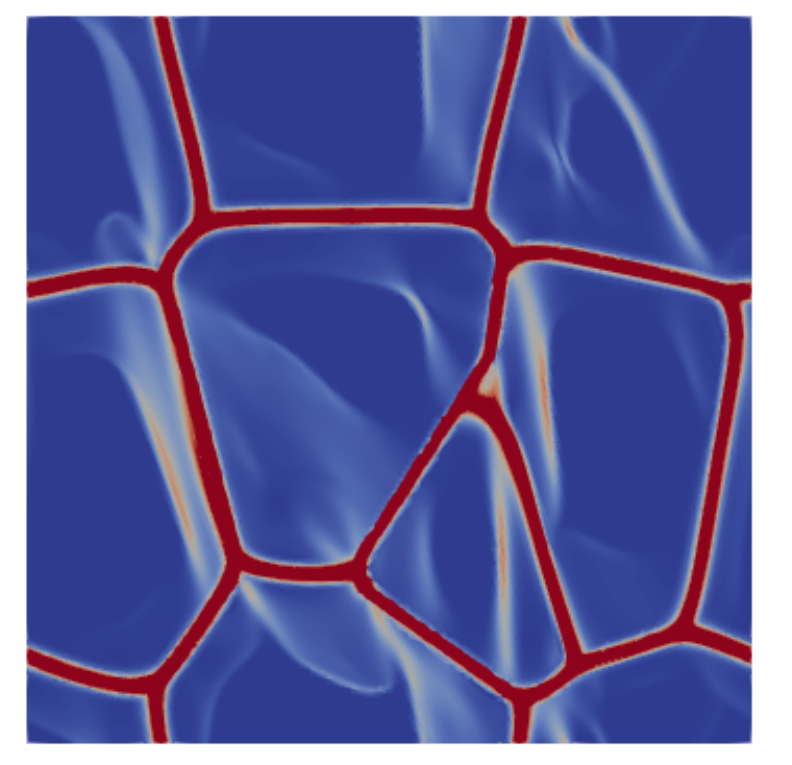}
	\includegraphics[width=0.29\textwidth]{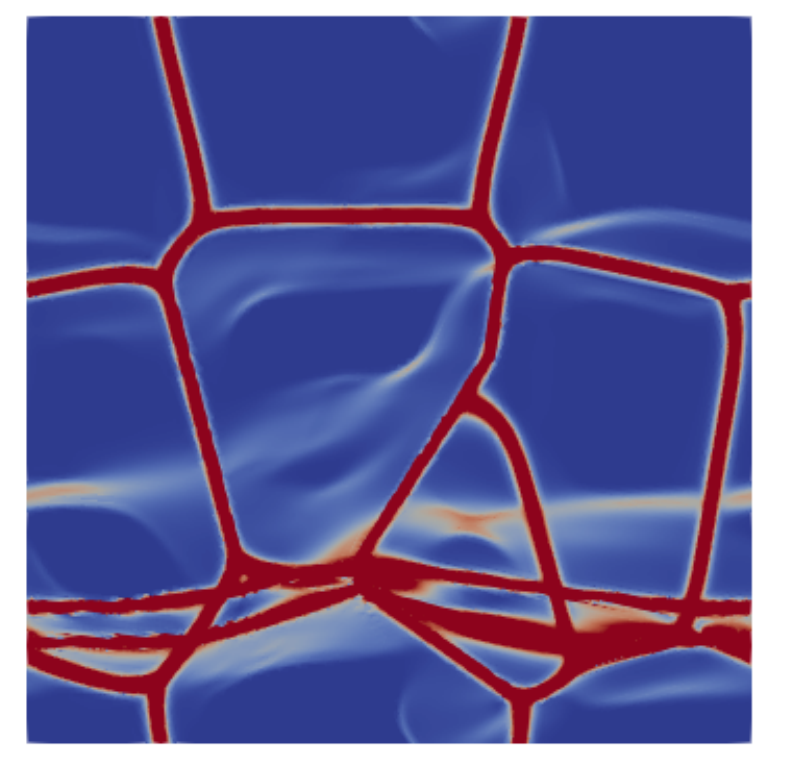}
	\includegraphics[width=0.375\textwidth]{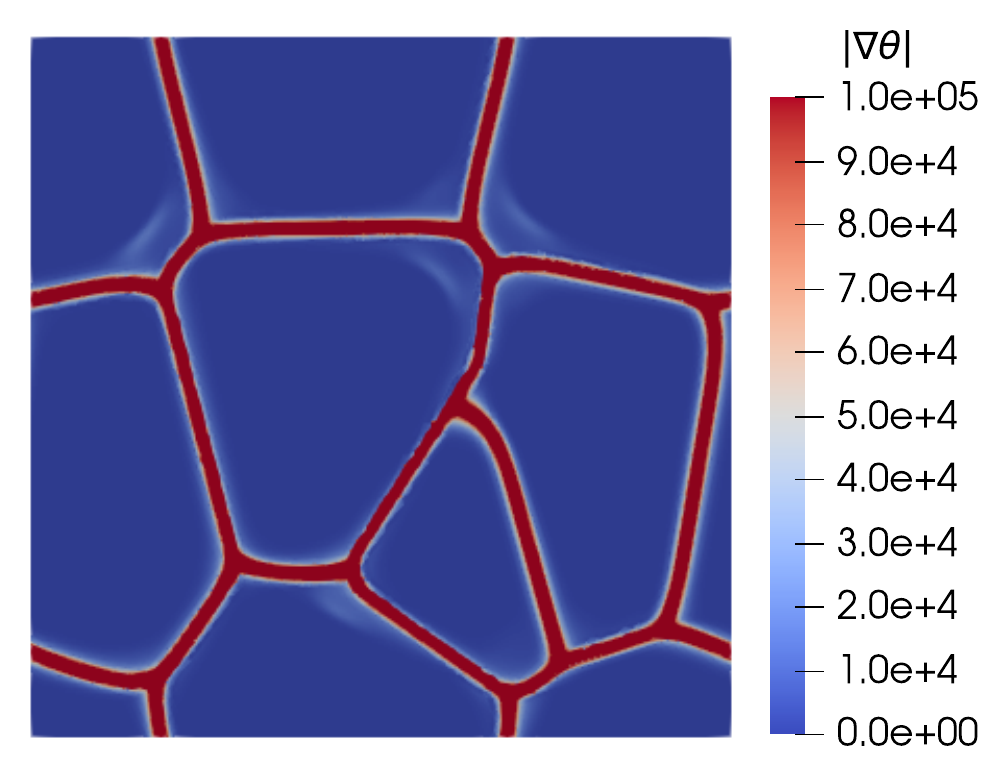}
	\vspace*{-2mm}\hspace*{-1cm}
	\begin{subfigure}[t]{0.32\textwidth}
		\caption{Slip system 1}\label{fig:6grainsload55_suba}
	\end{subfigure}
	\hspace*{-11mm}
	\begin{subfigure}[t]{0.32\textwidth}
		\caption{Slip system 2}\label{fig:6grainsload55_subb}
	\end{subfigure}
	\hspace*{-8mm}
	\begin{subfigure}[t]{0.32\textwidth}
		\caption{4 slips}\label{fig:6grainsload55_subc}
	\end{subfigure}
	\caption{Granular microstructure is loaded in shear with $B_{12}=0.055$ in 5.5 s, for the cases where slip system 1 (a), slip system 2 (b) or 4 slip systems (c) are activated. From top to bottom orientation $\theta$, order parameter $\eta$ and norm of the curvature $|\nabla\theta|$ are shown at the deformed state. \MB{The arrows show localized kink bands.}}
	\label{fig:6grainsload55}
\end{figure}
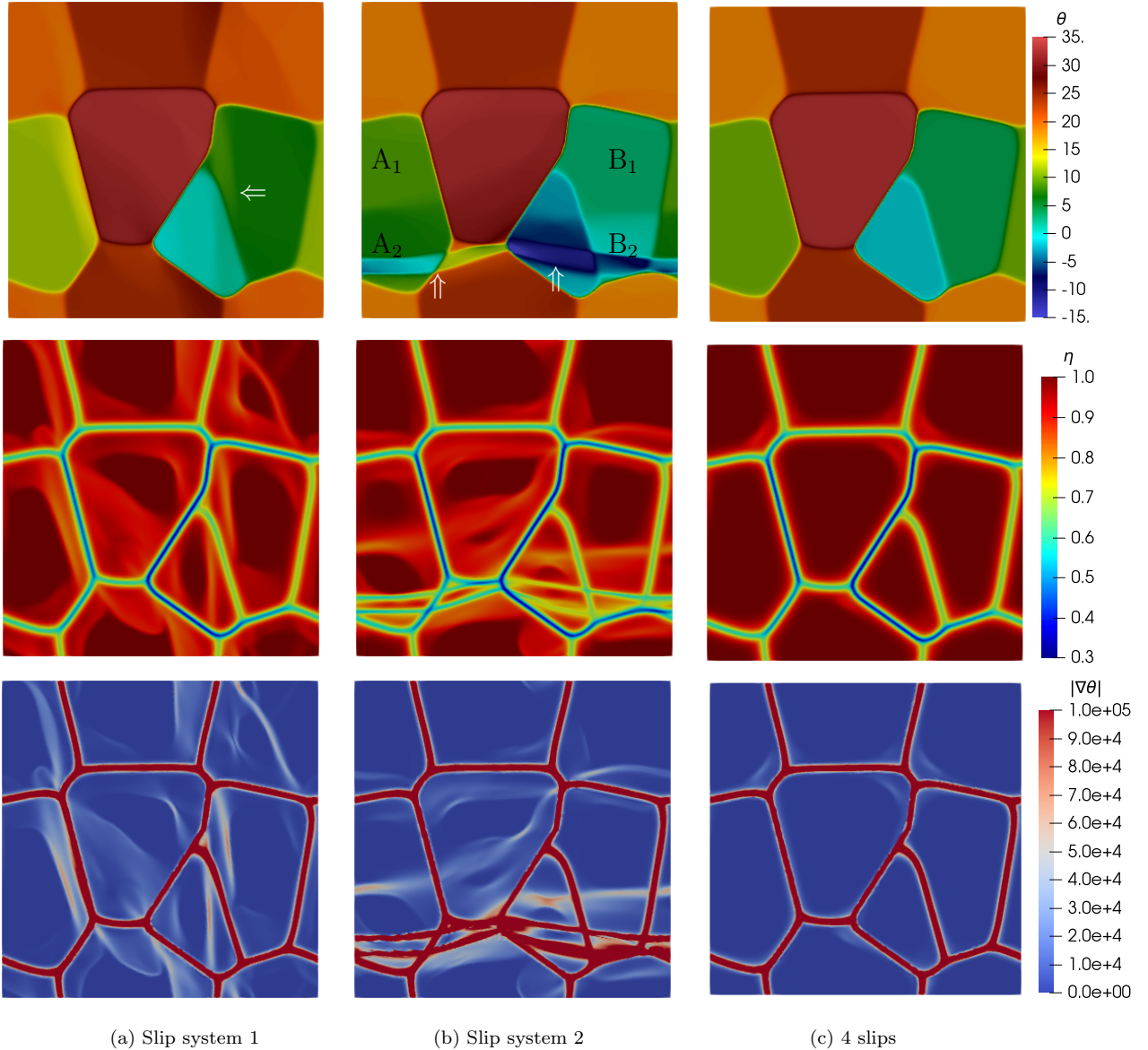

\MB{\subsubsection*{Mechanical loading - 6 grains}}

\begin{figure}[ht!]
	\centering
	\includegraphics[width=0.35\textwidth]{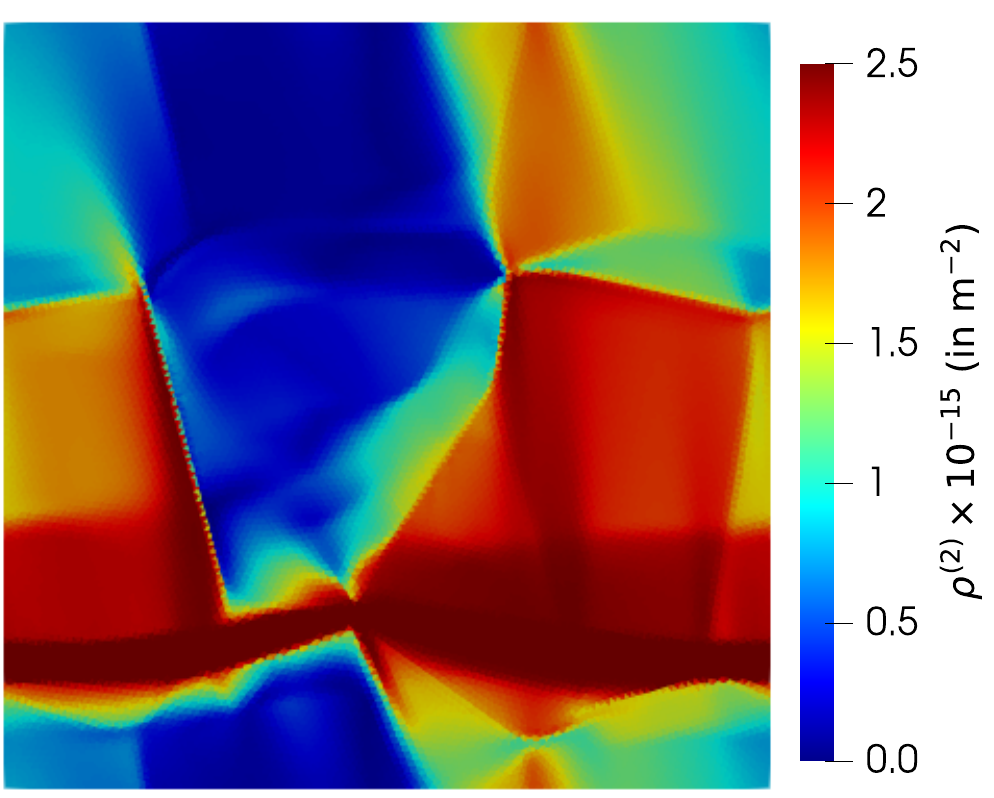}
	\includegraphics[width=0.35\textwidth]{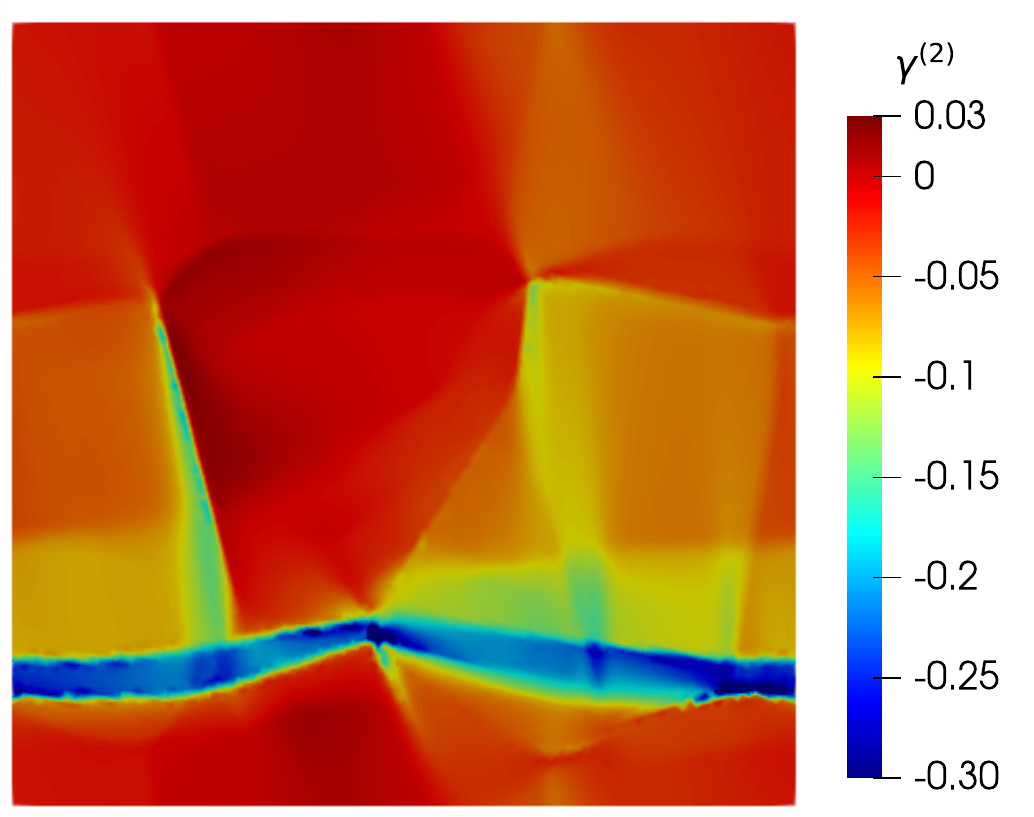}
	\caption{Granular microstructure is loaded in shear with $B_{12}=0.055$ in 5.5 s, where slip system 2 is active. Statically stored dislocation density $\rho^{(2)}$ (left) and viscoplastic slip $\gamma^{(2)}$ (right) are shown at the deformed state.}
	\label{fig:6grainsload55_ssdslip}
\end{figure}

Fig. \ref{fig:6grainsinit}a shows the lattice orientations $\theta$ of the granular microstructure, where the black and white arrows indicate the slip directions 1 and 2, respectively. Fig. \ref{fig:6grainsinit}b/c shows the state of $\theta$ and order parameter $\eta$, after the phase field is relaxed. We obtain diffuse grain boundaries and smoothed corners due to curvature driven GB migration. Fig. \ref{fig:6grainsload55} shows the deformed state \MB{with 5.5\% pre-strain} where either slip system 1 \MB{[\ref{fig:6grainsload55}a]} or 2 \MB{[\ref{fig:6grainsload55}b]}, or all four slip systems \MB{[\ref{fig:6grainsload55}c]} are activated. The Cosserat continuum formulation allows the formation of kink bands, when only a single slip system is active. Inside these bands of localized deformation, the material reorients significantly. We can clearly observe this in the orientation $\theta$ fields when slip system 2 is active (Fig. \ref{fig:6grainsload55_subb}) in the lower portion of the domain \MB{marked by white arrows}. When slip system 1 is active, it is not as significant, but still noticeable in \ref{fig:6grainsload55_suba}. In the 3rd row of Fig. \ref{fig:6grainsload55}, the norm of the orientation gradient $|\nabla\theta|$ is shown, which indicates the formation of new grain boundaries and regions of localized reorientation. Comparing the results for slip system 1 and 2, the localized reorientation regions are perpendicular to the slip direction. Moreover, in both single slip cases, we can see that subgrains form due to heterogeneous deformation inside the grains. For example, grains \MB{A and B} in Fig. \ref{fig:6grainsinit}b are divided into two subgrains \MB{A$_{1/2}$ and B$_{1/2}$ in Fig. \ref{fig:6grainsload55_subb}} with a small misorientation. When all four slips are active, we observe a bulk clockwise rotation of grains instead of localized deformation. Fig. \ref{fig:6grainsload55_ssdslip} shows the dislocation density and viscoplastic slip when slip system 2 is active. Compared to Fig. \ref{fig:6grainsload55_subb}, the regions of localized deformation and reorientation coincide. This confirms that the formation of kink band and subgrains is indeed caused by the plastic deformation. In addition, bands with localized strain in the slip direction are observed.

\begin{figure}[ht!]
	\centering
	\hspace*{0mm}
	\raisebox{1.8cm}{\includegraphics[width=0.02\textwidth]{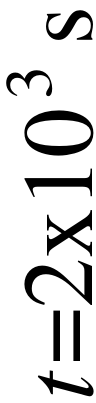}}
    \begin{tikzpicture}
	\draw(0,0)node[inner sep=0]{\includegraphics[width=0.285\textwidth]{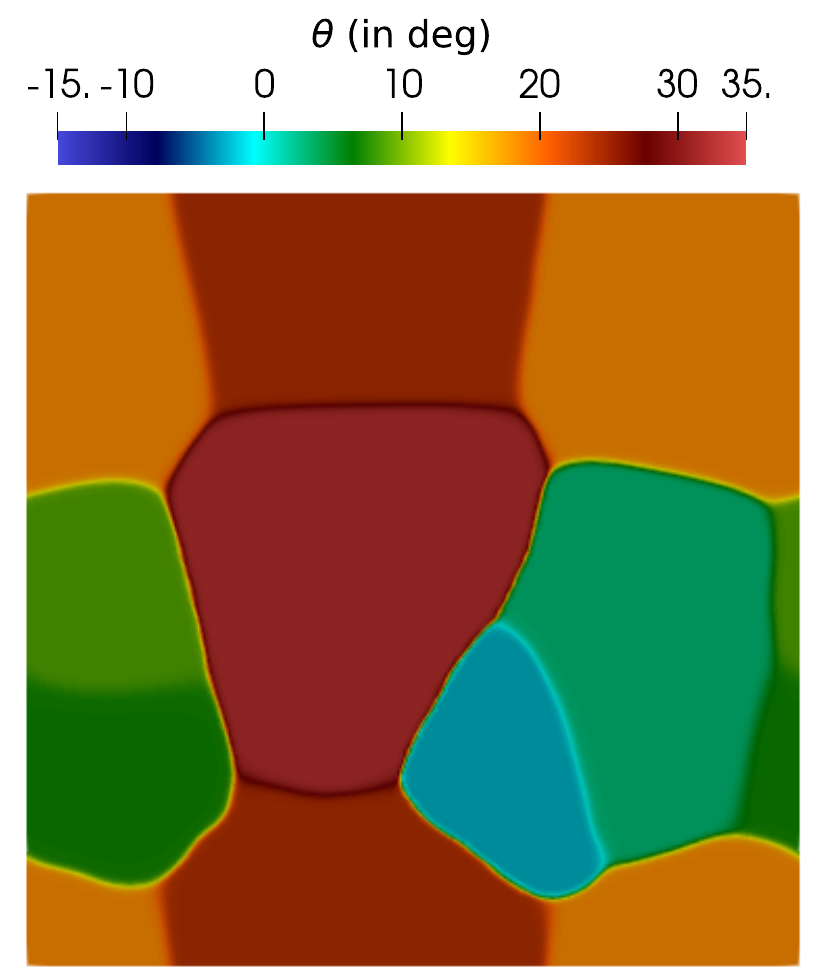}};
    \draw(-2,-0.5)node{A$_1$}; \draw(-2,-1.75)node{A$_2$}; \draw(-0.5,-0.5)node{C}; \draw(0.5,-1.5)node{D}; \draw(1.5,-1)node{B};
    \end{tikzpicture}
    \includegraphics[width=0.285\textwidth]{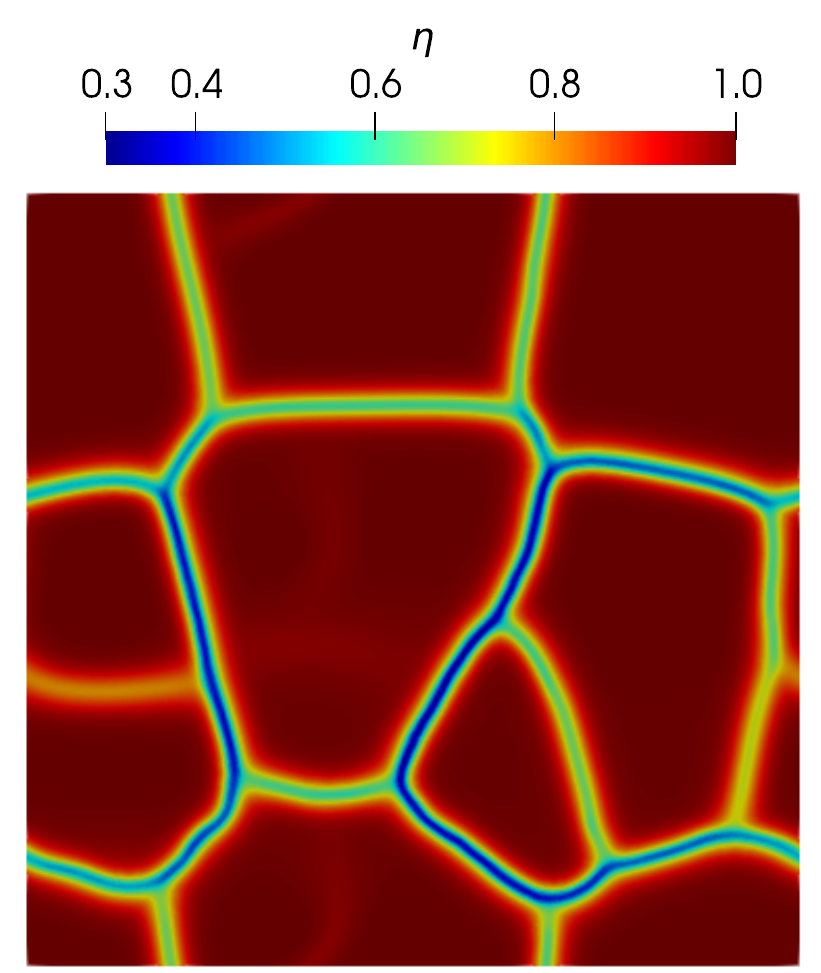}
	\includegraphics[width=0.285\textwidth]{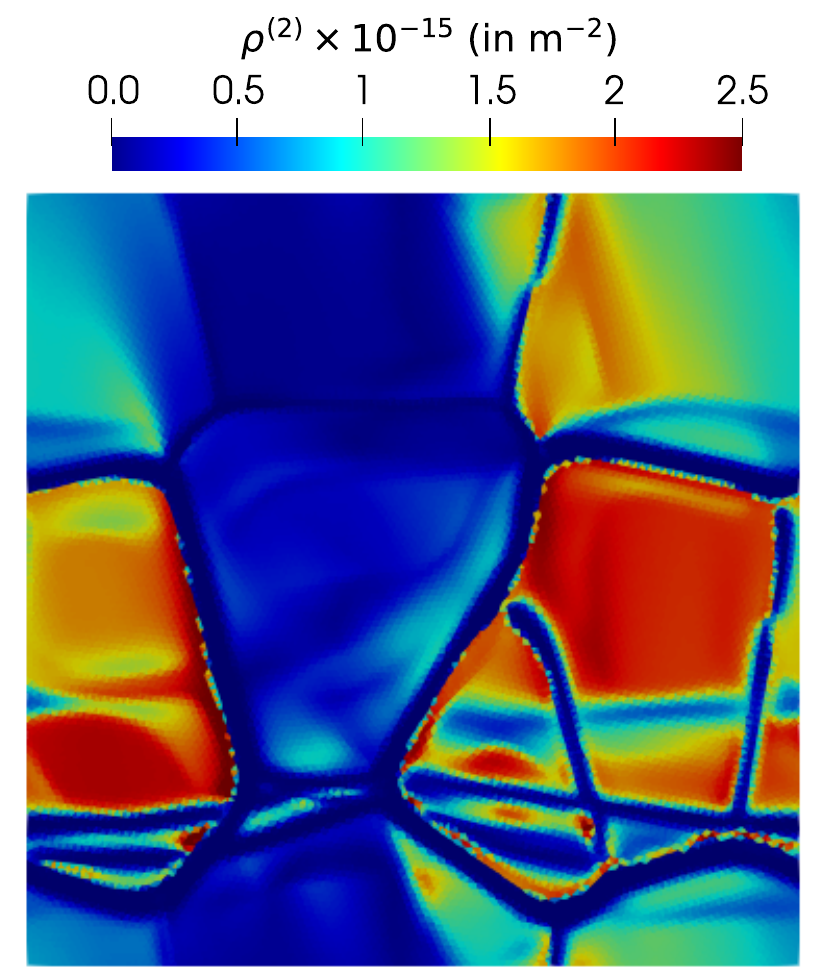}
	
    \hspace*{0mm}
	\raisebox{1.8cm}{\includegraphics[width=0.02\textwidth]{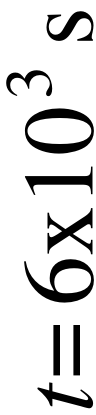}}
	\includegraphics[width=0.285\textwidth,trim={0 0 0 3cm},clip]{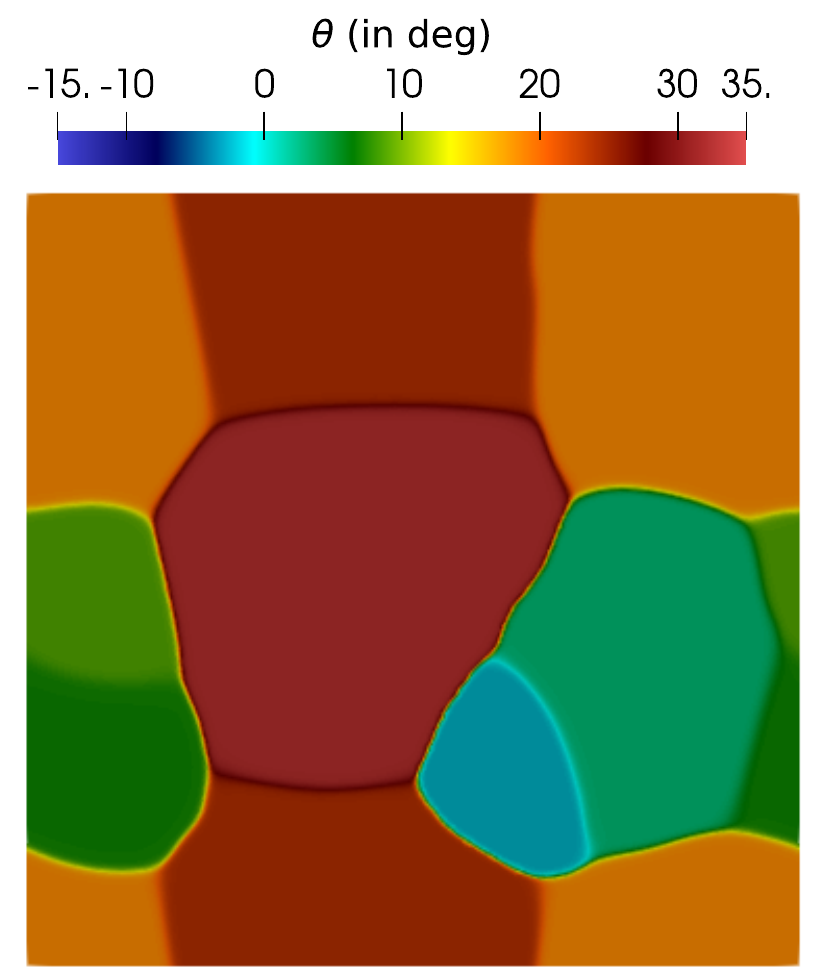}
	\includegraphics[width=0.285\textwidth,trim={0 0 0 3cm},clip]{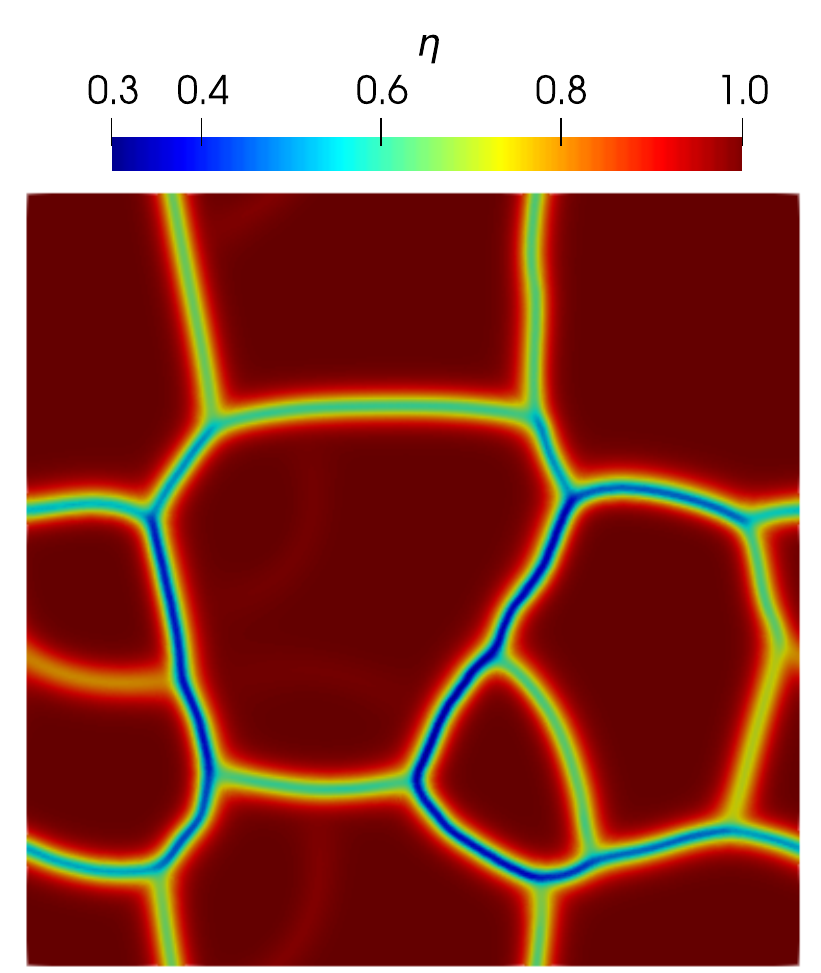}
	\includegraphics[width=0.285\textwidth,trim={0 0 0 3cm},clip]{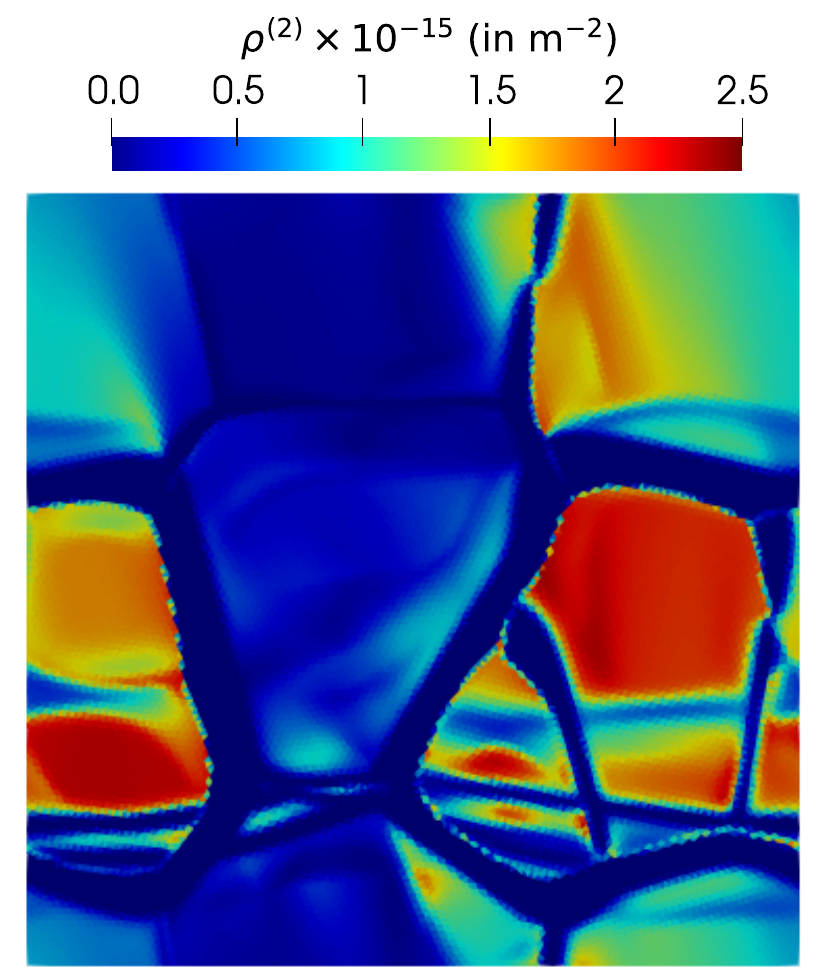}
	\caption{The deformed structure in Fig. \ref{fig:6grainsload55_subb} and Fig. \ref{fig:6grainsload55_ssdslip} is allowed to recrystallize with $c_2=0.7$. From left to right orientation $\theta$, order parameter $\eta$ and statically stored dislocation density $\rho^{(2)}$ are shown at different times.}
	\label{fig:6grainsload55_c2_07}
\end{figure}

\begin{figure}[ht!]
	\centering
	\hspace*{0mm}
	\raisebox{1.8cm}{\scalebox{1}{\includegraphics[width=0.02\textwidth]{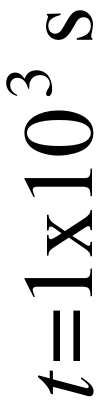}}}
    \begin{tikzpicture}
	\draw(0,0)node[inner sep=0]{\includegraphics[width=0.285\textwidth]{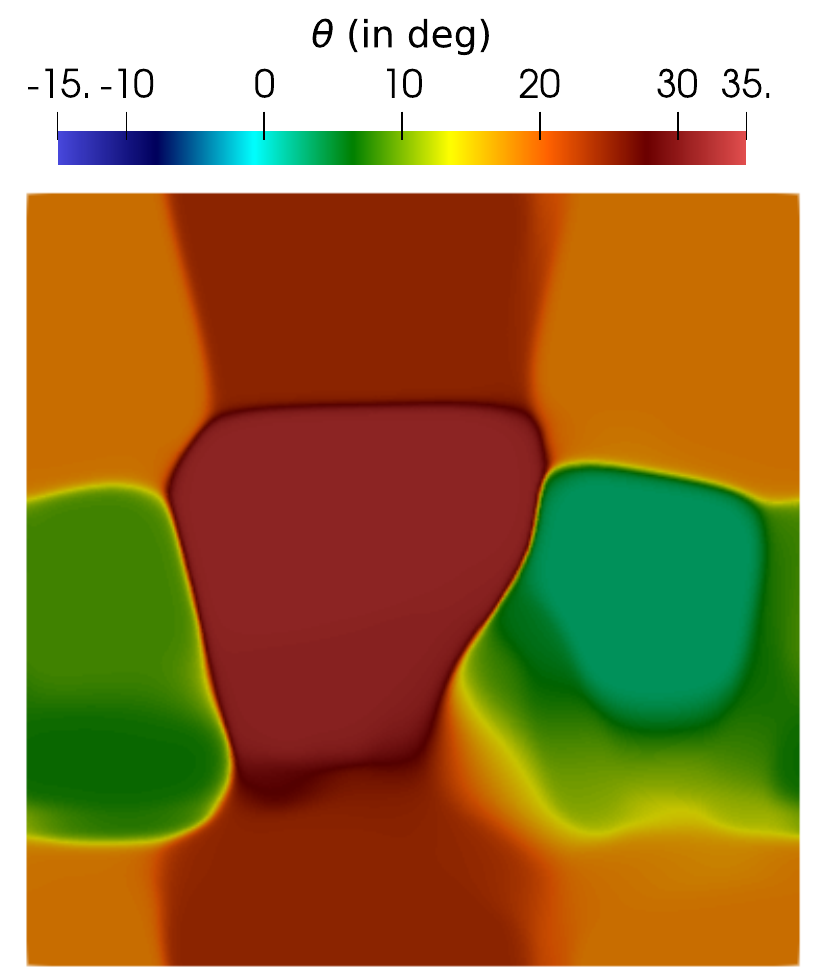}};
    \draw(1.5,-1)node{B}; \draw(1.5,1)node{E};
    \end{tikzpicture}
    \begin{tikzpicture}
	\draw(0,0)node[inner sep=0]{\includegraphics[width=0.285\textwidth]{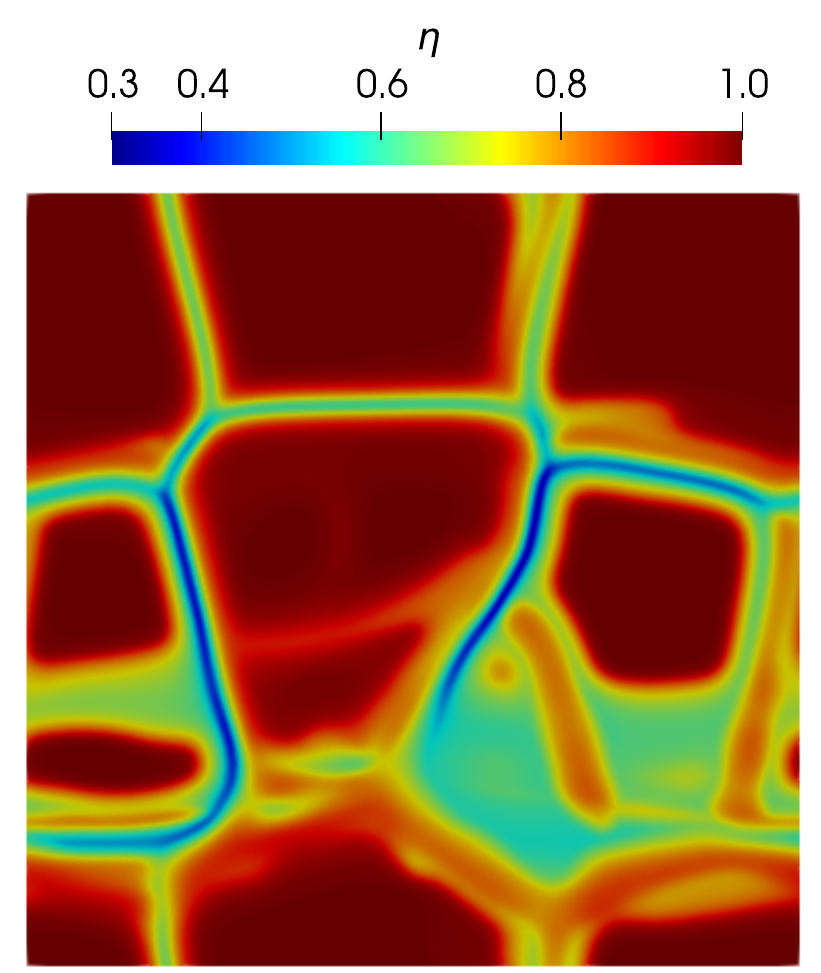}};
    \draw(0.6,-1.5)node{\textcolor{white}{$\Rightarrow$}}; \draw(1.7,-1.5)node{\textcolor{white}{$\Rightarrow$}};
    \end{tikzpicture}
    \begin{tikzpicture}
	\draw(0,0)node[inner sep=0]{\includegraphics[width=0.285\textwidth]{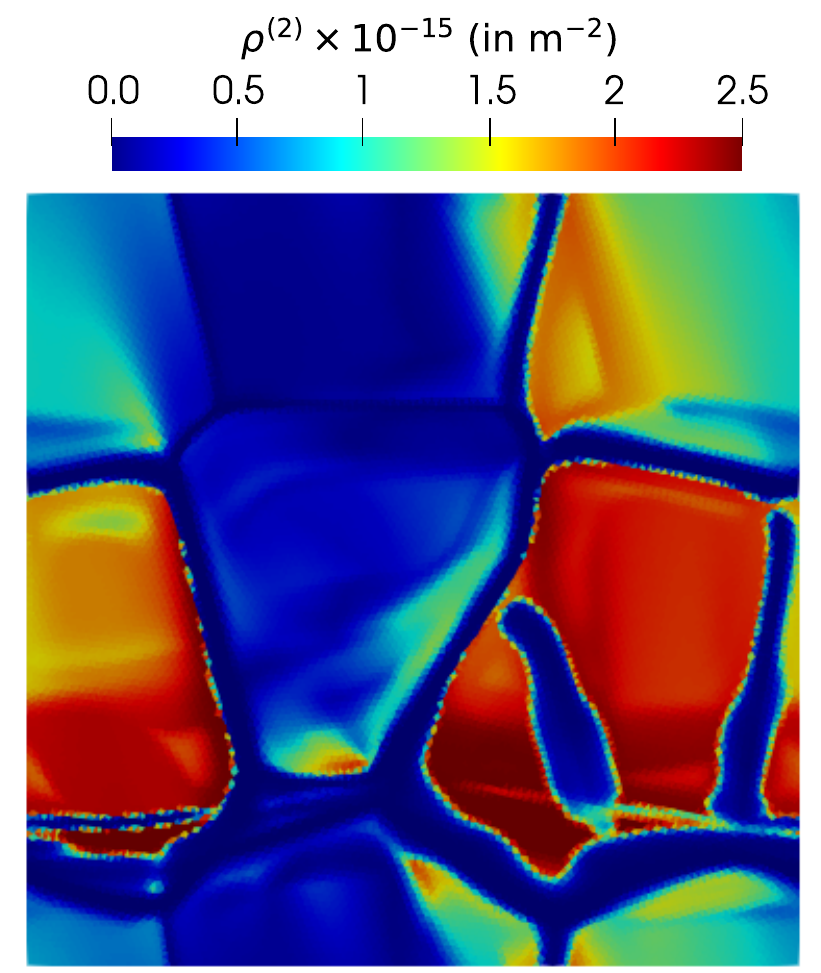}};
    \draw(0.6,-1.5)node{\textcolor{white}{$\Rightarrow$}}; \draw(1.7,-1.5)node{\textcolor{white}{$\Rightarrow$}};
    \end{tikzpicture}
 
	\hspace*{0mm}
	\raisebox{1.8cm}{\includegraphics[width=0.02\textwidth]{figure_t2000.pdf}}
    \begin{tikzpicture}
	\draw(0,0)node[inner sep=0]{\includegraphics[width=0.285\textwidth,trim={0 0 0 3cm},clip]{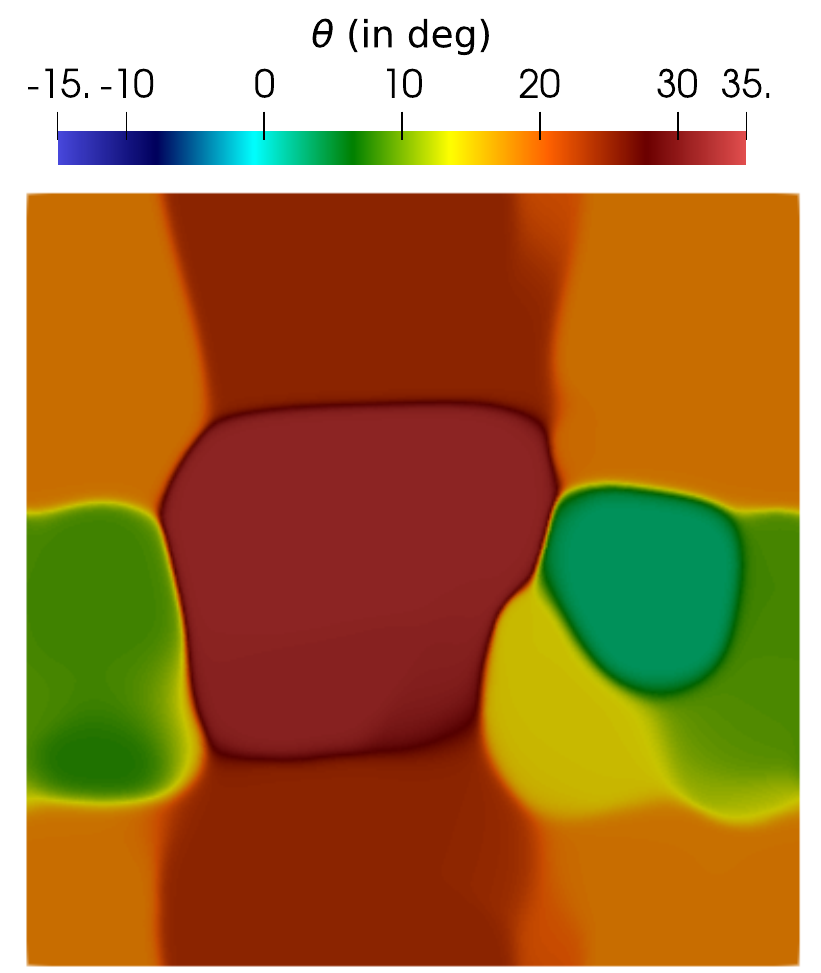}};
    \draw(1,-1)node{G};
    \end{tikzpicture}
	\includegraphics[width=0.285\textwidth,trim={0 0 0 31mm},clip]{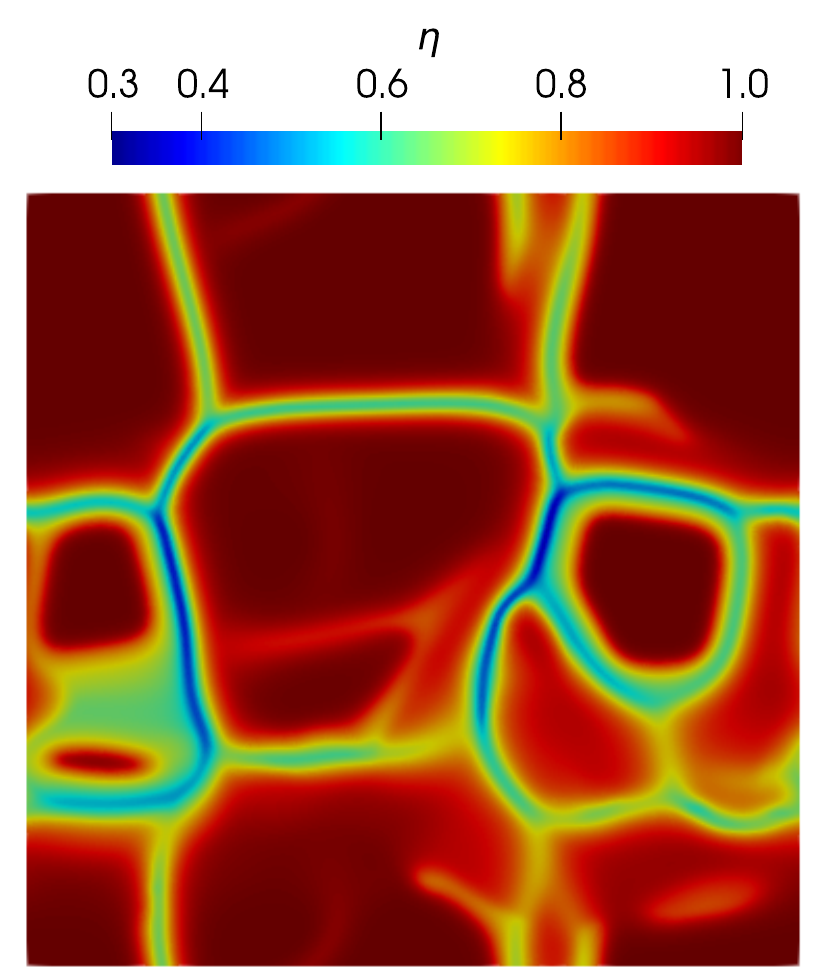}
	\includegraphics[width=0.285\textwidth,trim={0 0 0 31mm},clip]{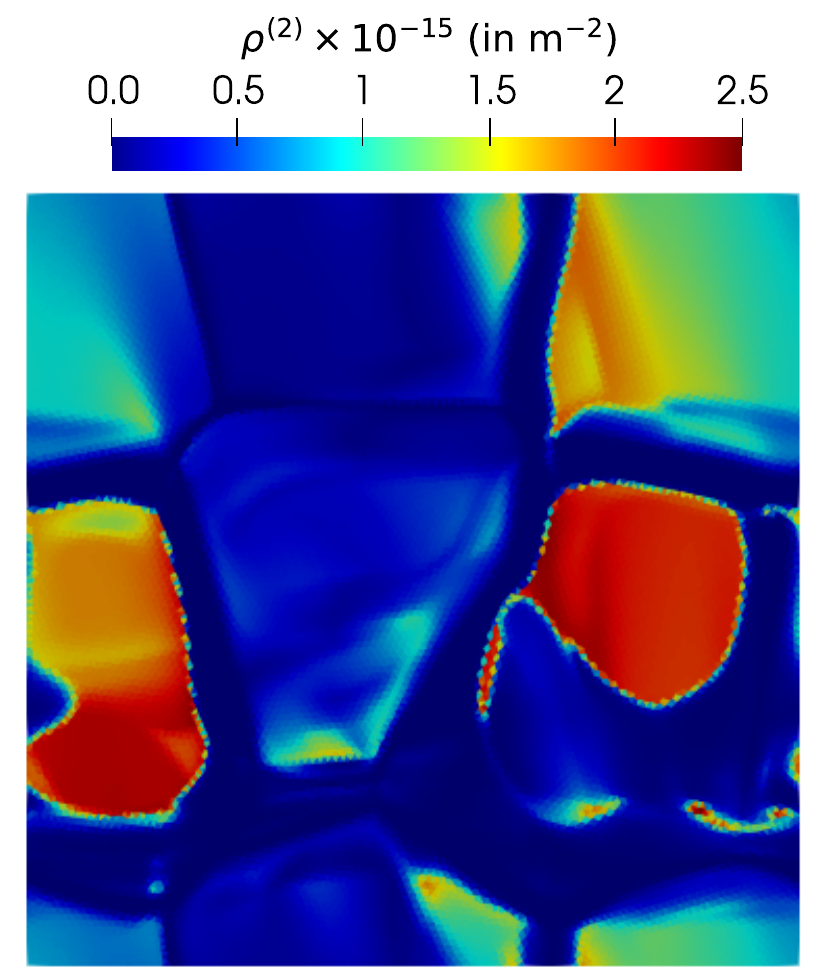}
 
	\hspace*{0mm}
	\raisebox{1.8cm}{\includegraphics[width=0.02\textwidth]{figure_t6000.pdf}}
    \begin{tikzpicture}
	\draw(0,0)node[inner sep=0]{\includegraphics[width=0.285\textwidth,trim={0 0 0 3cm},clip]{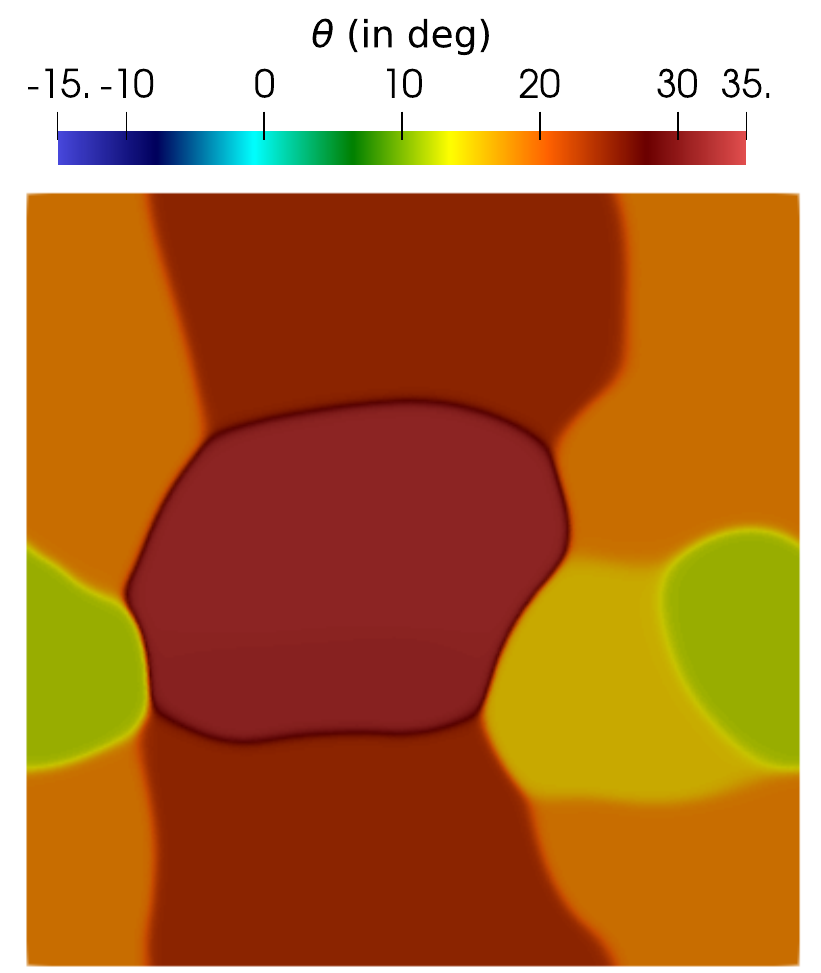}};
    \draw(1.25,-0.8)node{G};
    \end{tikzpicture}
	\includegraphics[width=0.285\textwidth,trim={0 0 0 3cm},clip]{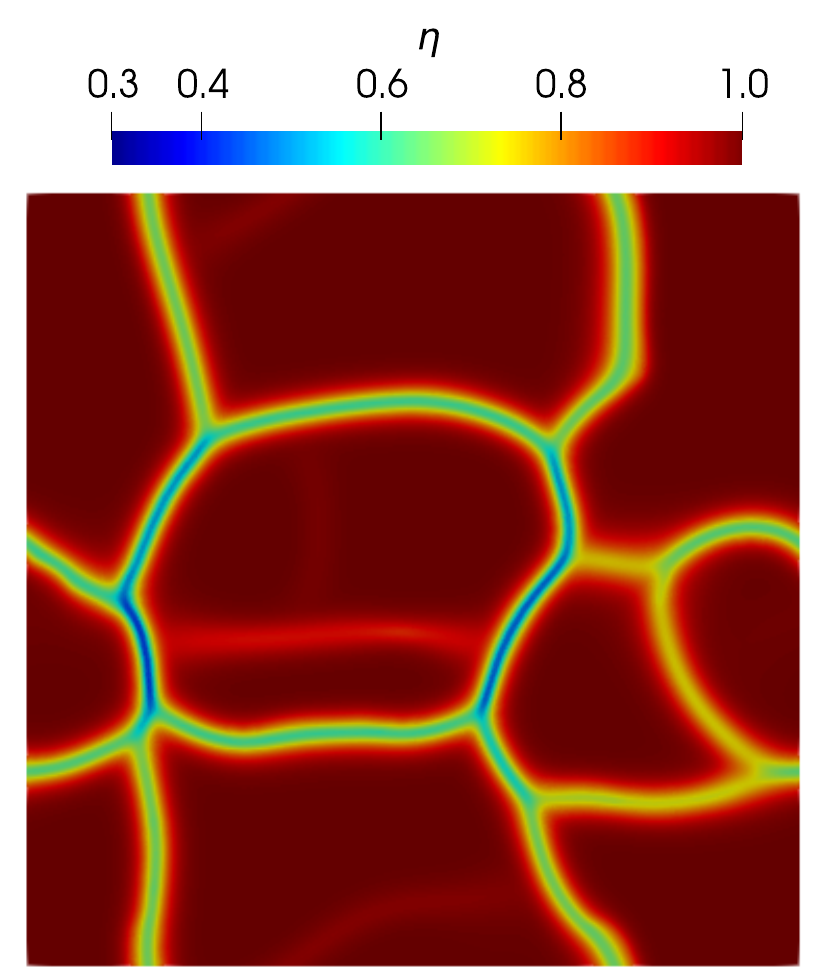}
	\includegraphics[width=0.285\textwidth,trim={0 0 0 3cm},clip]{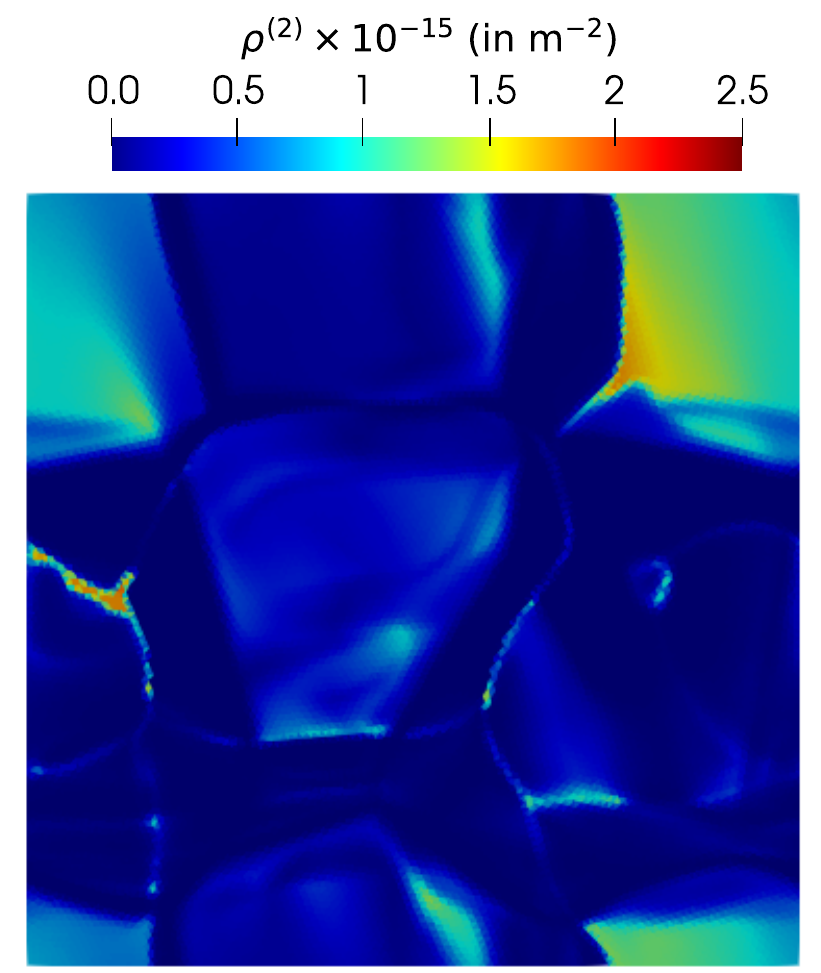}
	\caption{The deformed structure in Fig. \ref{fig:6grainsload55_subb} and Fig. \ref{fig:6grainsload55_ssdslip} is allowed to recrystallize with $c_2=0.9$ which activates grain nucleation. From left to right orientation $\theta$, order parameter $\eta$ and statically stored dislocation density $\rho^{(2)}$ are shown at different times. \MB{The arrows at $t=1\times 10^3$\;s show nucleation sites.}}
	\label{fig:6grainsload55_c2_09}
\end{figure}

\MB{\subsubsection*{Relaxation phase - 6 grains}}

After the loading phase, displacements are held constant, and the heat treatment phase is simulated by activating GB migration with $c_2=0.7$ or $0.9$ and dislocation recovery with $C_\mathrm{D}=100$. The grain boundaries can migrate due to both curvature and stored dislocations. We first look at the single slip case, where \MB{slip system} $\alpha=2$ was active. Fig. \ref{fig:6grainsload55_c2_07} shows the evolution of the microstructure at $2\times 10^3$\;s and $6\times 10^3$\;s when $c_2$ is set to 0.7. Since $c_2$ is small, we expect SSD driven GB migration, but no grain nucleation. Indeed, a mixture of curvature and SSD driven migration with dislocation recovery at the wake is observed, where the middle grain \MB{C} expands due to stored dislocations on its left and right boundaries. The small grain \MB{D} next to it shrinks due to curvature driven migration. The two subgrains \MB{A$_1$ and A$_2$} at the left, which formed due to the localized deformation, have stabilized, i.e., the order parameter $\eta$ inside is 1, and \MB{subgrain A$_1$} shrinks while \MB{subgrain A$_2$} expands due to curvature and stored energy, showing subgrain growth. The kink band that formed during deformation in Fig. \ref{fig:6grainsload55_subb} is quickly restored during the initial phase of the heat treatment. This is expected considering the model's formulation, which allows grain rotation if $\eta<1$. Since the kink band has not expanded wide enough to form a bulk \MB{region} inside, it rotates back toward the surrounding grains. We note that the convergence of the model deteriorates when the band starts to expand during deformation. 
%A finite deformation setting could remedy this and open up new possibilities. 
Similarly, the subgrains \MB{B$_1$ and B$_2$ that} formed during deformation (see Fig. \ref{fig:6grainsload55_subb}) are recovered quickly, because the subgrain \MB{B$_2$} had \MB{not stabilized} ($\eta<1$) in the bulk due to deformation and it rotated back during heat treatment, which resembles a subgrain coalescence mechanism.  

Next, Fig. \ref{fig:6grainsload55_c2_09} shows the case when $c_2$ is set to 0.9, which increases the GB velocity significantly and also activates the grain nucleation mechanism. Hence, at $t=1\times 10^3$\;s nucleation is triggered at the left and right boundaries of \MB{grain B}. The nucleation locations can be identified in the order parameter $\eta$ field \MB{(marked by arrows in Fig. \ref{fig:6grainsload55_c2_09}}), as the regions of light red strips surfacing inside green areas. At the same locations, blue strips are observed in the dislocation density $\rho$ field, indicating full recovery at nucleation sites. At $t=2\times 10^3$\;s, the nuclei on the left formed a dislocation-free new grain \MB{G}, which then expands into the neighboring grains as seen at $t=6000$\;s (Fig. \ref{fig:6grainsload55_c2_09}). We can also observe that compared to the case of $c_2=0.7$, the driving force \MB{due to stored} dislocations is much more dominant compared to curvature effects. This causes \MB{grain E in Fig. \ref{fig:6grainsload55_c2_09}} to evolve into a concave shape. Such shapes can be observed during abnormal grain growth \citep{bozzolo2013strain,jin2021strain}. If the specimen is allowed to evolve further after the dislocations are recovered, the migration direction is reversed due to curvature taking over. It should also be possible to adjust the relative strength of both forces with the model parameter $c_2$, but this requires further investigation. Another complexity in the polycrystal example compared to the simpler bicrystal example is that the deformation can generate gradients of $\theta$ inside the bulk of grains, which causes $\eta$ to \MB{deviate} from 1. It can be further decreased if dislocations are present in that region since $f_{\eta_4}$ is activated. Such a region is not treated by the model as the bulk of a grain, and it is allowed to rotate. This is not exactly the same as the nucleation of a grain which expands only by migration of its new GBs, but rather a mixture of expansion and rotation. However, it is still driven by deformation and stored dislocations, which represent a more complex nucleation mechanism that is not limited to the grain boundaries.

\begin{figure}[ht!]
	\centering
	\hspace*{0mm}
	\raisebox{1.8cm}{\includegraphics[width=0.02\textwidth]{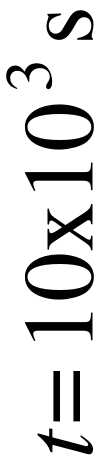}}
    \begin{tikzpicture}
	\draw(0,0)node[inner sep=0]{\includegraphics[width=0.28\textwidth,trim={0 0 3cm 0},clip]{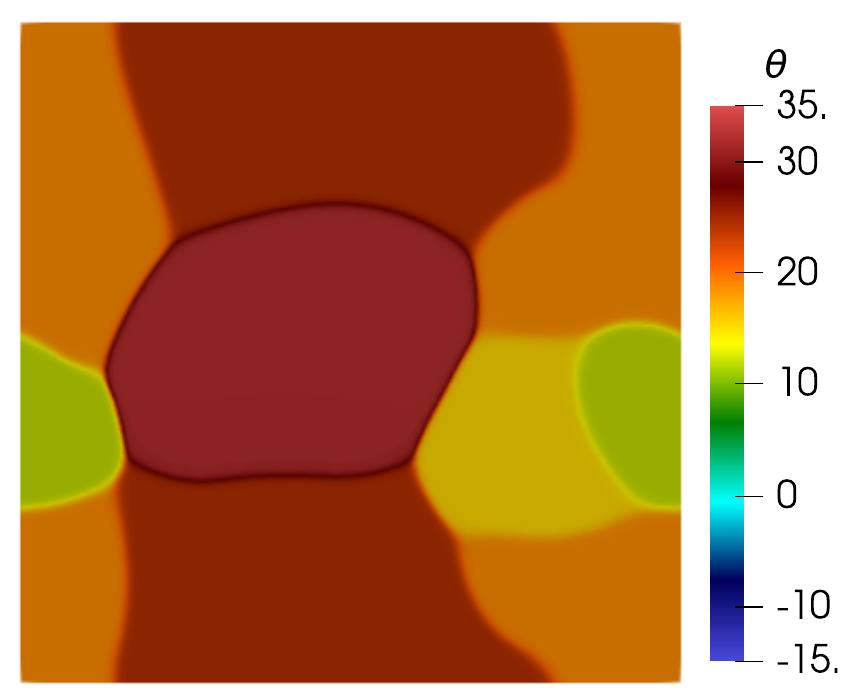}};
    \draw(2,1)node{E}; \draw(1.25,-0.5)node{G};
    \end{tikzpicture}
    \begin{tikzpicture}
	\draw(0,0)node[inner sep=0]{\includegraphics[width=0.28\textwidth,trim={0 0 3cm 0},clip]{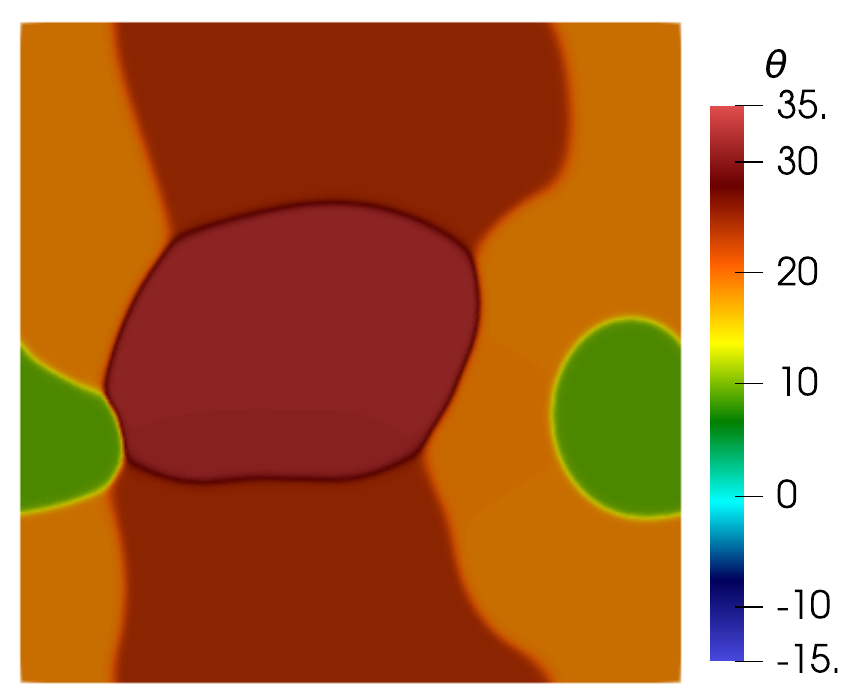}};
    \draw(2,1)node{E};
    \end{tikzpicture}
	\vspace*{-3mm}
    \begin{tikzpicture}
	\draw(0,0)node[inner sep=0]{\includegraphics[width=0.352\textwidth,trim={0 0 0 0},clip]{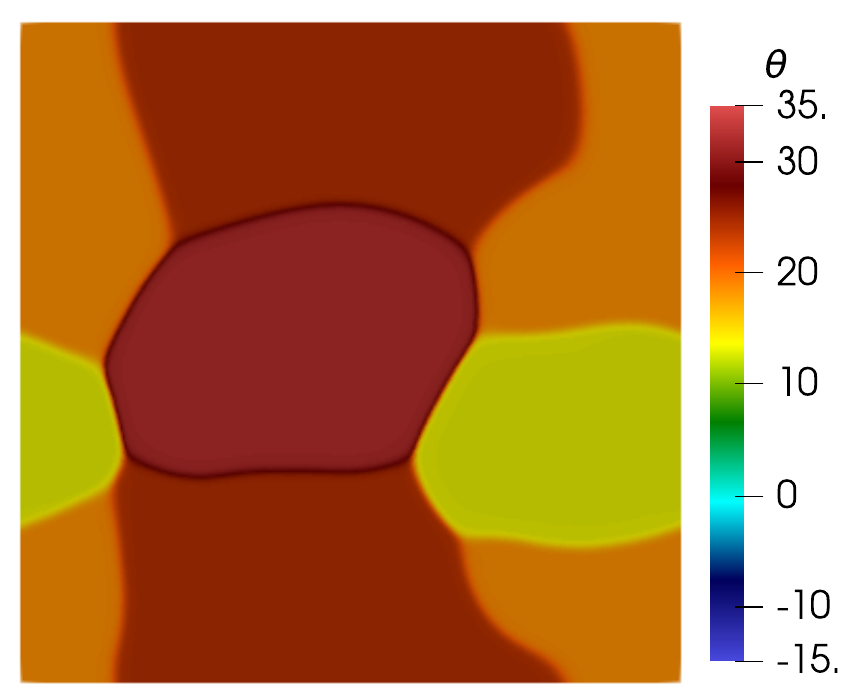}};
    \draw(1.4,1)node{E}; \draw(1.,-0.5)node{G};
    \end{tikzpicture}
	\vspace*{-2mm}\hspace*{-5mm}
	\begin{subfigure}[t]{0.32\textwidth}
		\caption{$B_{12}=0.055$}
	\end{subfigure}
	\hspace*{-11mm}
	\begin{subfigure}[t]{0.32\textwidth}
		\caption{$B_{12}=0.050$}
	\end{subfigure}
	\hspace*{-8mm}
	\begin{subfigure}[t]{0.32\textwidth}
		\caption{$B_{12}=0.060$}
	\end{subfigure}
	\caption{The structure is deformed by 5\%, 5.5\% or 6\%, then heat treated for 10000\;s with $c_2=0.9$. The resulting orientation fields are shown for different amounts of applied deformation.}
	\label{fig:6grainsloadcomp_c2_09}
\end{figure}
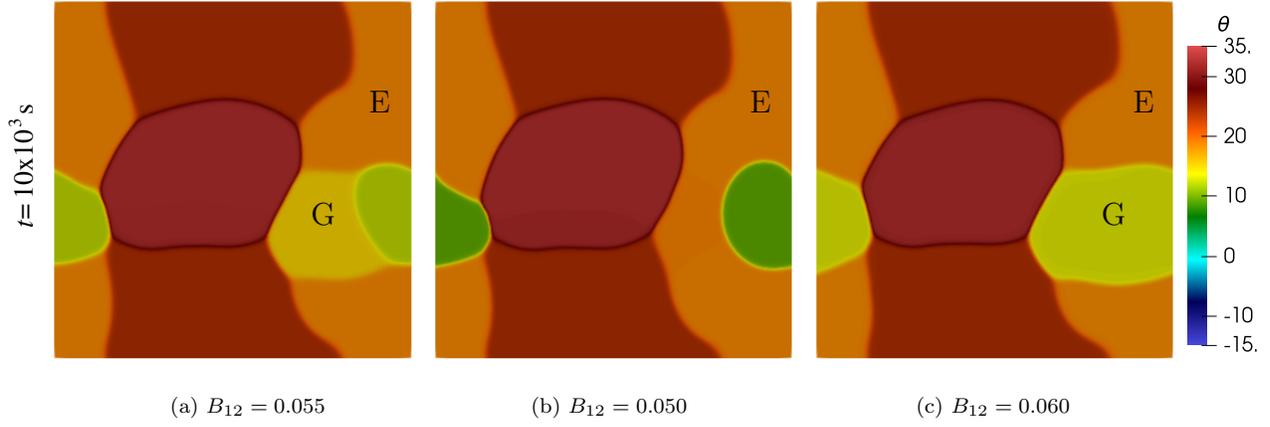

\begin{figure}[ht!]
	\centering
	\hspace*{0mm}
	\includegraphics[width=0.27\textwidth,trim={0 0 4cm 0},clip]{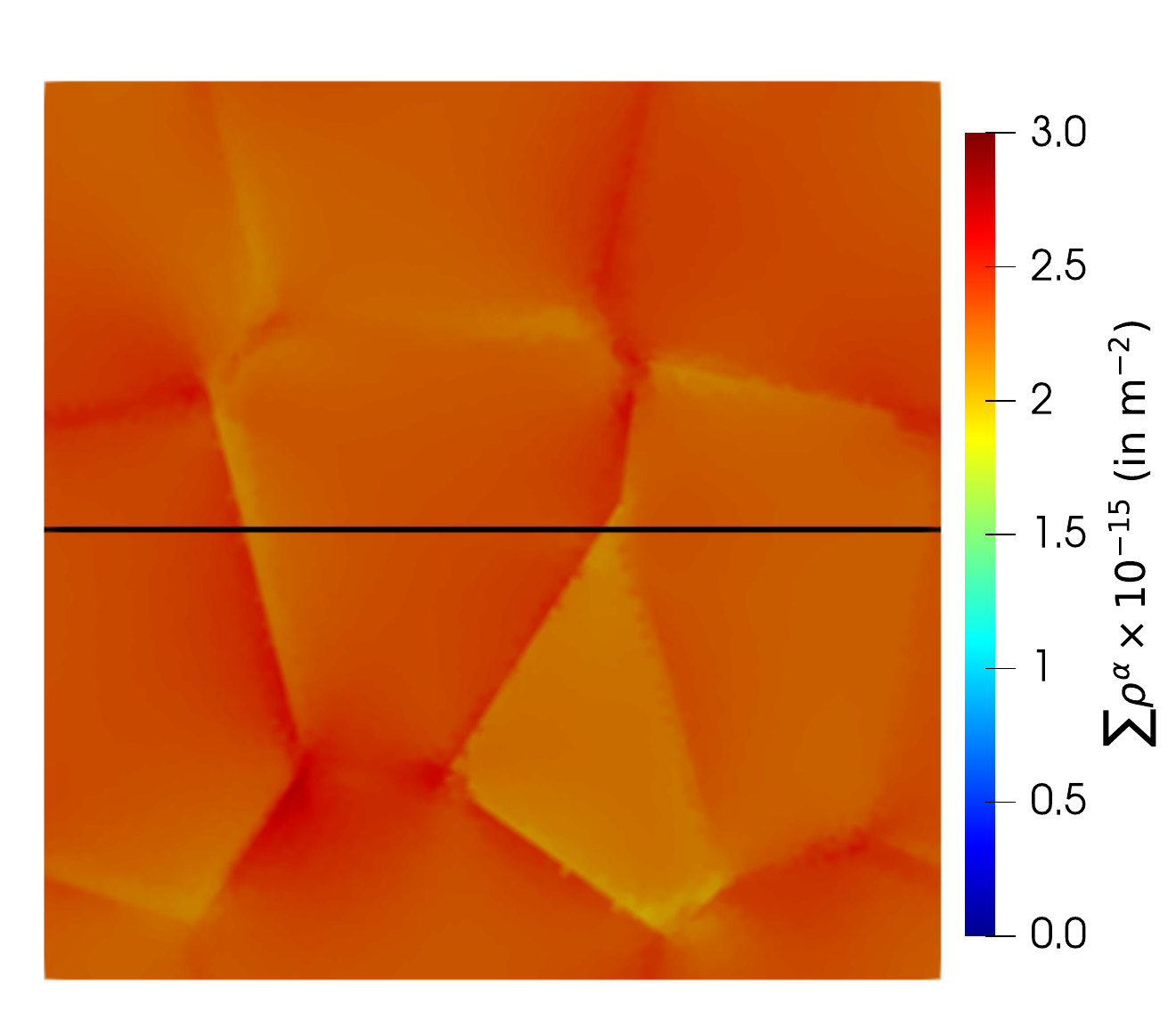}
	\includegraphics[width=0.33\textwidth,trim={0 0 0cm 0},clip]{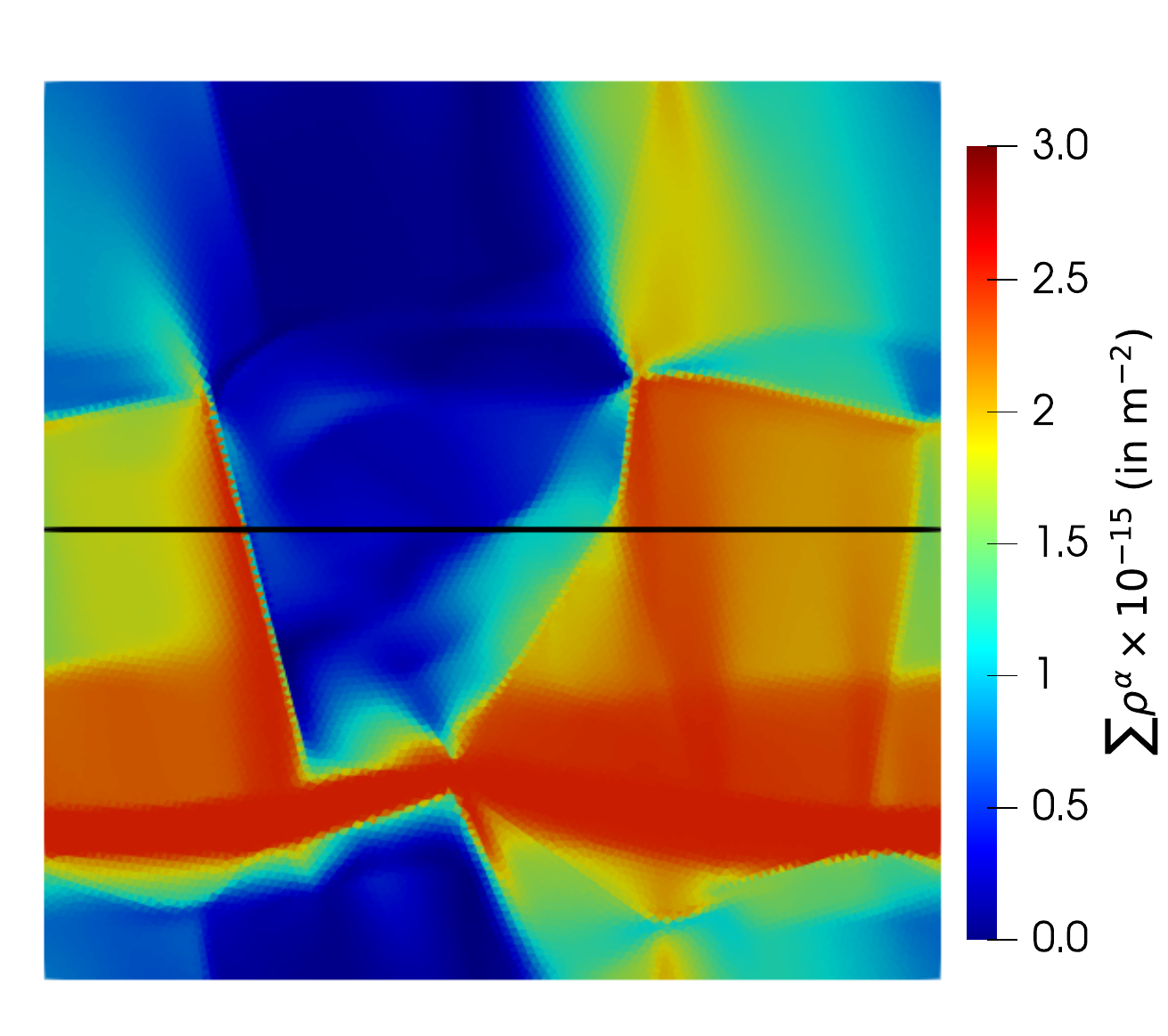}
	\vspace*{-4mm}
	\includegraphics[width=0.36\textwidth,trim={0 0 0 0},clip]{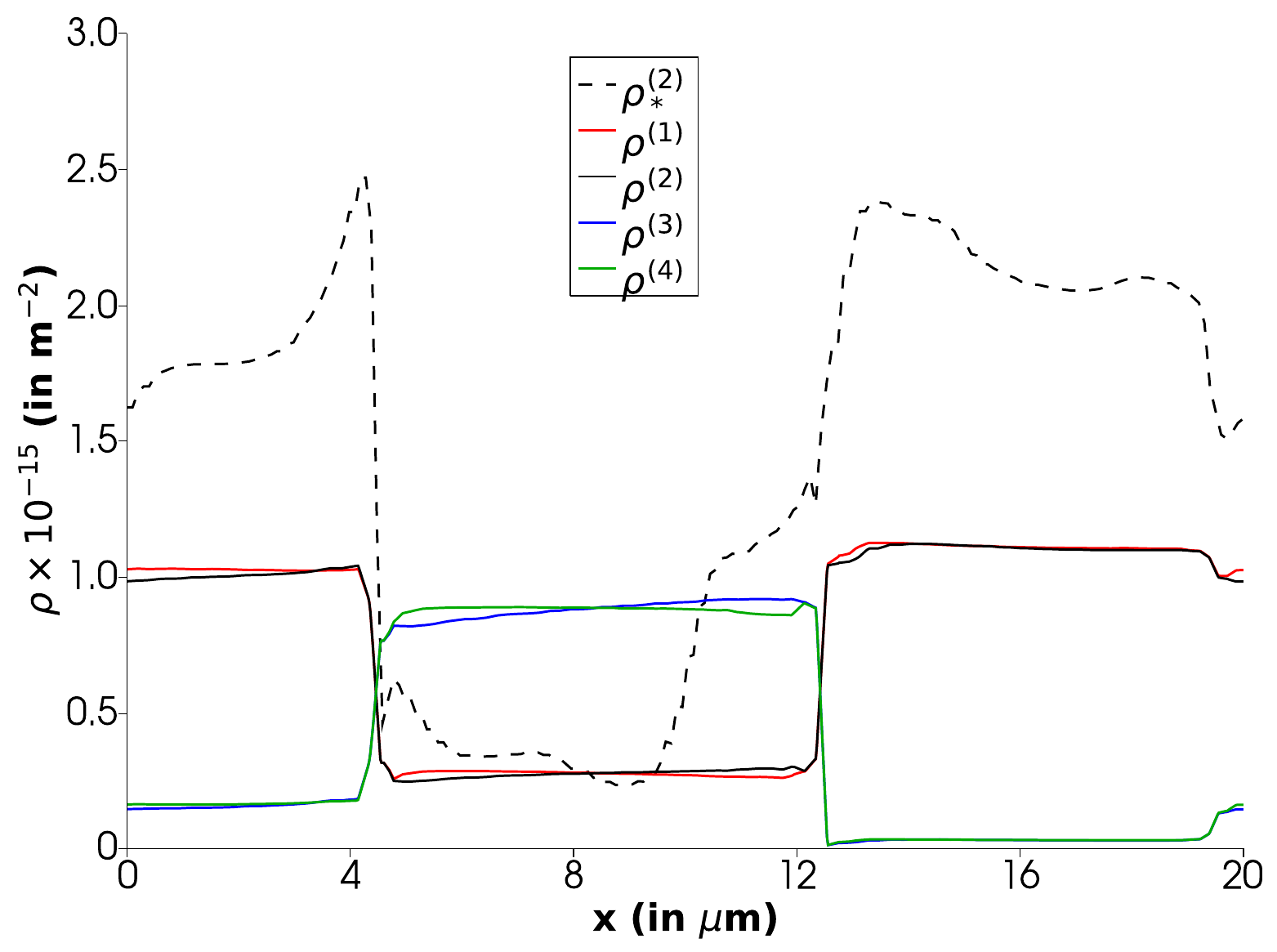}
	\vspace*{-2mm}\hspace*{-10mm}
	\begin{subfigure}[t]{0.32\textwidth}
		\caption{4 slip systems}\label{fig:4slipload55_suba}
	\end{subfigure}
	\hspace*{-11mm}
	\begin{subfigure}[t]{0.32\textwidth}
		\caption{Slip 2 only}
	\end{subfigure}
	\hspace*{5mm}
	\begin{subfigure}[t]{0.32\textwidth}
		\caption{Line plot}
	\end{subfigure}
	\caption{Granular microstructure is loaded in shear with $B_{12}=0.055$ in 5.5 s, where 4 slips systems (a) or only slip system 2 (b) is active. The contours show the total dislocation density. Corresponding dislocation densities are shown across the specimen (c), where dashed line is the single slip case.}
	\label{fig:6grains_4slipcomp}
\end{figure}

Fig. \ref{fig:6grainsloadcomp_c2_09} compares the effect of the amount of \MB{pre-strain} on nucleation, where loading is 5\%, 5.5\% and 6\% before heat treatment. The orientations after heat treatment at $t=10000$\;s are shown. Clearly, the nucleation behavior differs significantly depending on the deformation. In the 5\% case the nucleated grain \MB{G} was absorbed by the neighboring grain \MB{E} before expanding, while in the 6\% case it expanded but has a different final orientation compared to the 5.5\% case.

\begin{figure}[ht!]
	\centering
	\hspace*{0mm}
	\raisebox{1.8cm}{\includegraphics[width=0.02\textwidth]{figure_t2000.pdf}}
	\includegraphics[width=0.285\textwidth,trim={1cm 0 1cm 0cm},clip]{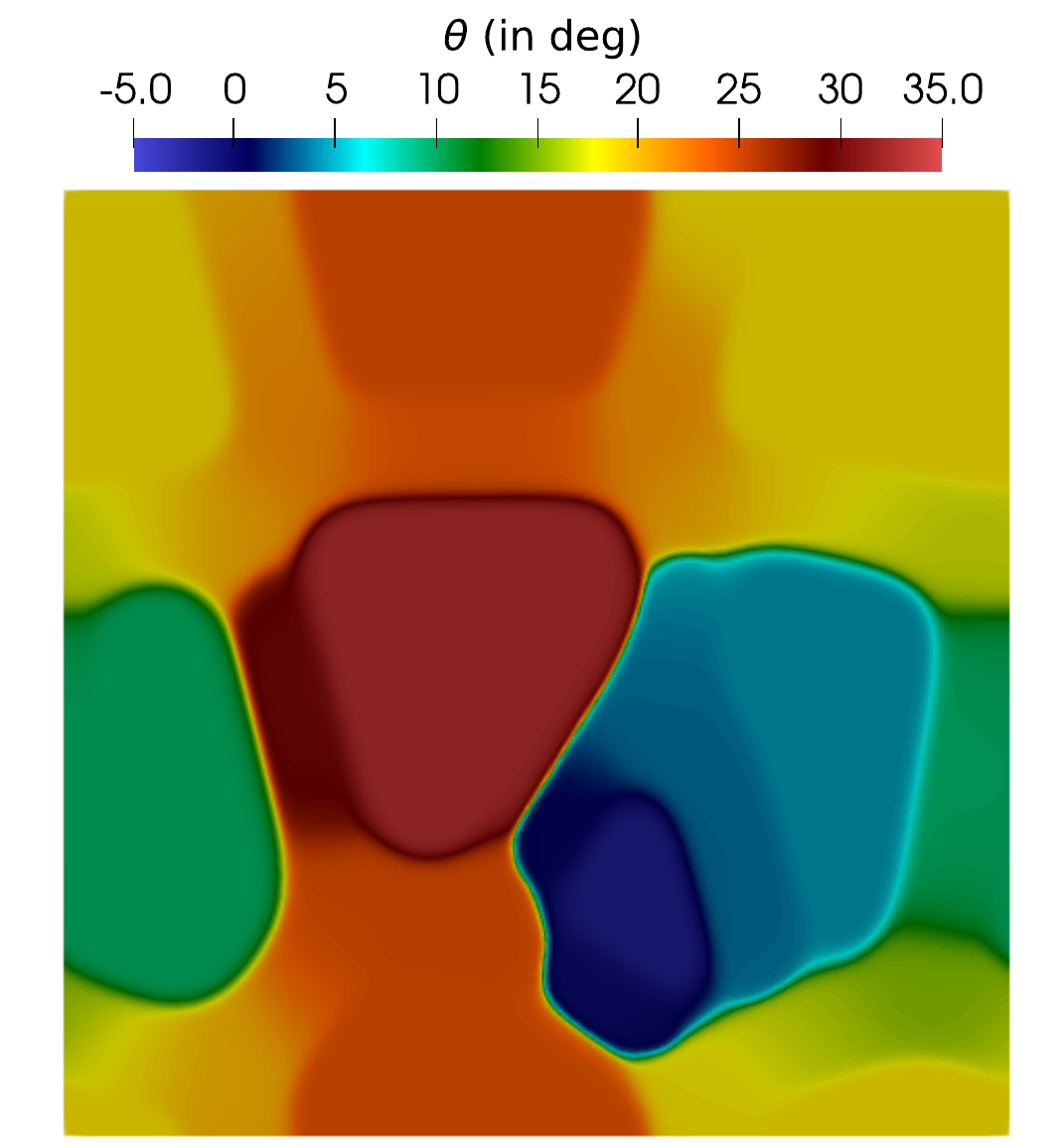}
	\includegraphics[width=0.285\textwidth,trim={1cm 0 1cm 0cm},clip]{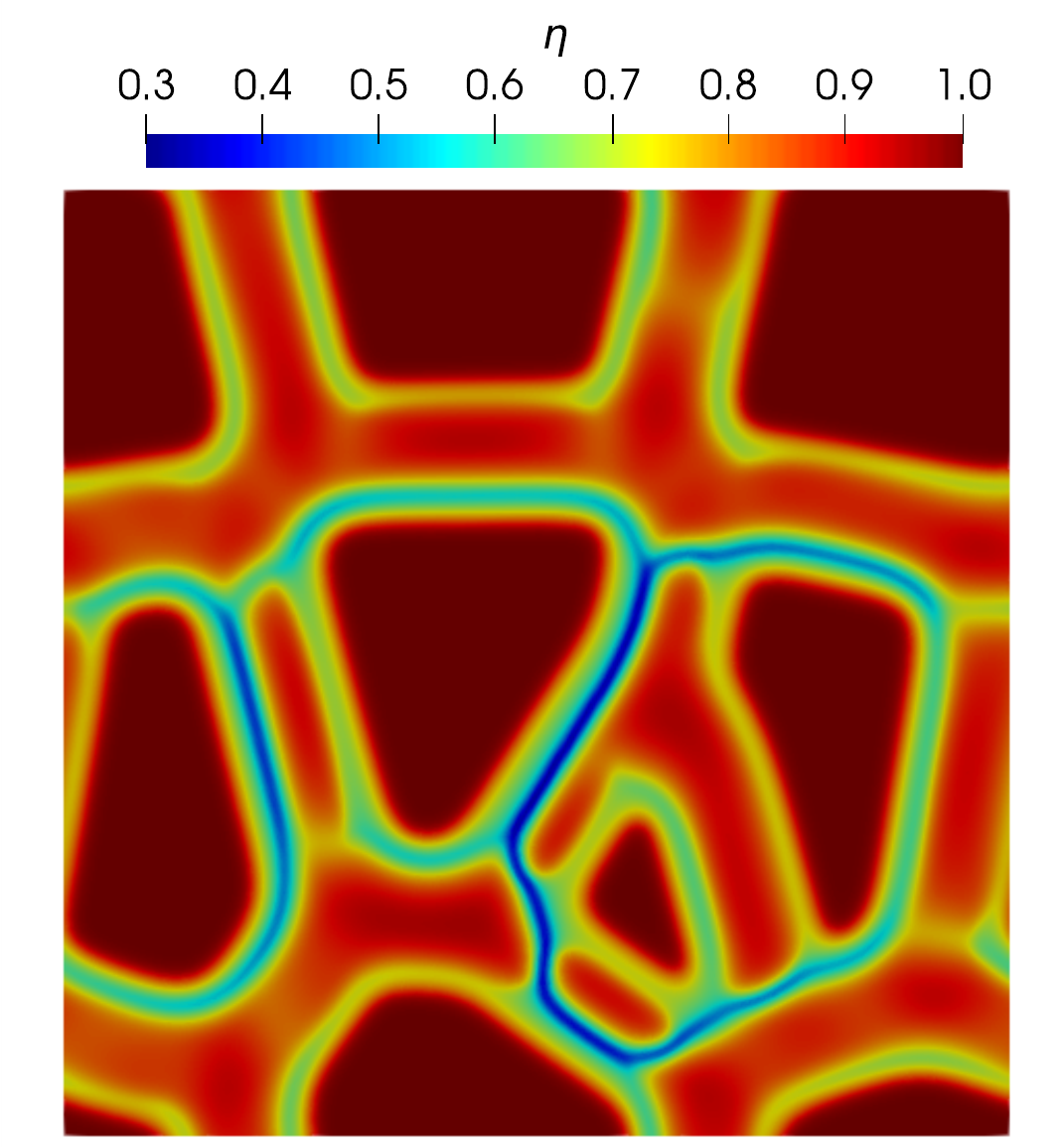}
	\includegraphics[width=0.285\textwidth,trim={1cm 0 1cm 0cm},clip]{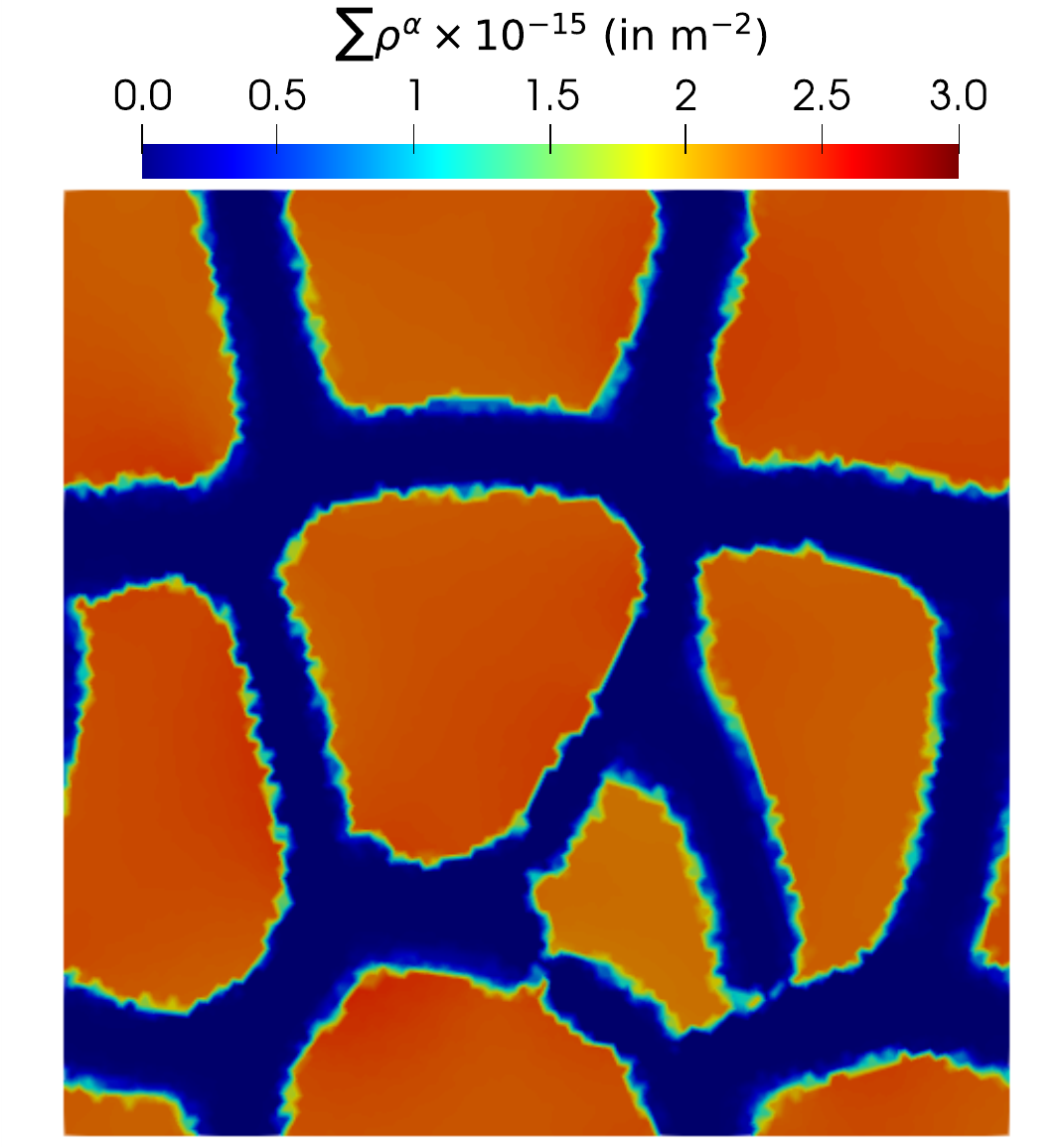}
 
	\hspace*{0mm}
	\raisebox{1.8cm}{\includegraphics[width=0.02\textwidth]{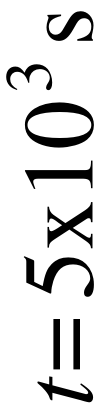}}
	\includegraphics[width=0.285\textwidth,trim={1cm 0 1cm 3.1cm},clip]{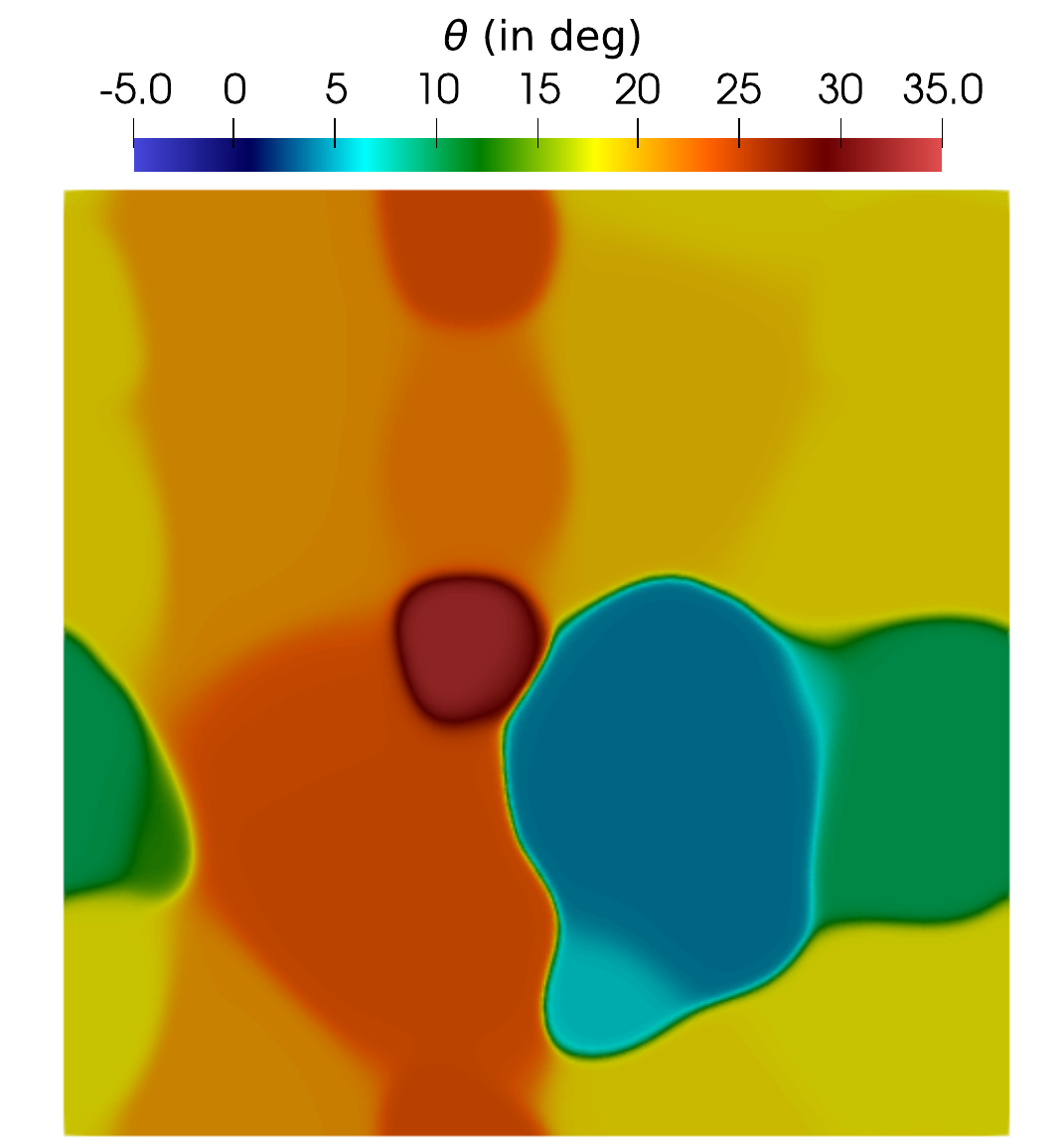}
	\includegraphics[width=0.285\textwidth,trim={1cm 0 1cm 3.1cm},clip]{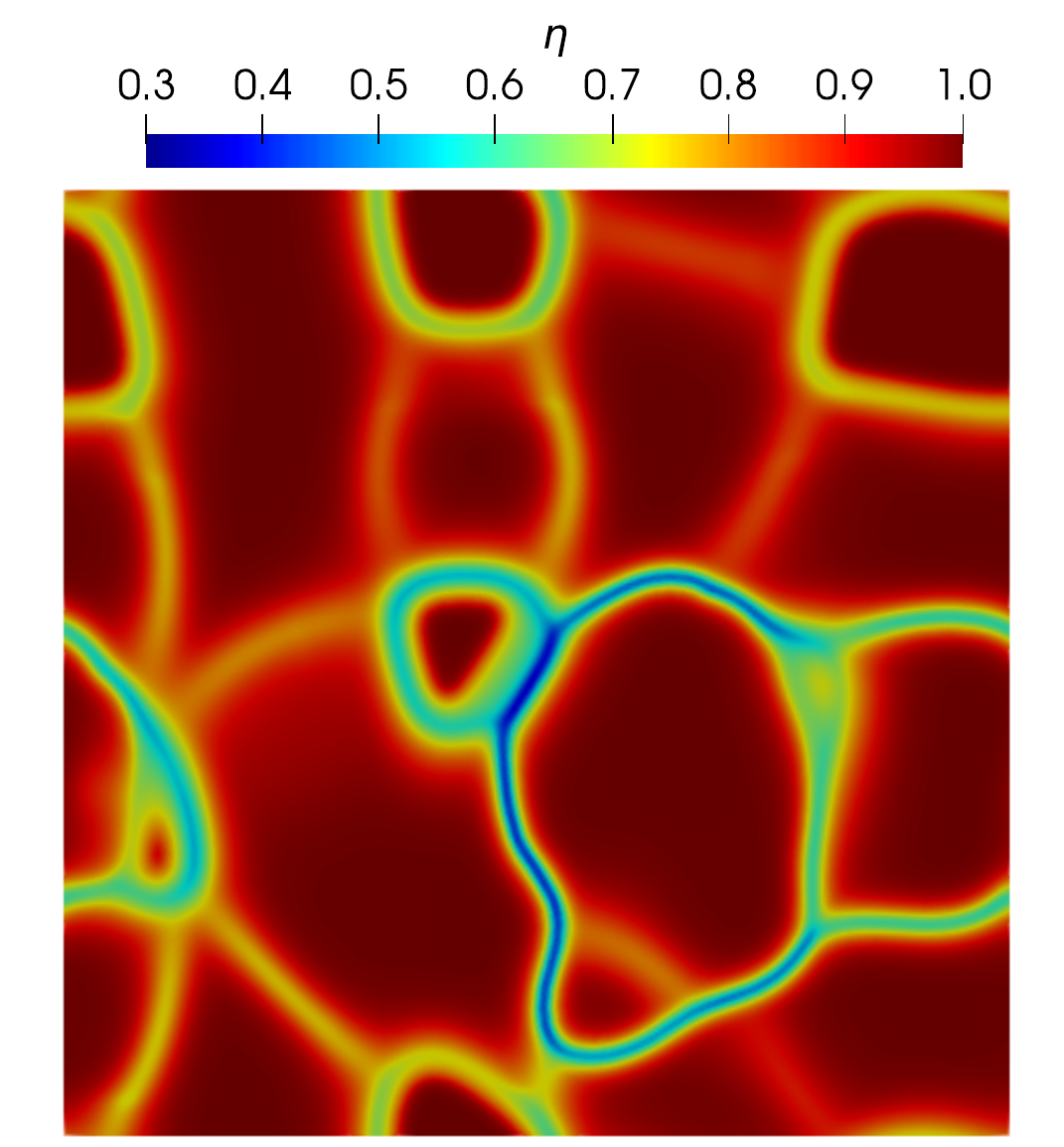}
	\includegraphics[width=0.285\textwidth,trim={1cm 0 1cm 3.1cm},clip]{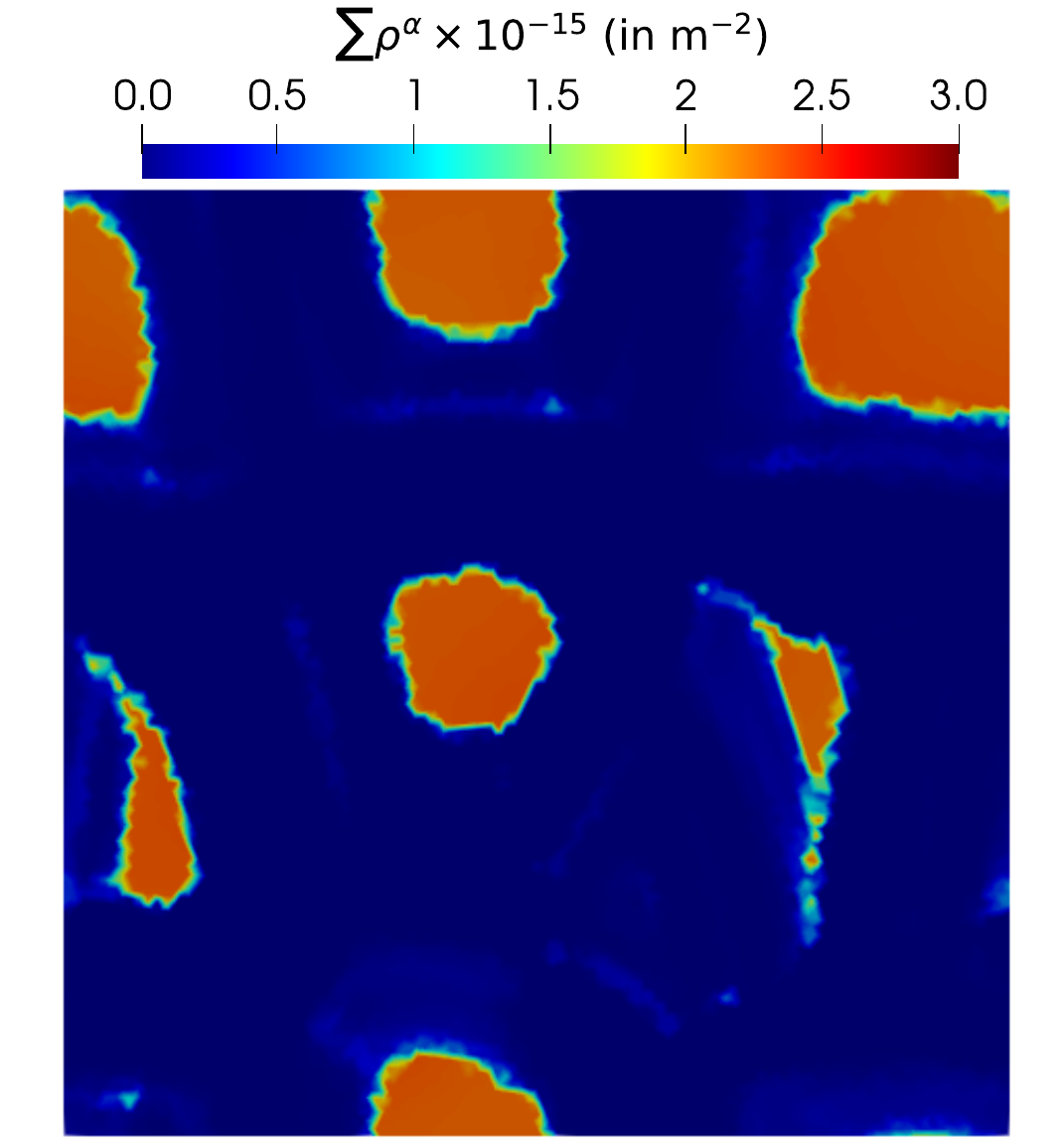}
 
	\hspace*{0mm}
	\raisebox{1.8cm}{\includegraphics[width=0.02\textwidth]{figure_t10000.pdf}}
	\includegraphics[width=0.285\textwidth,trim={1cm 0 1cm 3.1cm},clip]{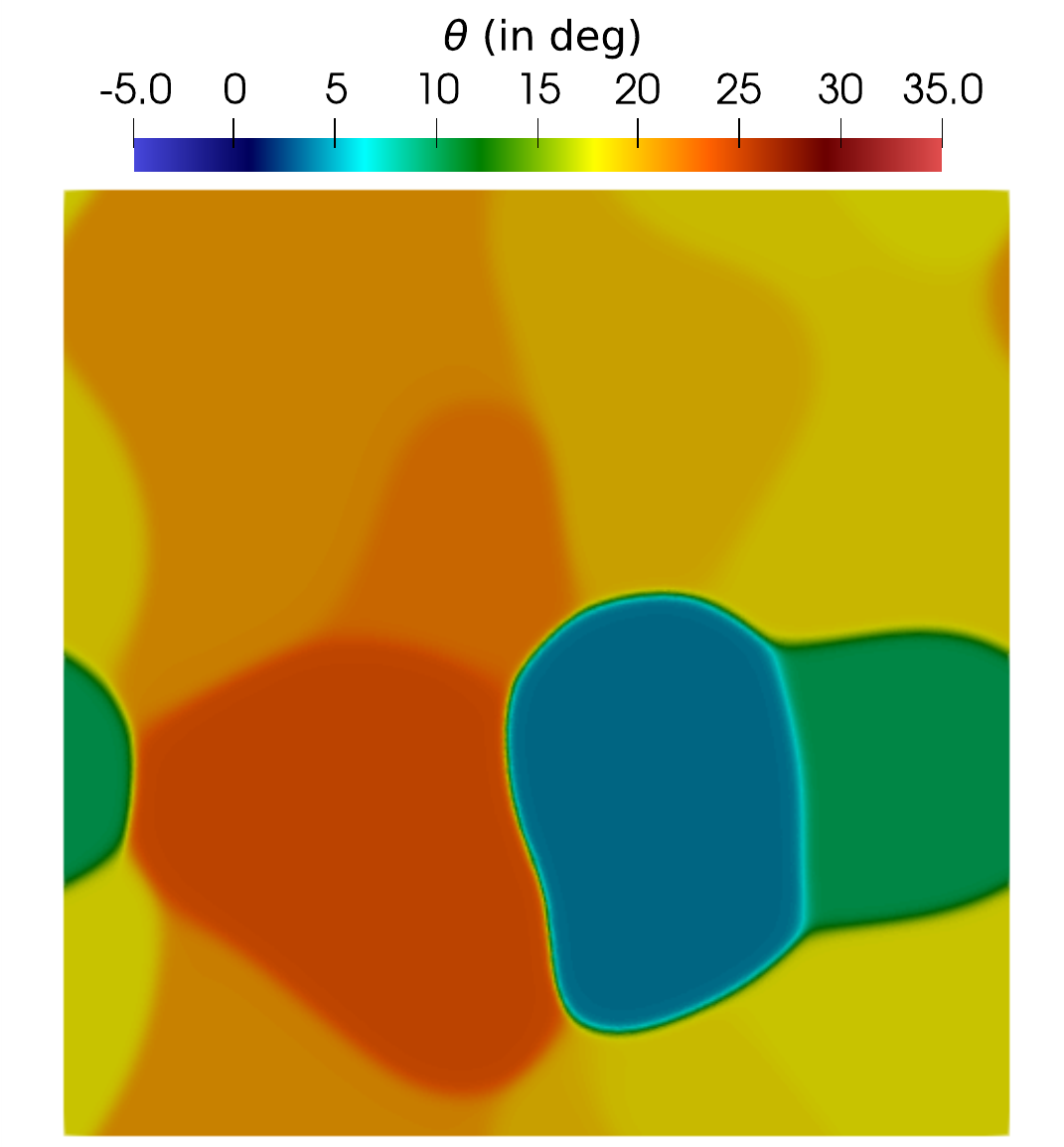}
	\includegraphics[width=0.285\textwidth,trim={1cm 0 1cm 3.1cm},clip]{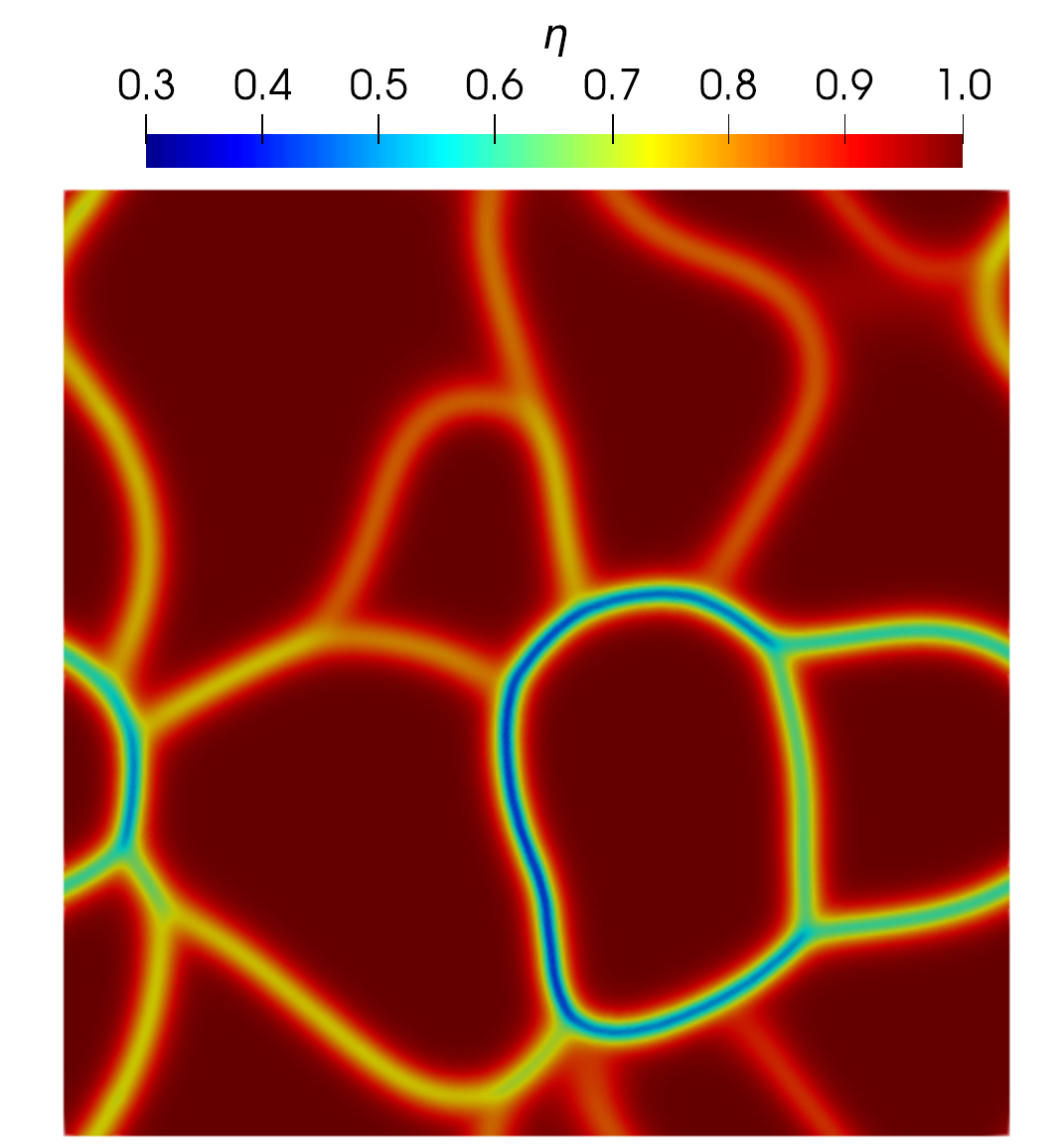}
	\includegraphics[width=0.285\textwidth,trim={1cm 0 1cm 3.1cm},clip]{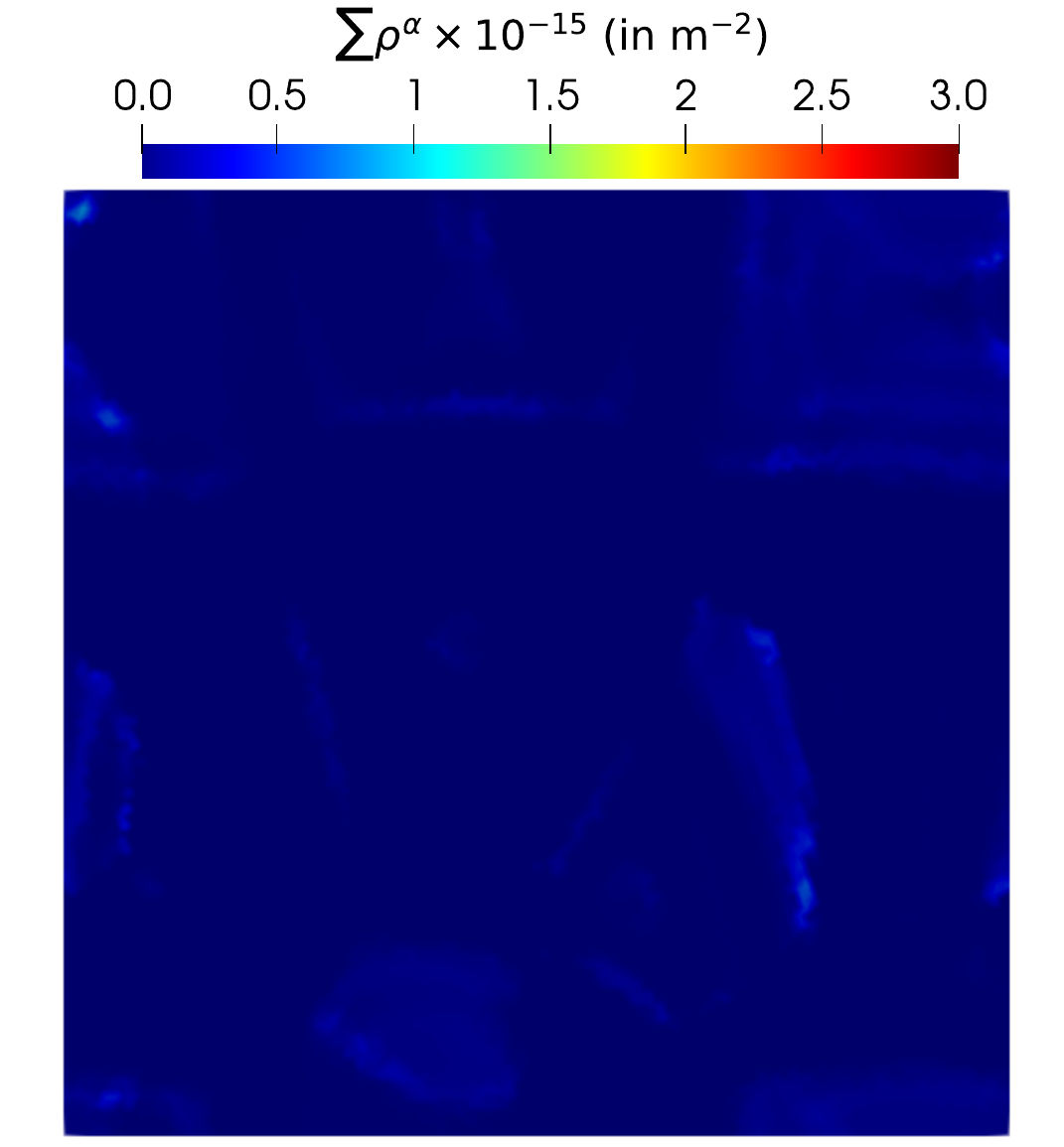}
	\caption{The deformed structure in Fig. \ref{fig:6grainsload55_subc} and Fig. \ref{fig:4slipload55_suba} is allowed to recrystallize with $c_2=0.9$ which activates grain nucleation. From left to right orientation $\theta$, order parameter $\eta$ and total statically stored dislocation density $\sum\rho^{\alpha}$ are shown at different times.}
	\label{fig:6grainsload55_4slip_c2_09}
\end{figure}

The localized deformation is pronounced, when only a single slip system is active. When all four \MB{systems} are activated the deformation becomes much more homogeneous inside the grains, as seen in Fig. \ref{fig:6grains_4slipcomp}a, where the total dislocation density of all 4 slip systems is plotted. Fig. \ref{fig:6grains_4slipcomp}c compares the individual dislocation densities along the center line shown in Fig. \ref{fig:6grains_4slipcomp}a/b. The deformation is evenly distributed in the bulk, but varies from grain to grain. While the individual values of $\rho^\alpha$ are lower compared to the case of single slip, the total value of dislocation density is higher, and it is the total value that enters into the driving \MB{force for $\eta$} in Eq. \eqref{eqn:hmpetadot}. Fig. \ref{fig:6grainsload55_4slip_c2_09} shows the evolution during heat treatment for the case of 4 \MB{active} slip systems. In this case, nucleation is triggered at almost all of the grain boundaries, since the total $\rho$ is high and more evenly distributed. Most of the nuclei expand and a new microstructure with new orientations is formed. However, the new orientations are \MB{bounded by} those of the adjacent \MB{parent} grains. Dislocations are almost completely recovered within the new grains.

\begin{figure}[bp]
	\centering
	\hspace*{0mm}
	\raisebox{2cm}{\includegraphics[width=0.020\textwidth]{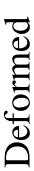}}
    \begin{tikzpicture}
	\draw(0,0)node[inner sep=0]{\includegraphics[width=0.285\textwidth,trim={5mm 0 5mm 0cm},clip]{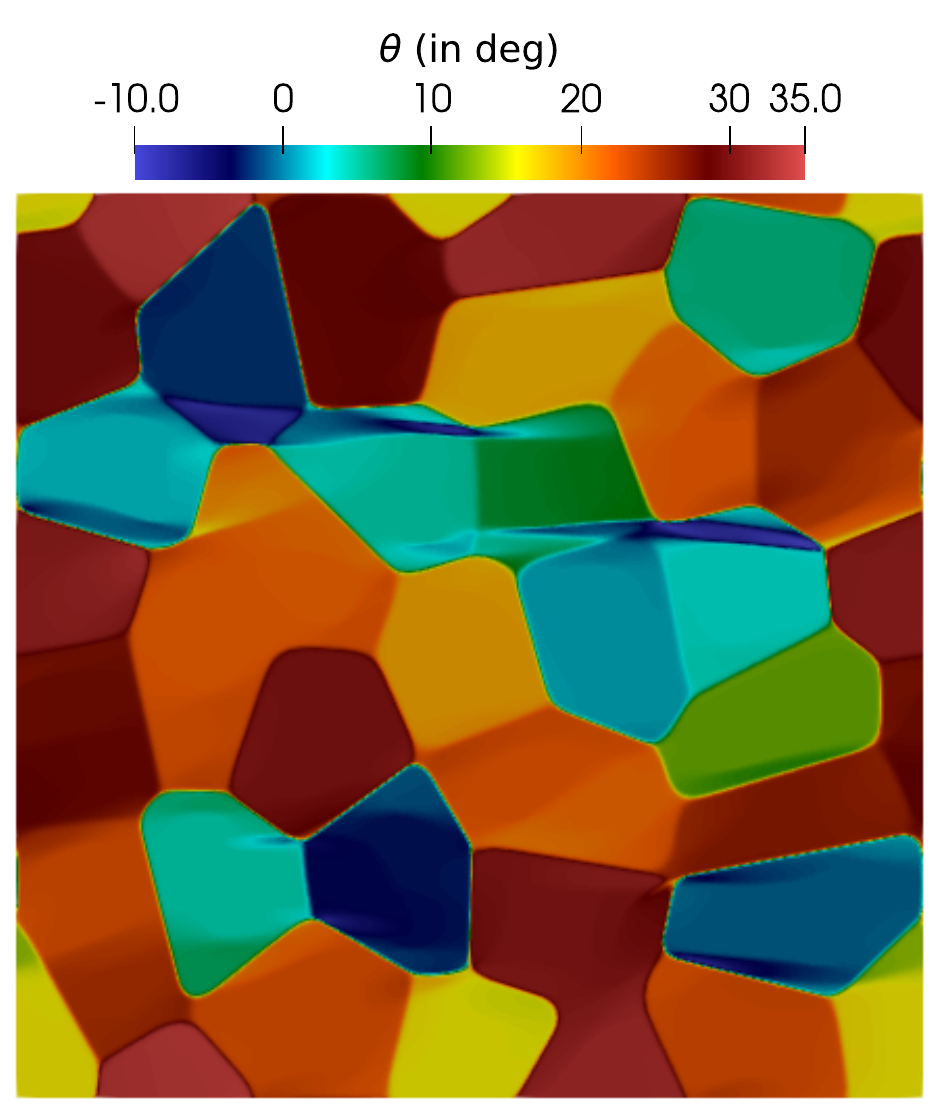}};
    \draw(-2.1,-1.9)node{\scalebox{0.8}{A$_1$}}; \draw(-1.9,-2.5)node{\scalebox{0.8}{A$_2$}};
    \draw(0.3,-1.2)node{\scalebox{0.8}{B$_1$}}; \draw(0.8,-1.5)node{\scalebox{0.8}{B$_2$}};
    %\draw(-2.1,-0.7)node{\scalebox{0.8}{C$_1$}}; \draw(-2,-1.2)node{\scalebox{0.8}{C$_2$}};
    \draw(-1.5,-0.3)node{\scalebox{0.8}{C$_1$}}; \draw(-1.55,-1.1)node{\scalebox{0.8}{C$_2$}};
    \end{tikzpicture}
	\includegraphics[width=0.285\textwidth,trim={5mm 0 5mm 0cm},clip]{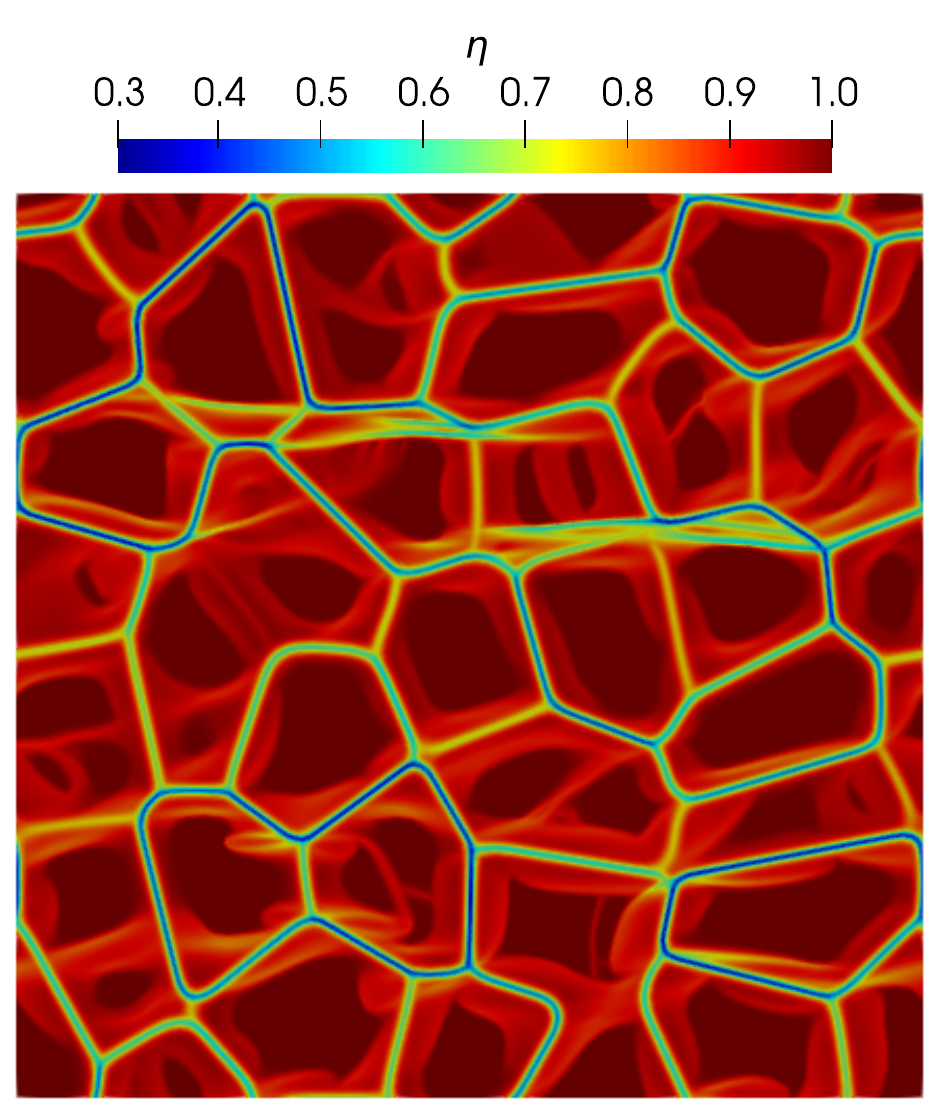}
	\includegraphics[width=0.285\textwidth,trim={5mm 0 5mm 0cm},clip]{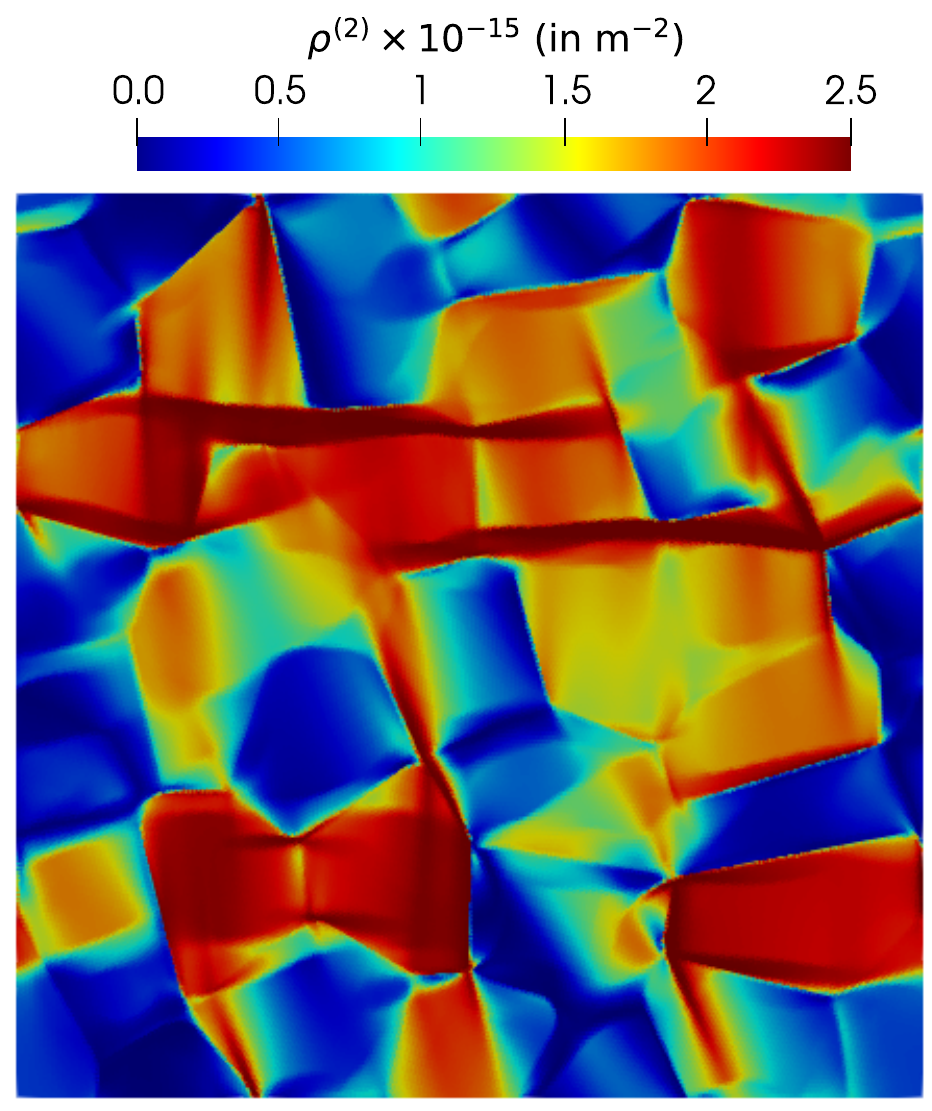}

    \hspace*{0mm}
	\raisebox{2cm}{\includegraphics[width=0.020\textwidth]{figure_t2000.pdf}}
    \begin{tikzpicture}
	\draw(0,0)node[inner sep=0]{\includegraphics[width=0.285\textwidth,trim={5mm 0 5mm 3.2cm},clip]{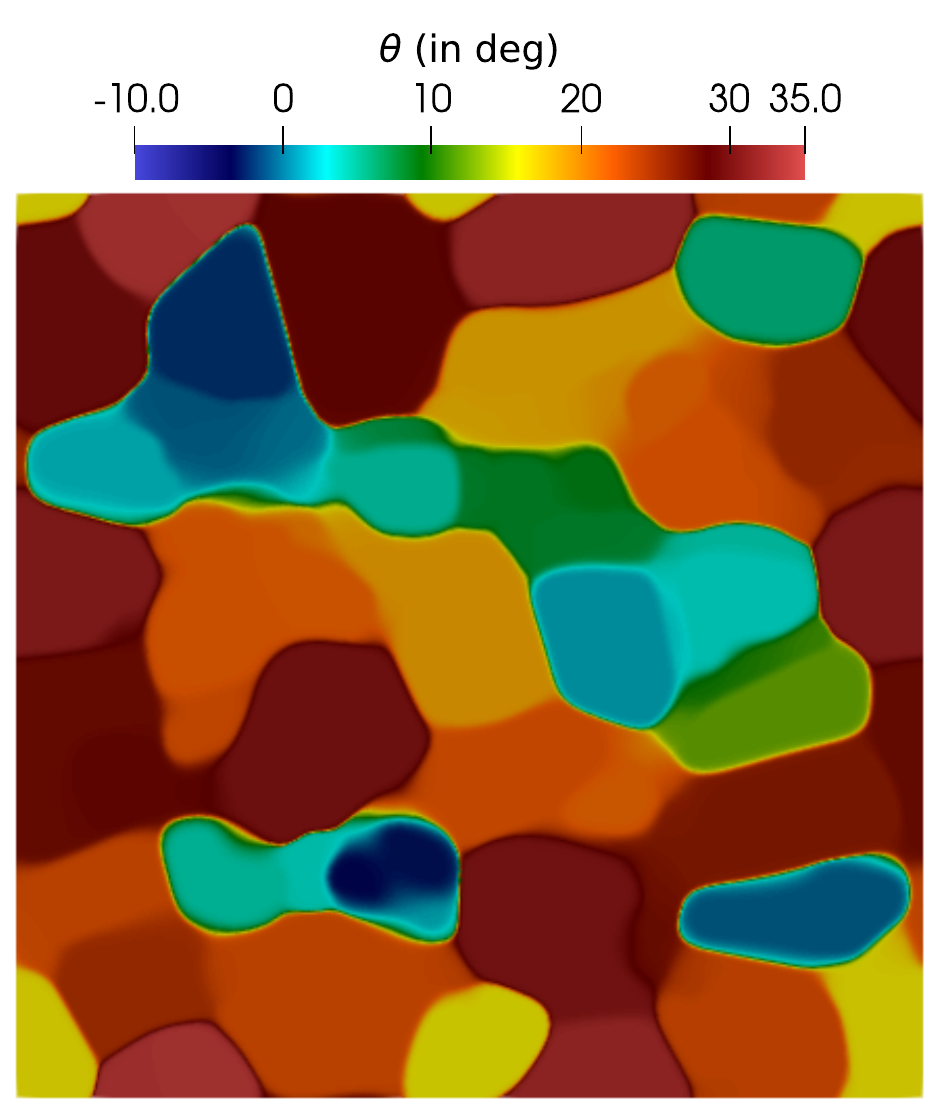}};
    \draw(-1.4,0.8)node{\textcolor{white}{$\Uparrow$}}; \draw(-0.9,-1.7)node{\textcolor{white}{$\Uparrow$}}; \draw(1.7,-0.4)node{\textcolor{white}{$\Uparrow$}};
    \draw(-1.2,0.6)node{\textcolor{white}{\scalebox{0.8}{1}}}; \draw(-0.7,-1.9)node{\textcolor{white}{\scalebox{0.8}{2}}}; \draw(1.9,-0.6)node{\textcolor{white}{\scalebox{0.8}{3}}};
    \end{tikzpicture}
    \begin{tikzpicture}
	\draw(0,0)node[inner sep=0]{\includegraphics[width=0.285\textwidth,trim={5mm 0 5mm 3.2cm},clip]{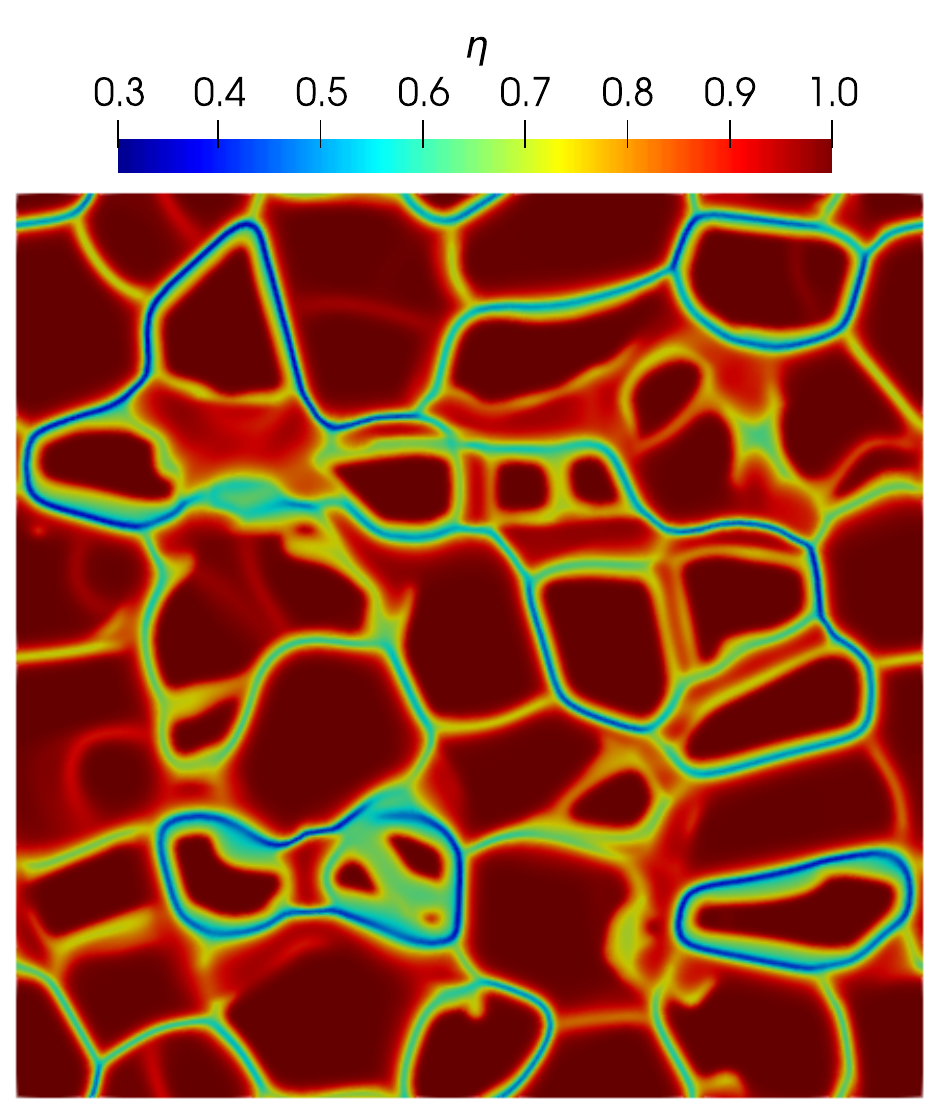}};
    \draw(-1.4,0.8)node{\textcolor{white}{$\Uparrow$}}; \draw(-0.9,-1.7)node{\textcolor{white}{$\Uparrow$}}; \draw(1.7,-0.4)node{\textcolor{white}{$\Uparrow$}};
    \end{tikzpicture}
    \begin{tikzpicture}
	\draw(0,0)node[inner sep=0]{\includegraphics[width=0.285\textwidth,trim={5mm 0 5mm 3.2cm},clip]{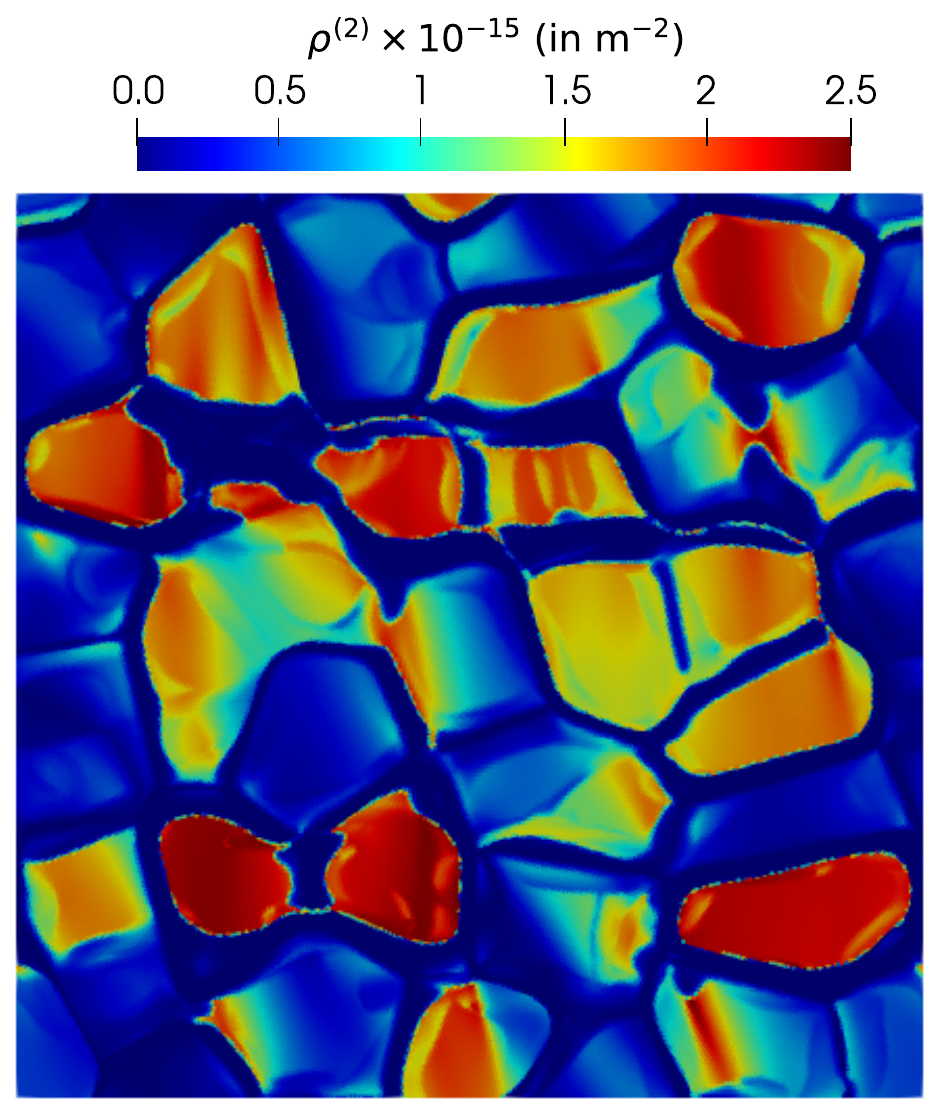}};
    \draw(-1.4,0.8)node{\textcolor{white}{$\Uparrow$}}; \draw(-0.9,-1.7)node{\textcolor{white}{$\Uparrow$}}; \draw(1.7,-0.4)node{\textcolor{white}{$\Uparrow$}};
    \end{tikzpicture}
 
	\hspace*{0mm}
	\raisebox{2cm}{\includegraphics[width=0.020\textwidth]{figure_t5000.pdf}}
	\includegraphics[width=0.285\textwidth,trim={5mm 0 5mm 3.2cm},clip]{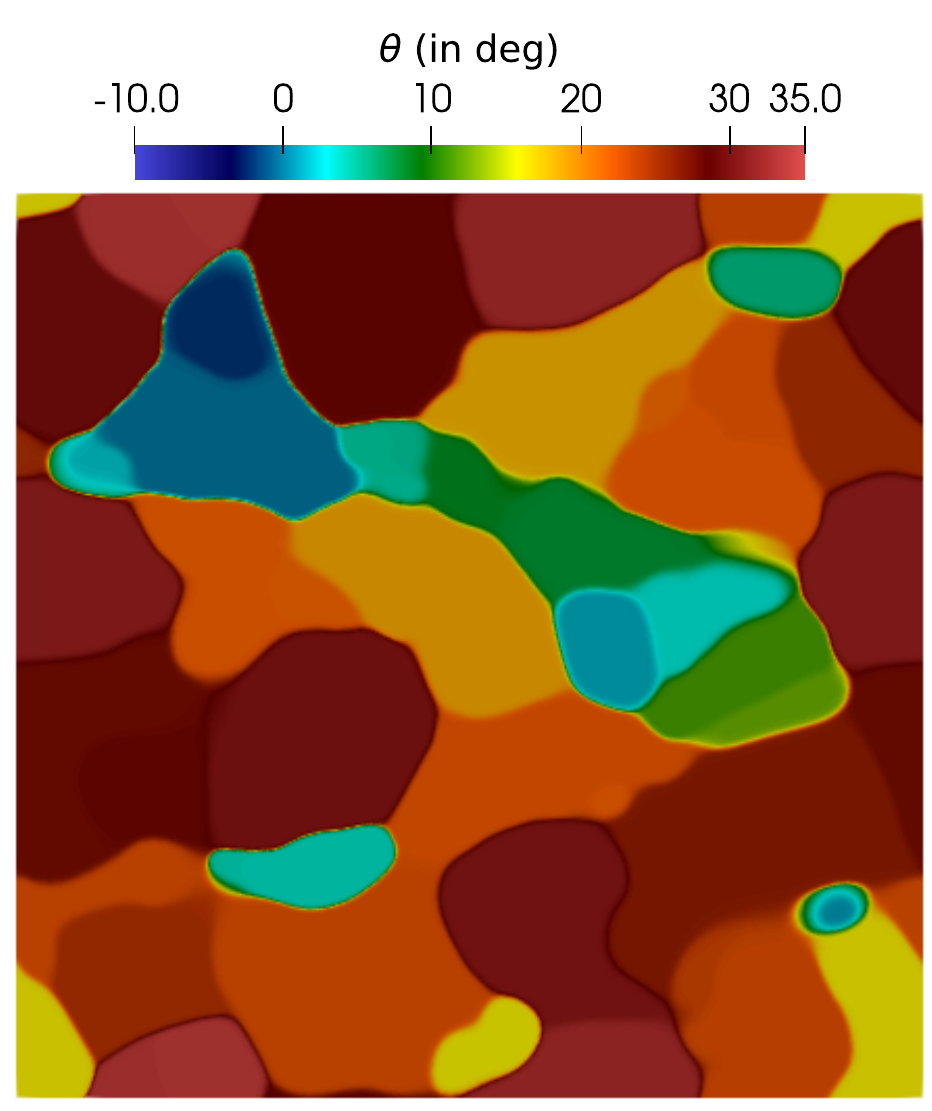}
	\includegraphics[width=0.285\textwidth,trim={5mm 0 5mm 3.2cm},clip]{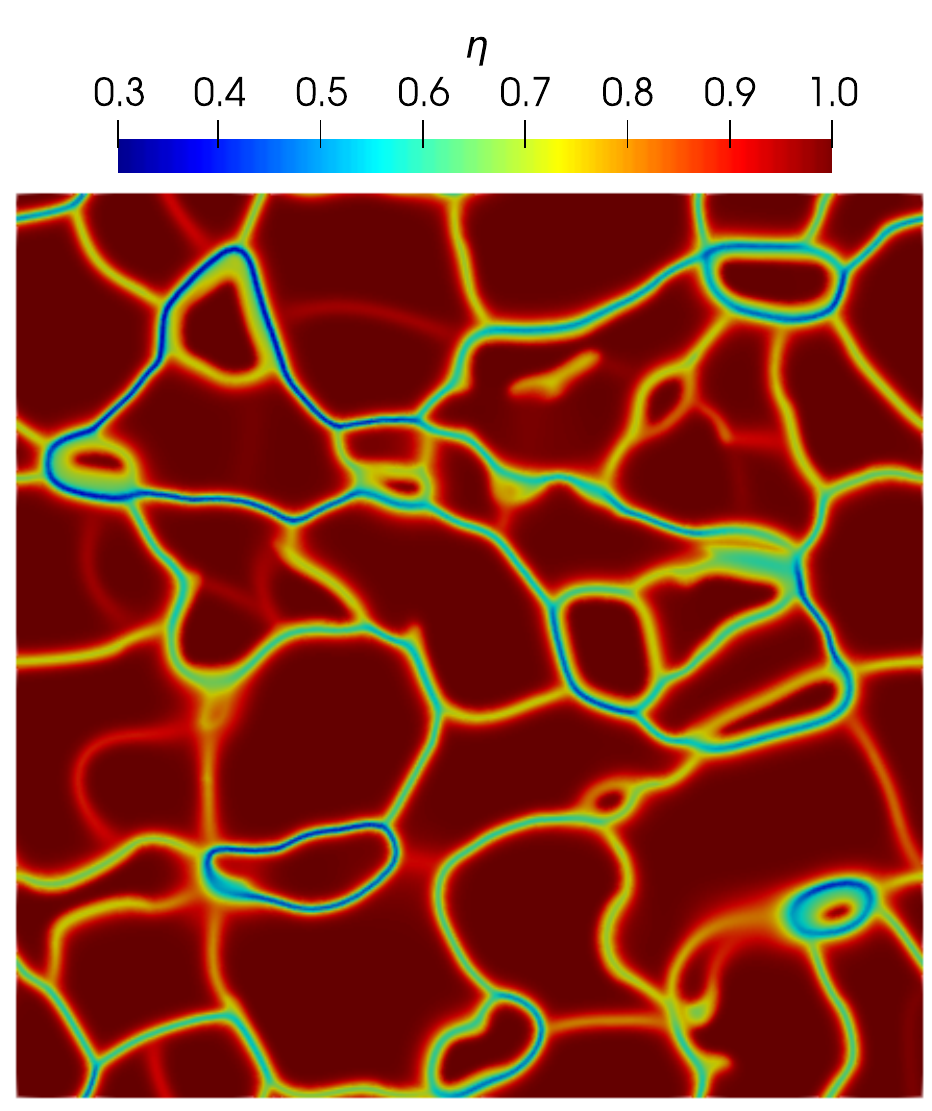}
	\includegraphics[width=0.285\textwidth,trim={5mm 0 5mm 3.2cm},clip]{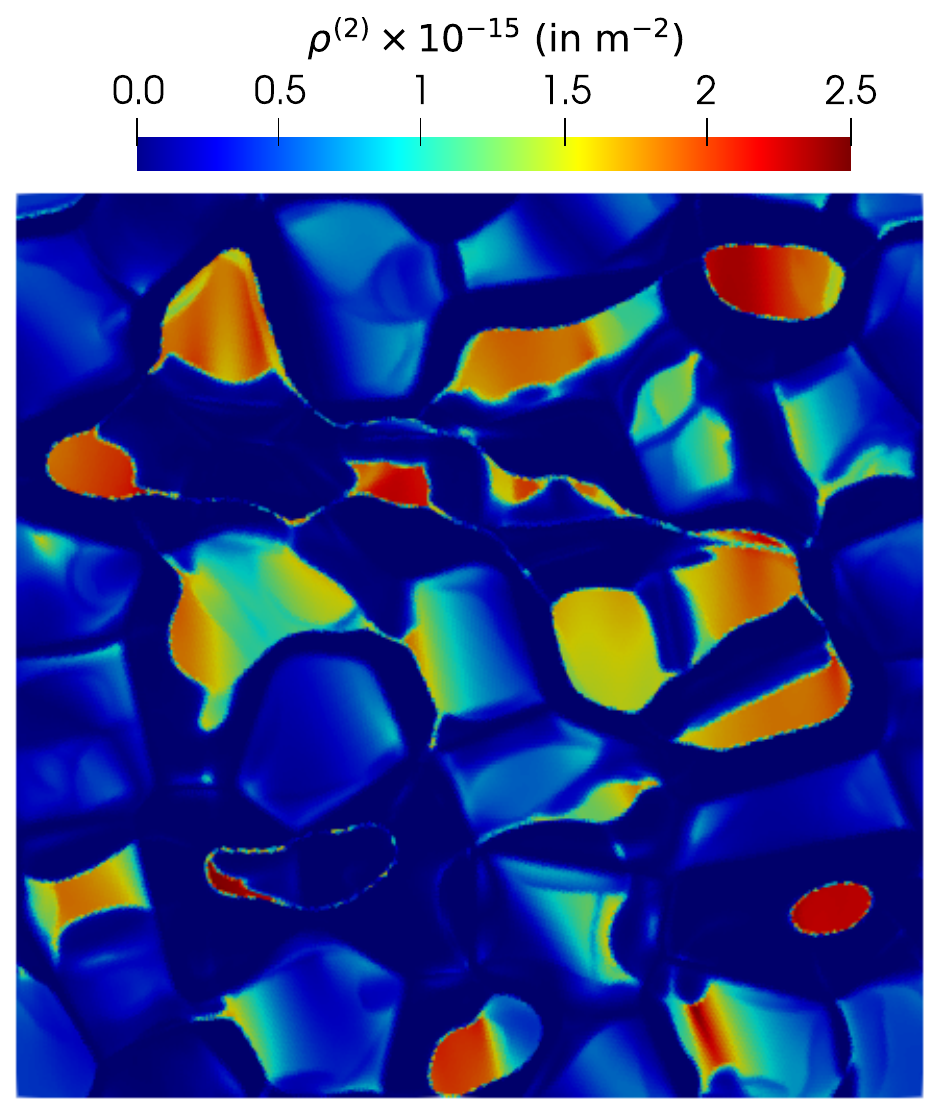}
	
	\hspace*{0mm}
	\raisebox{2cm}{\includegraphics[width=0.020\textwidth]{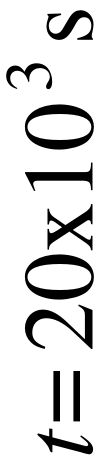}}
	\includegraphics[width=0.285\textwidth,trim={5mm 0 5mm 3.2cm},clip]{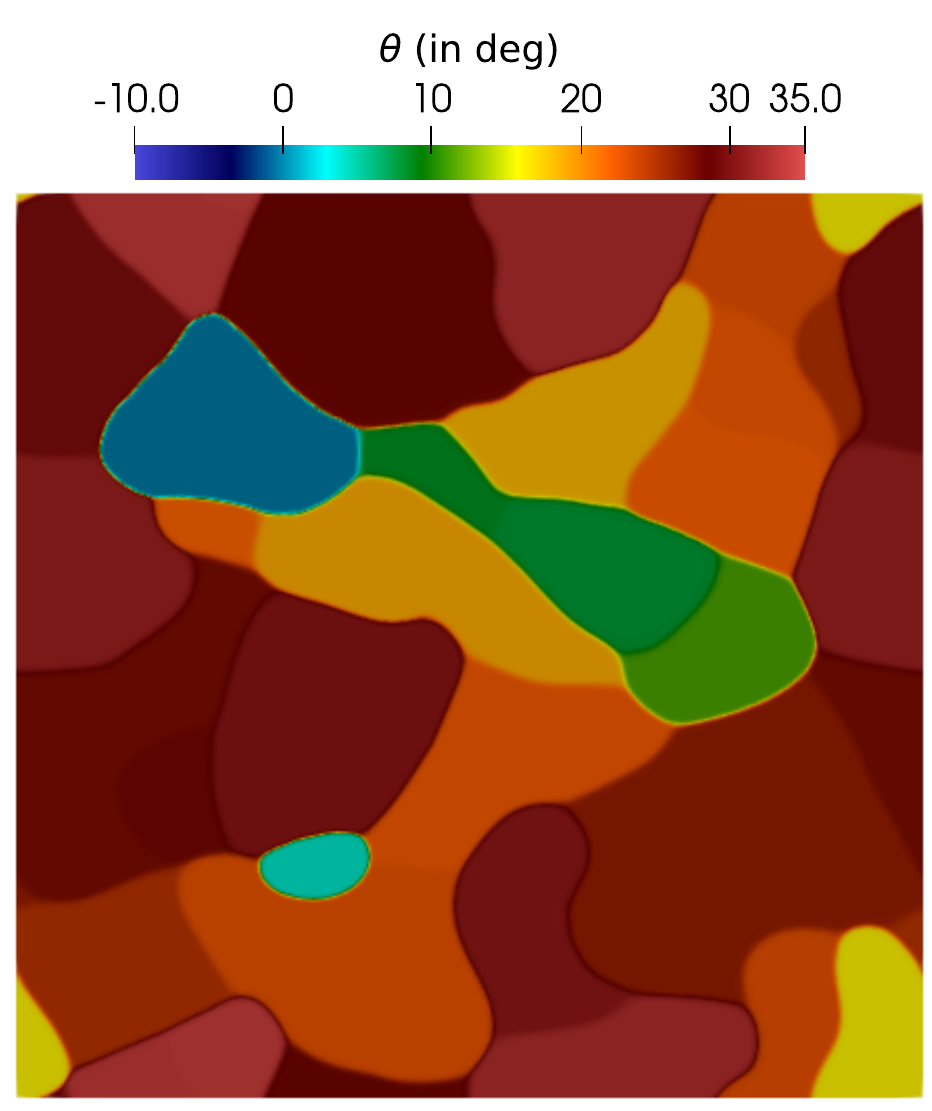}
	\includegraphics[width=0.285\textwidth,trim={5mm 0 5mm 3.2cm},clip]{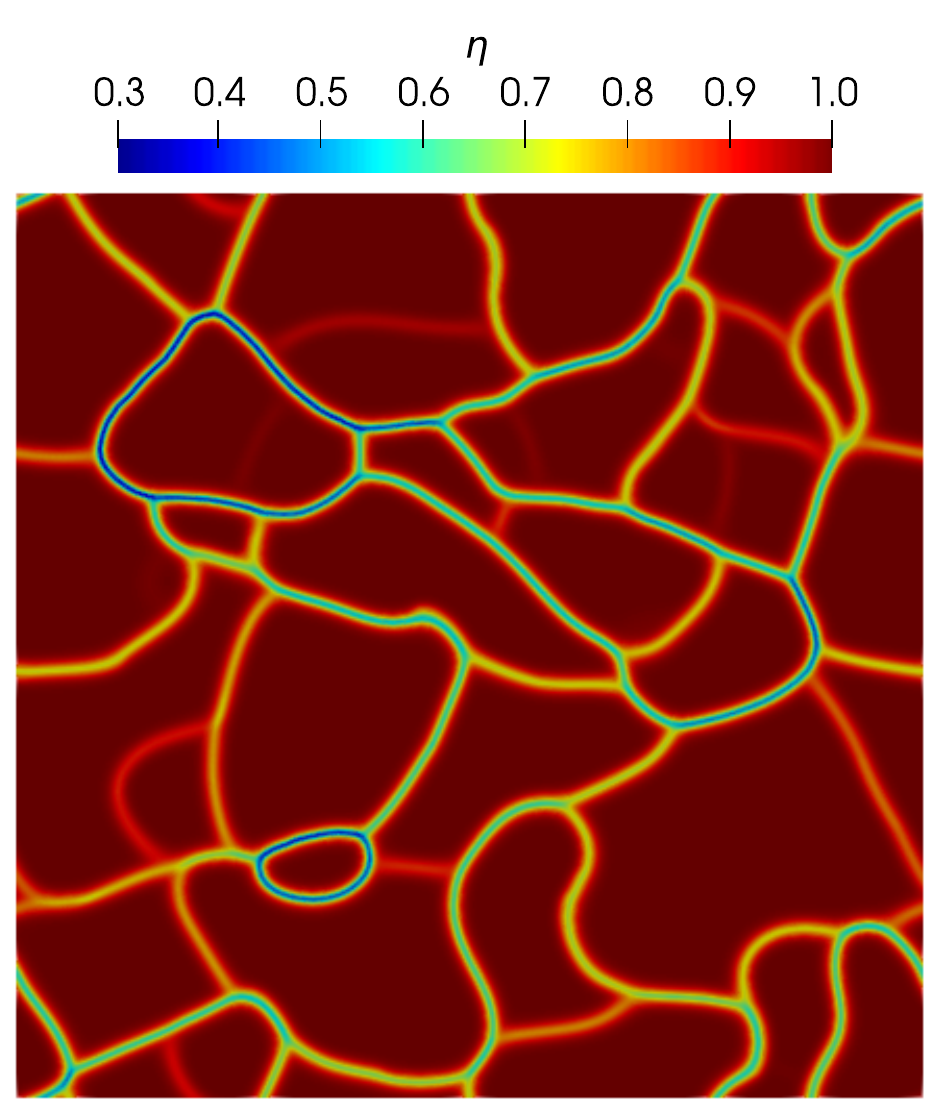}
	\includegraphics[width=0.285\textwidth,trim={5mm 0 5mm 3.2cm},clip]{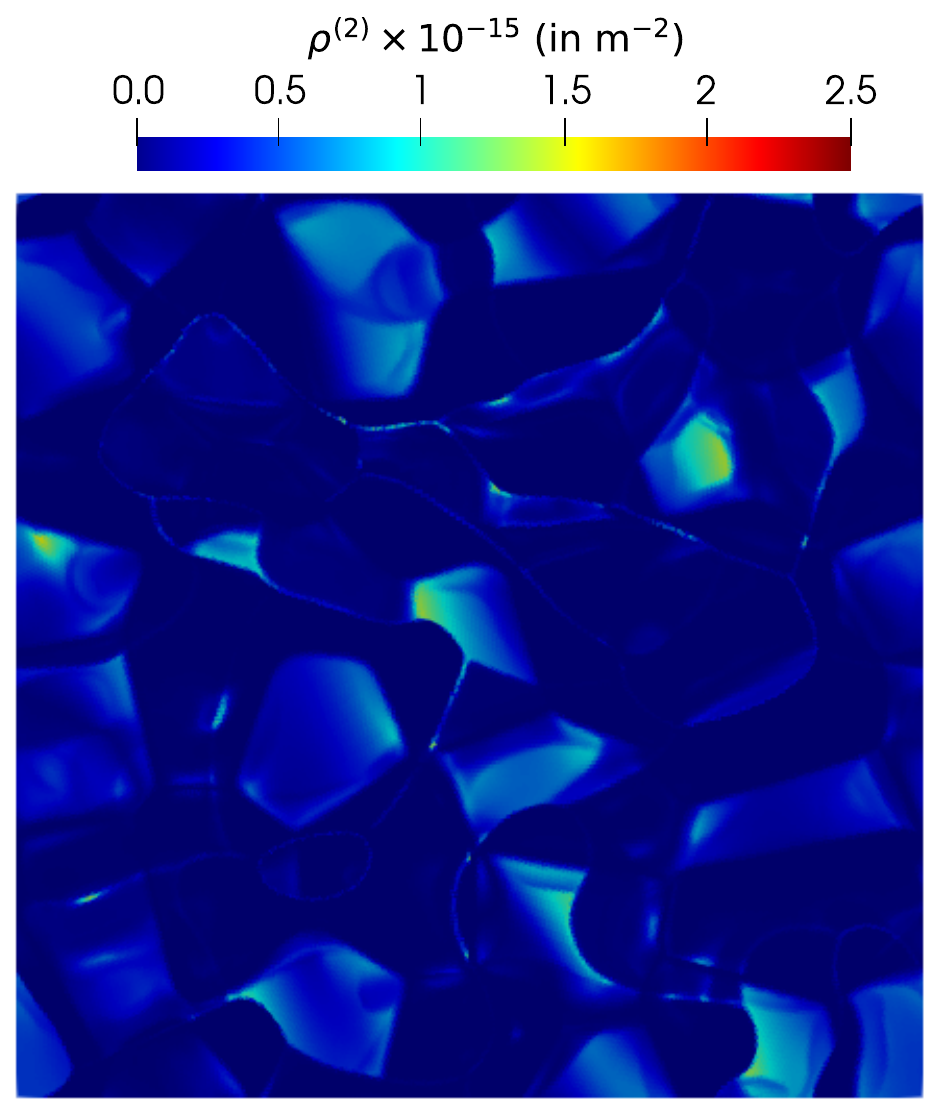}
	\caption{A granular structure with 32 grains are deformed in shear by $B_{12}=0.05$ with only slip system 2 active. Then, it is allowed to recrystallize with $c_2=0.9$ which activates grain nucleation, \MB{while holding displacements constant}. From left to right orientation $\theta$, order parameter $\eta$ and statically stored dislocation density $\rho^{(2)}$ are shown at different times. The arrows at $t=2\times 10^3$\;s show some of nucleation locations.}
	\label{fig:32grainsload5_slip2_c2_09}
\end{figure}

\begin{figure}[bp]
	\centering
    \hspace*{0mm}
	\raisebox{2cm}{\includegraphics[width=0.020\textwidth]{figure_t2000.pdf}}
	\includegraphics[width=0.285\textwidth,trim={5mm 0 5mm 0cm},clip]{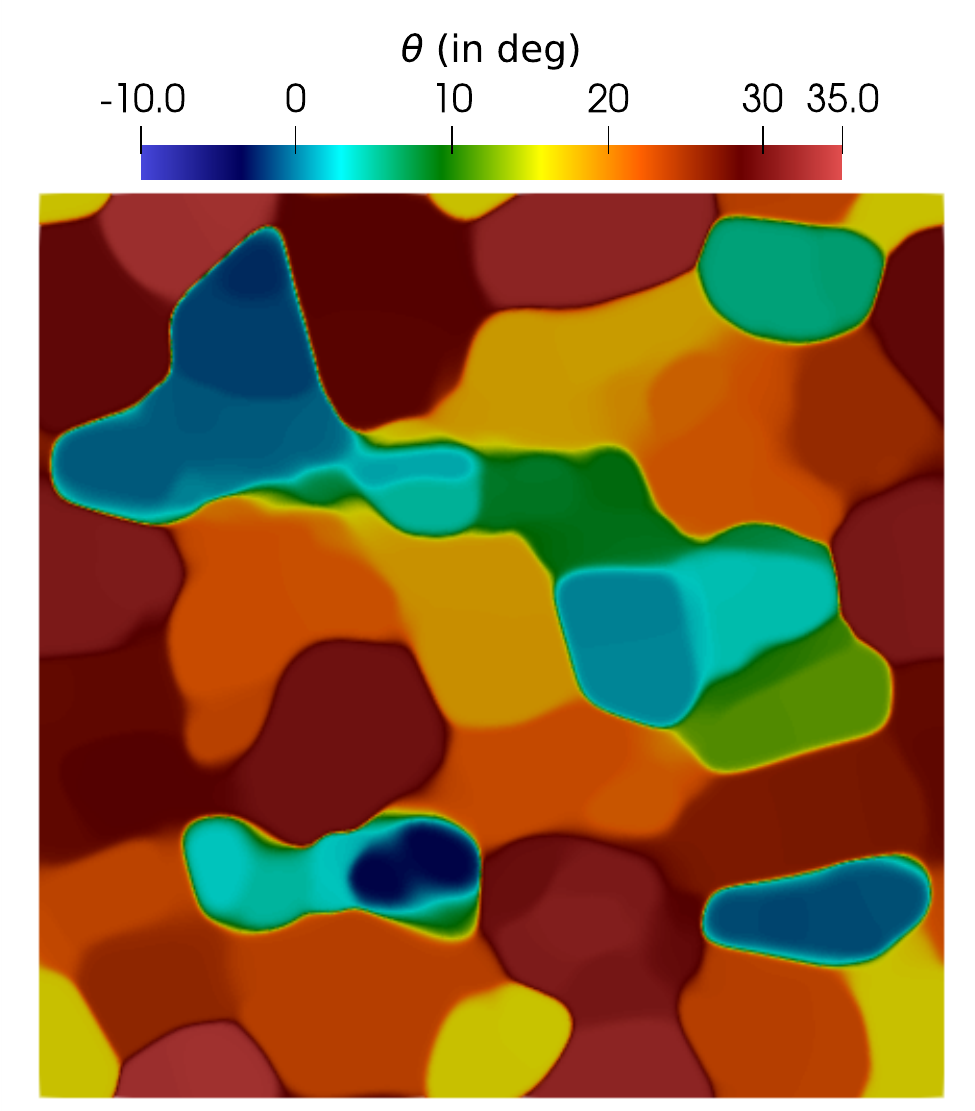}
	\includegraphics[width=0.285\textwidth,trim={5mm 0 5mm 0cm},clip]{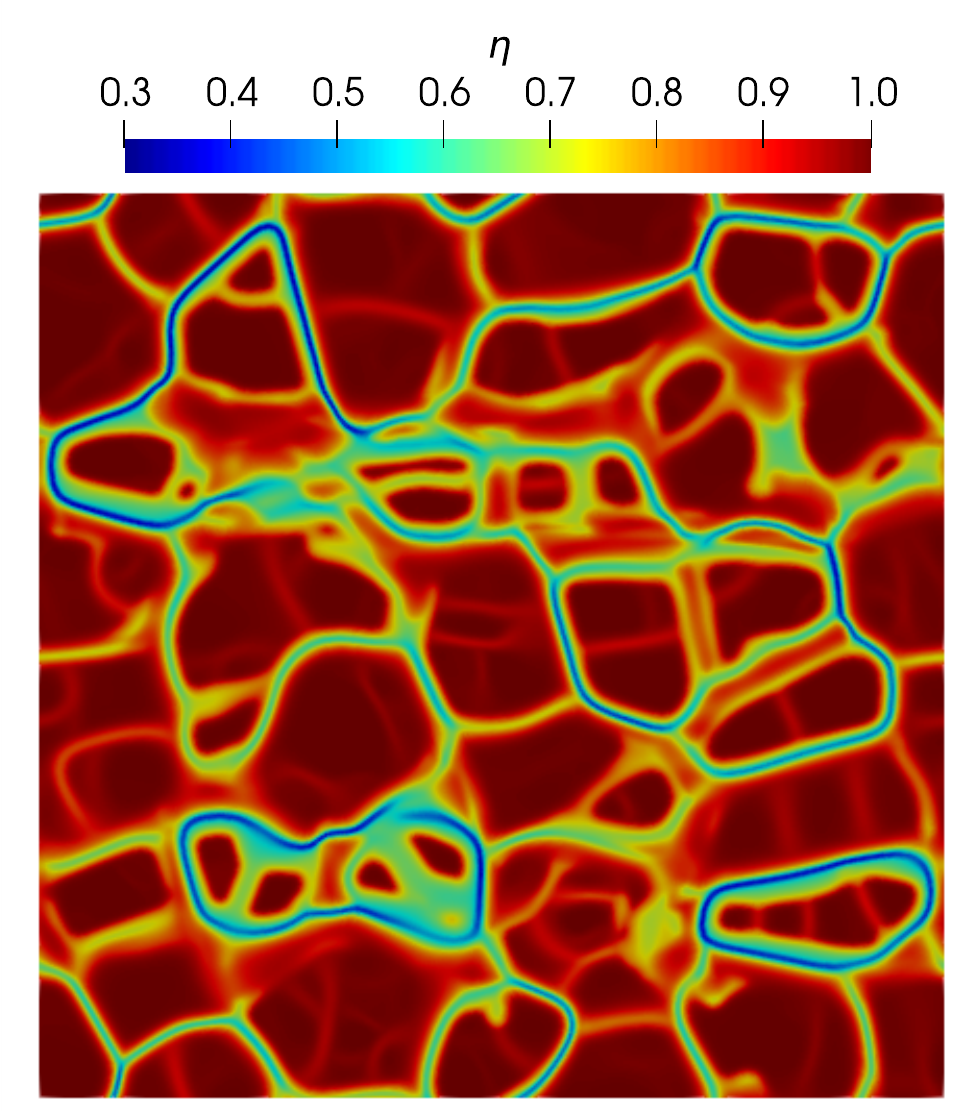}
	\includegraphics[width=0.285\textwidth,trim={5mm 0 5mm 0cm},clip]{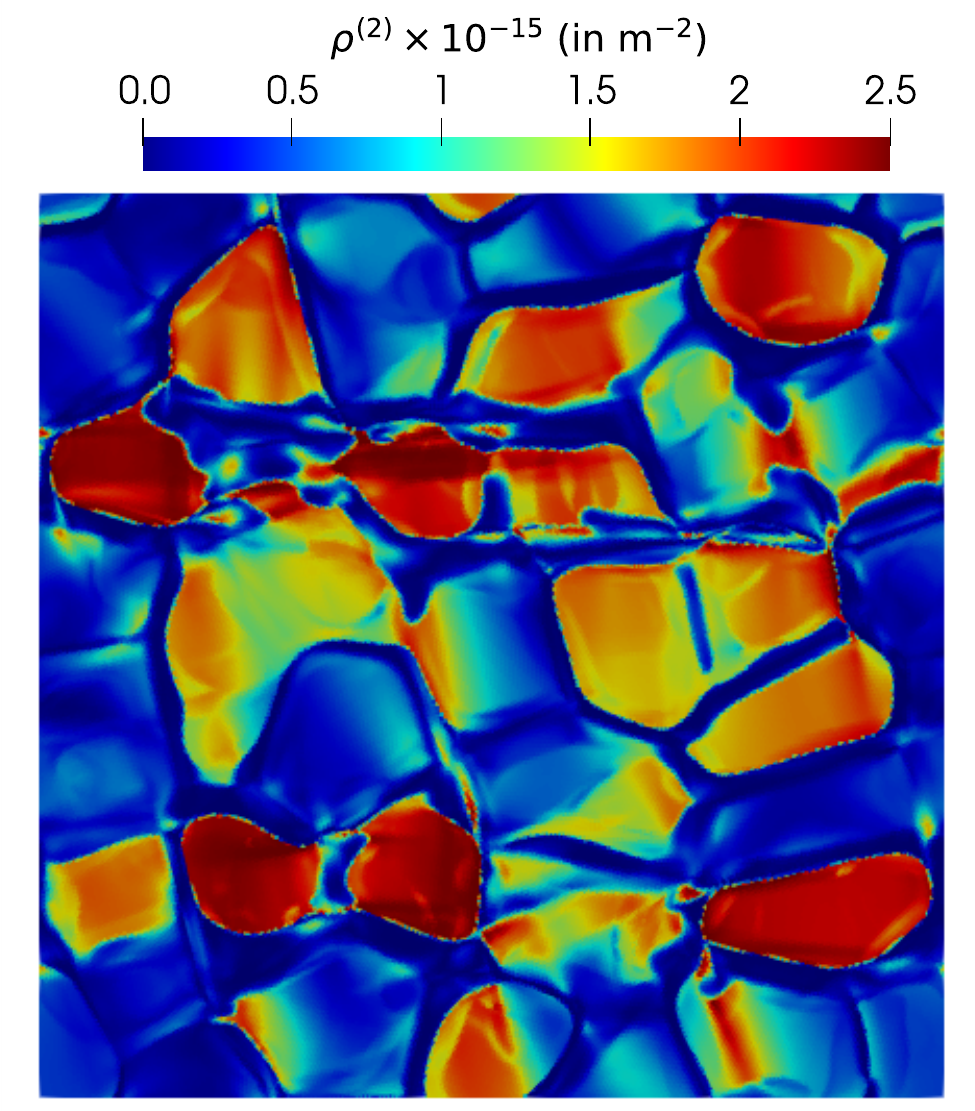}
 
	\hspace*{0mm}
	\raisebox{2cm}{\includegraphics[width=0.020\textwidth]{figure_t5000.pdf}}
	\includegraphics[width=0.285\textwidth,trim={5mm 0 5mm 3.2cm},clip]{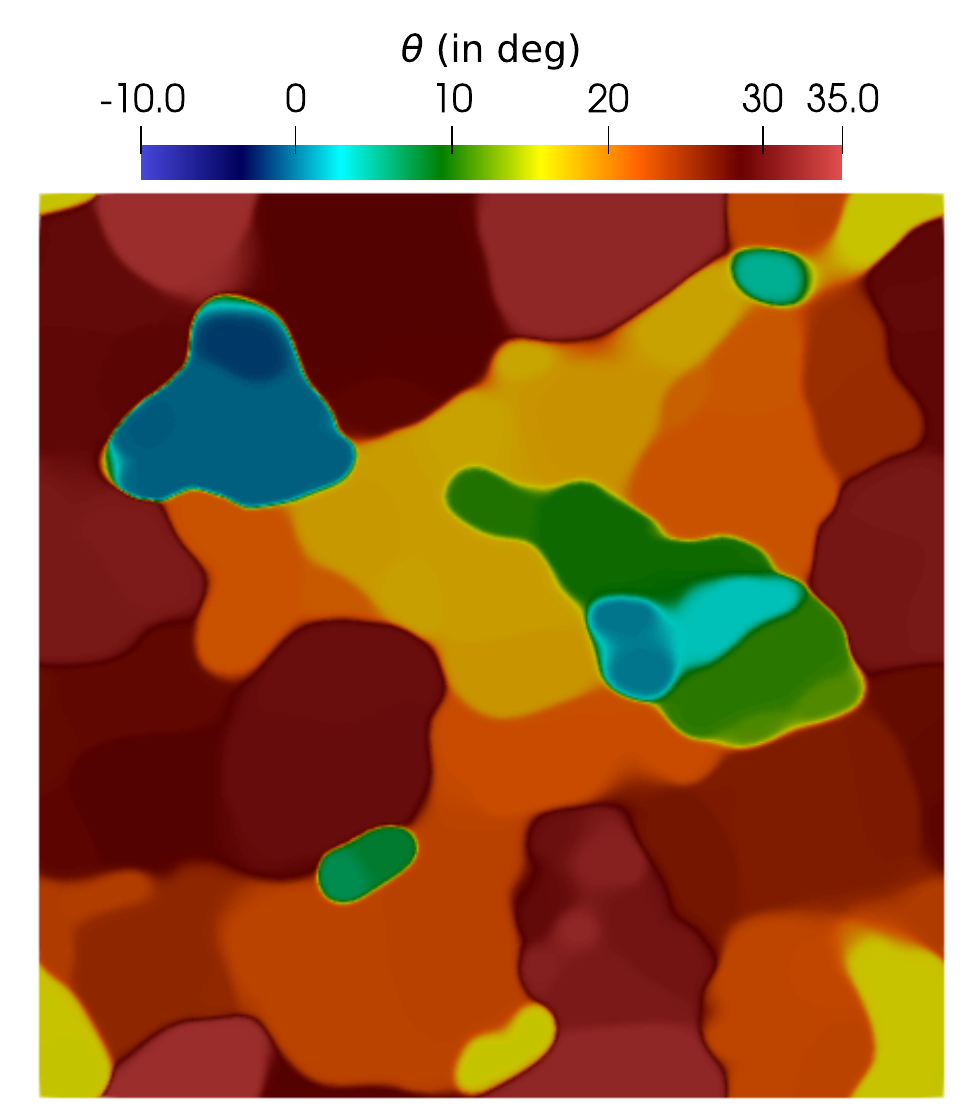}
	\includegraphics[width=0.285\textwidth,trim={5mm 0 5mm 3.2cm},clip]{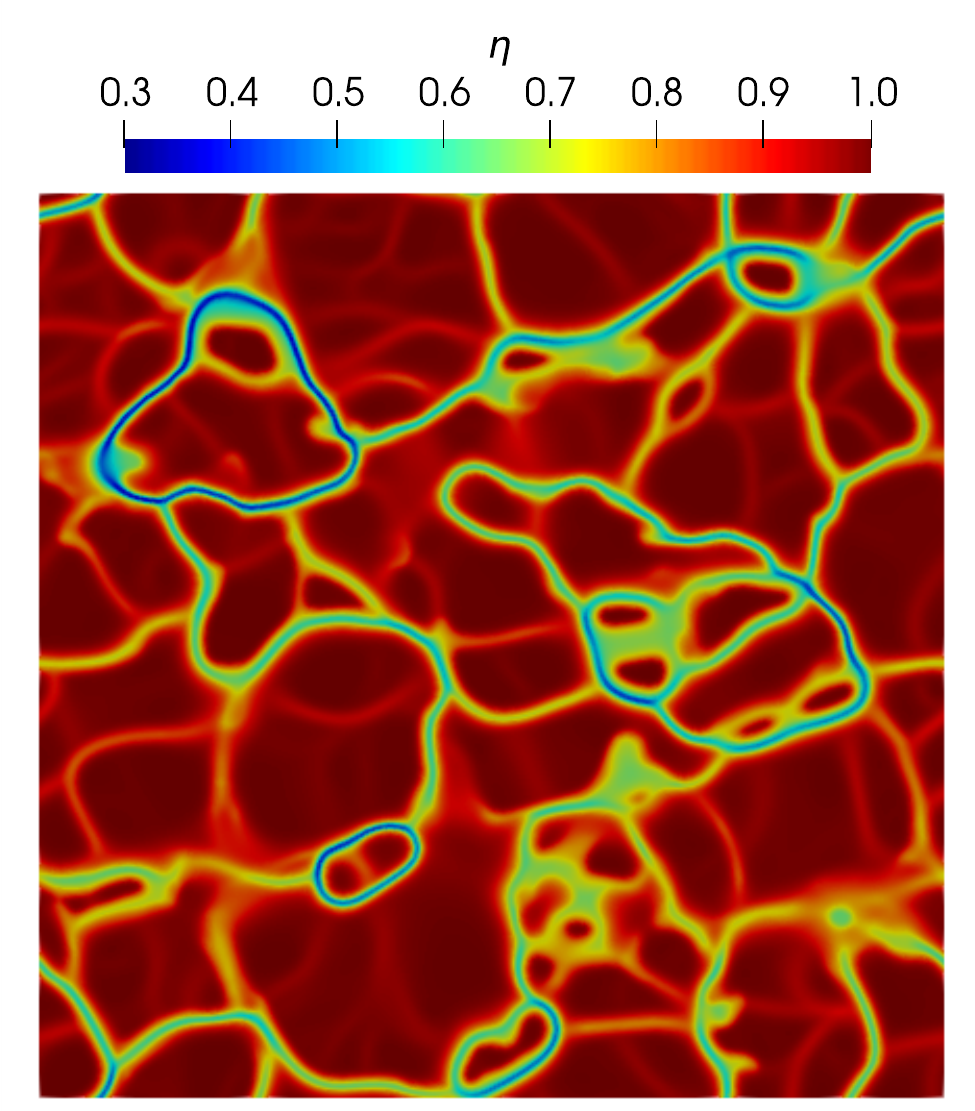}
	\includegraphics[width=0.285\textwidth,trim={5mm 0 5mm 3.2cm},clip]{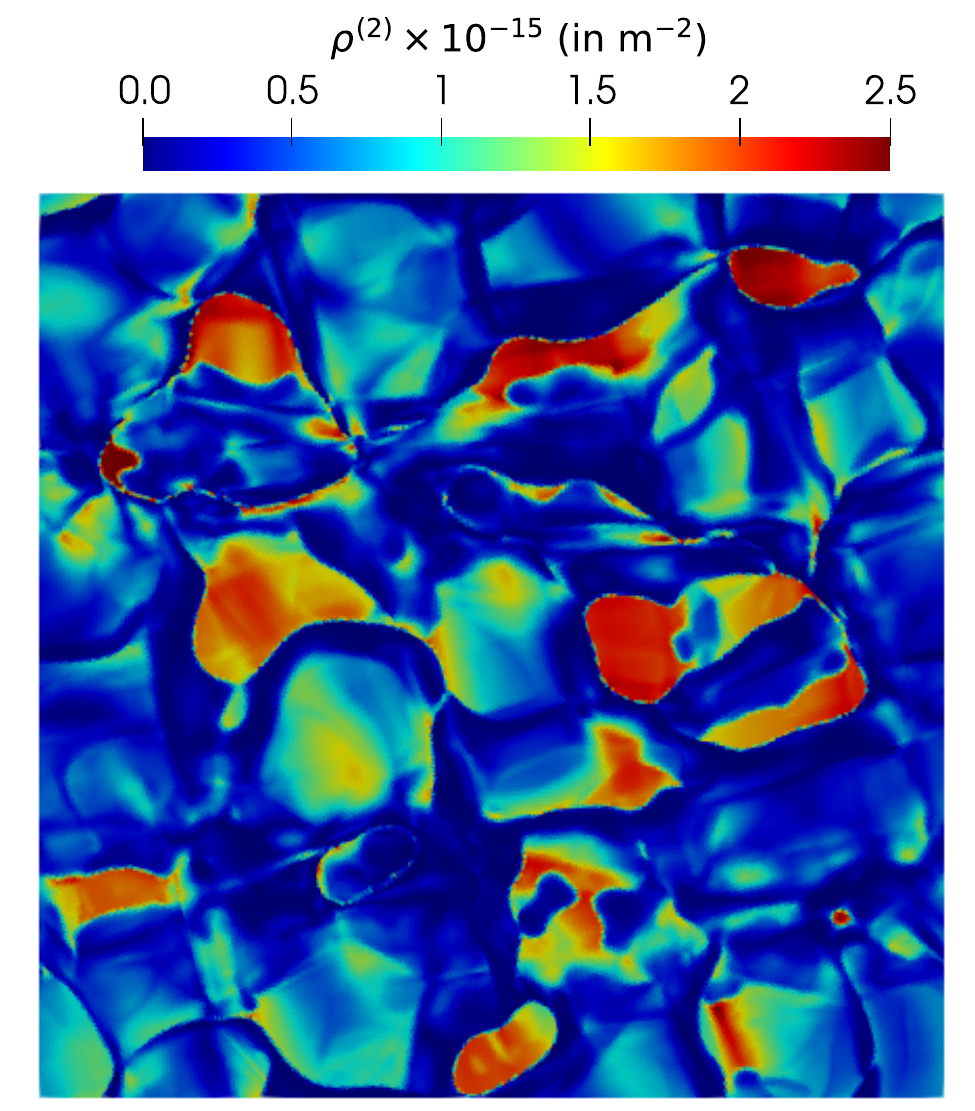}
	
	\hspace*{0mm}
	\raisebox{2cm}{\includegraphics[width=0.020\textwidth]{figure_t20000.pdf}}
	\includegraphics[width=0.285\textwidth,trim={5mm 0 5mm 3.2cm},clip]{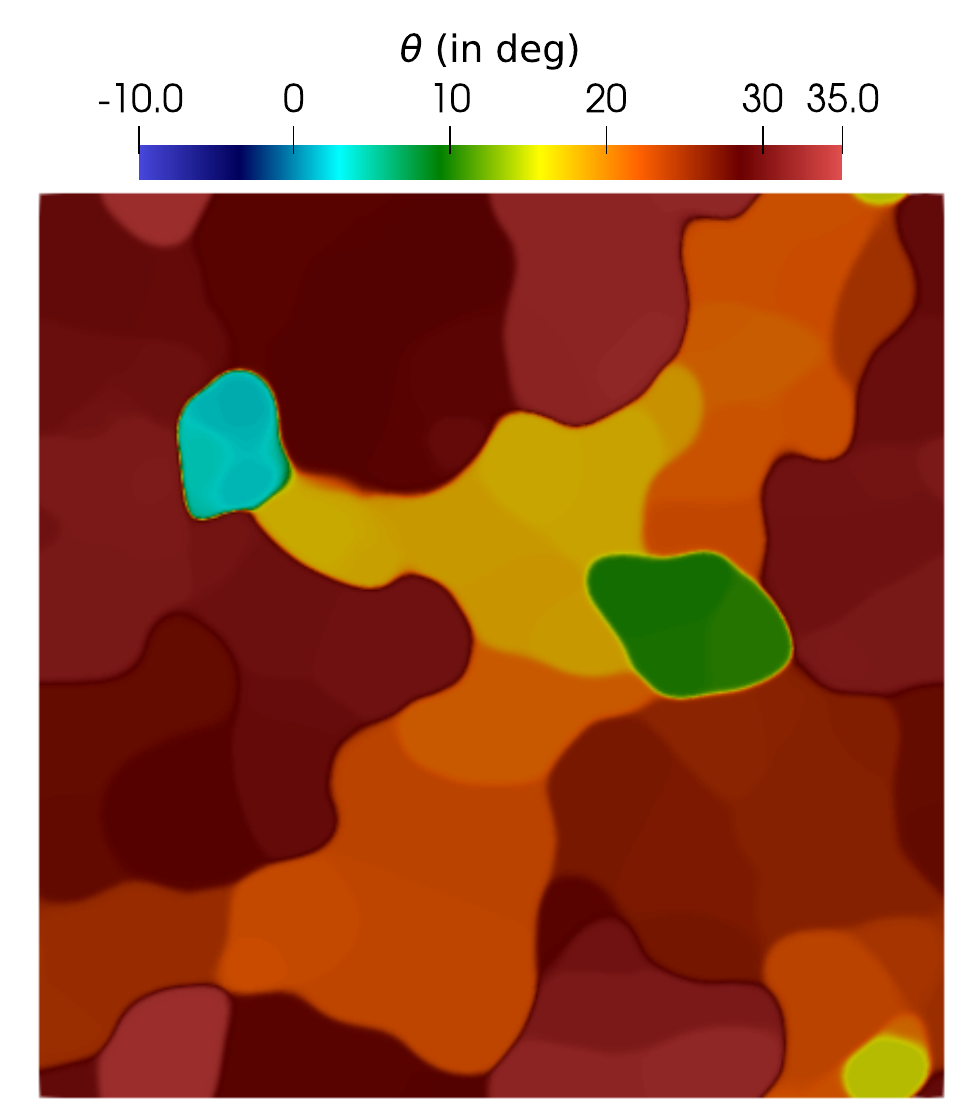}
    \includegraphics[width=0.285\textwidth,trim={5mm 0 5mm 3.2cm},clip]{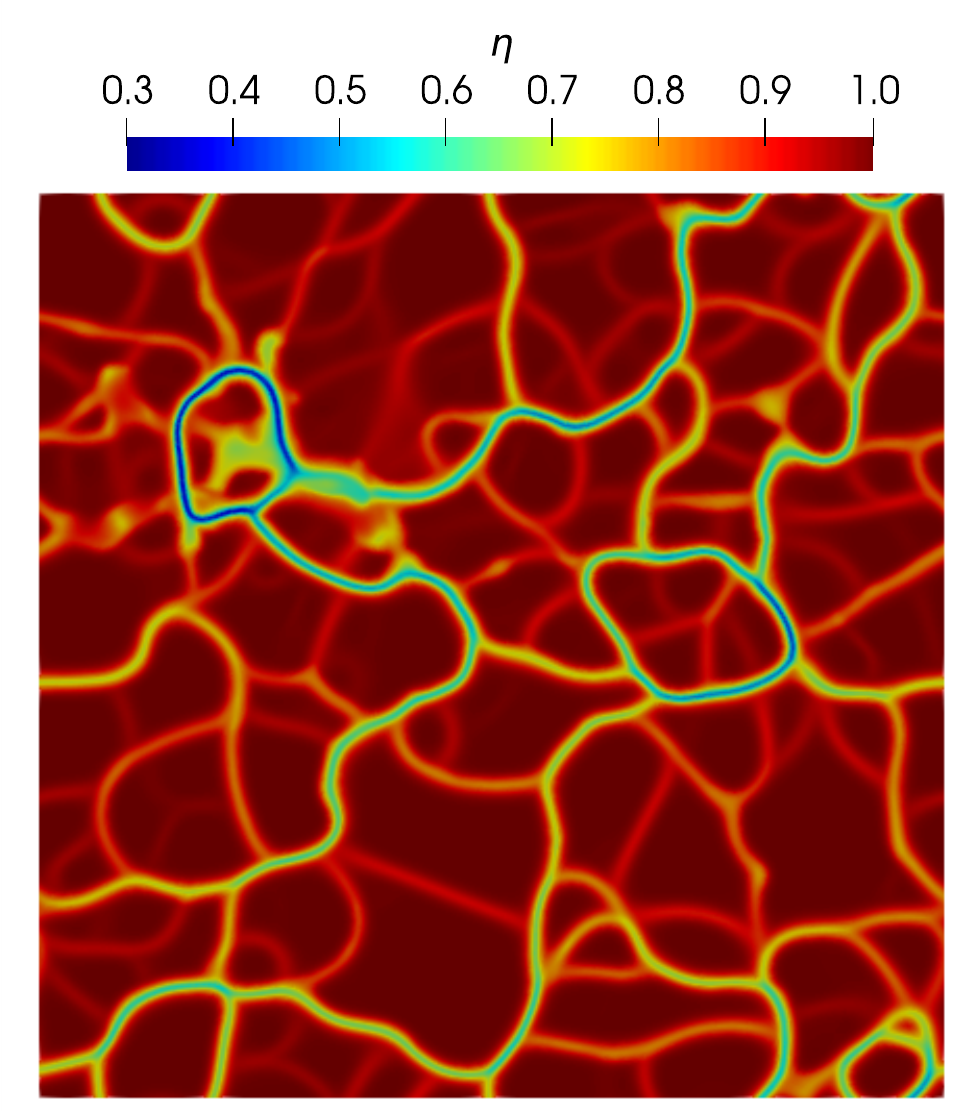}
	\includegraphics[width=0.285\textwidth,trim={5mm 0 5mm 3.2cm},clip]{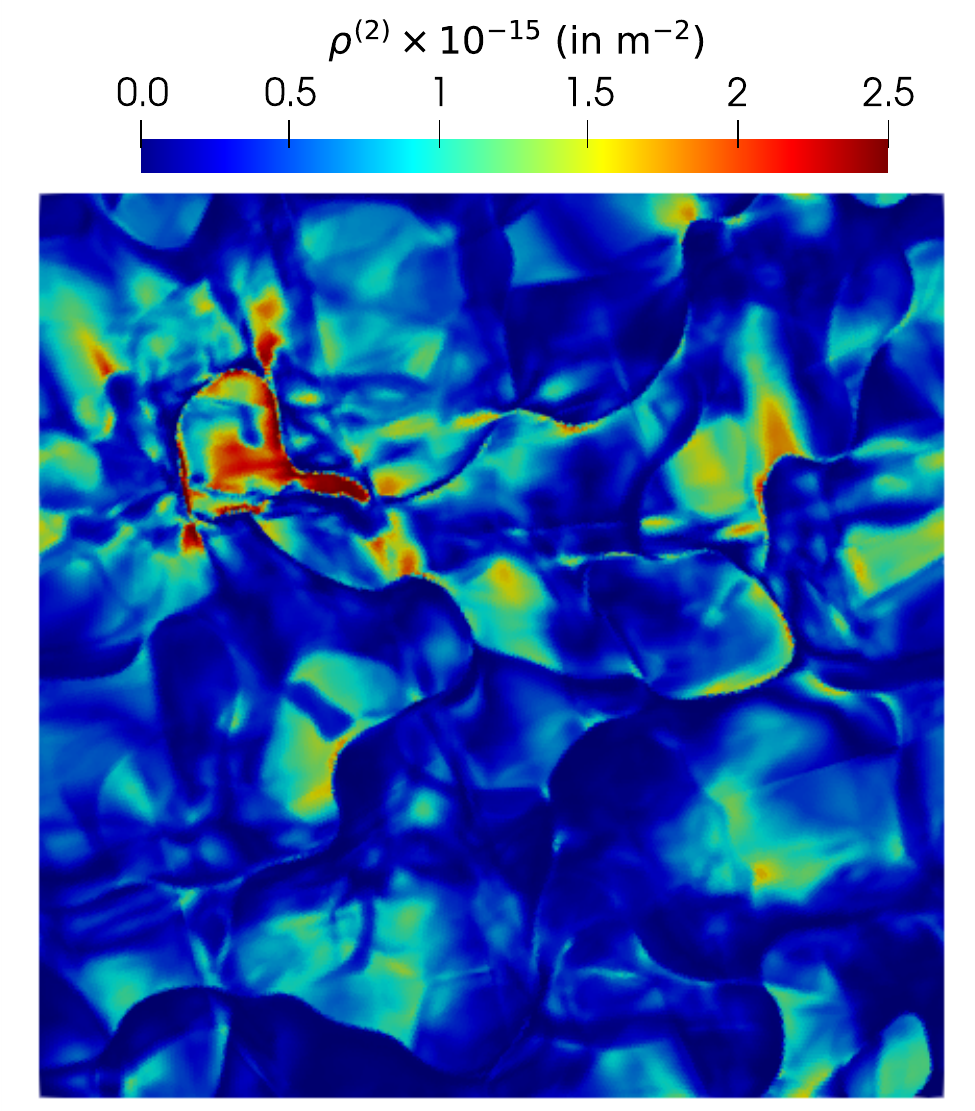}

    \hspace*{0mm}
	\raisebox{2cm}{\includegraphics[width=0.020\textwidth]{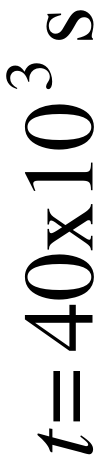}}
	\includegraphics[width=0.285\textwidth,trim={5mm 0 5mm 3.2cm},clip]{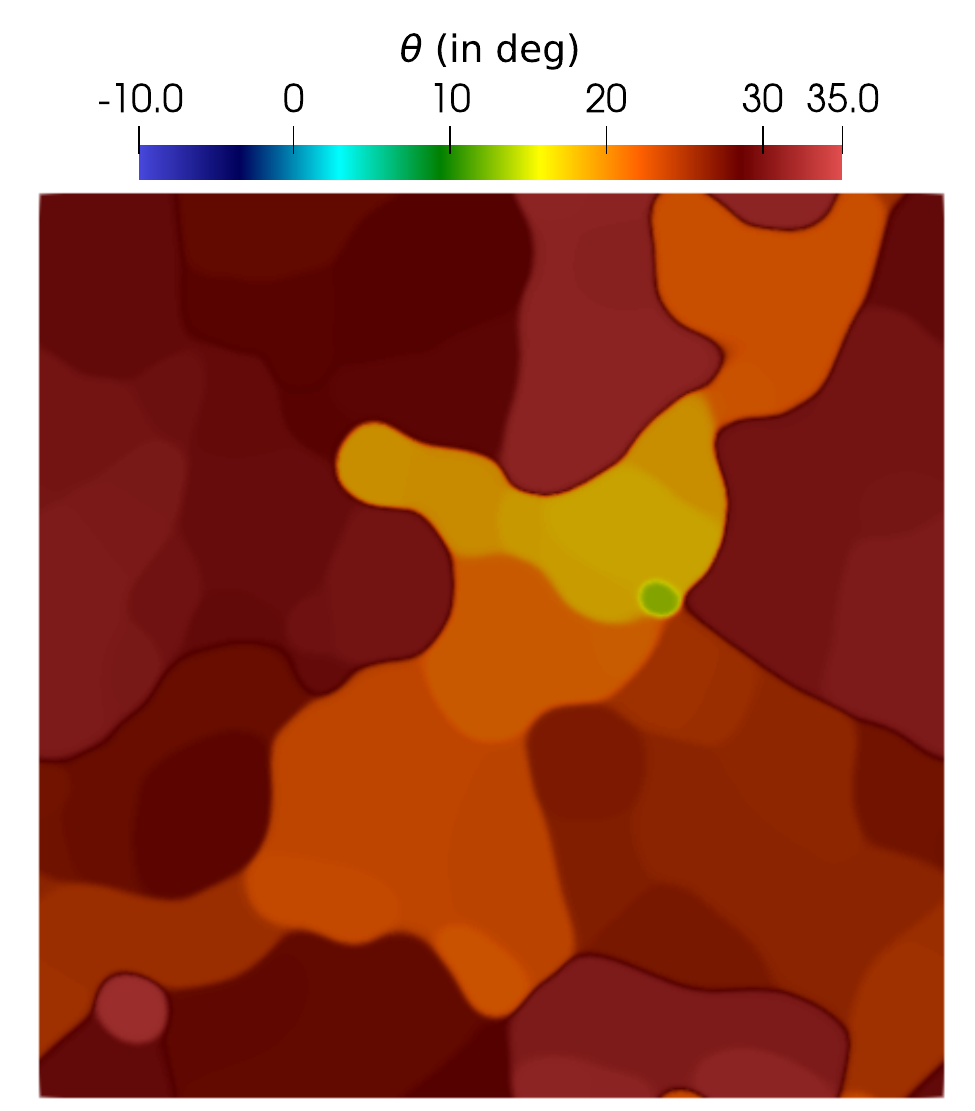}
    \includegraphics[width=0.285\textwidth,trim={5mm 0 5mm 3.2cm},clip]{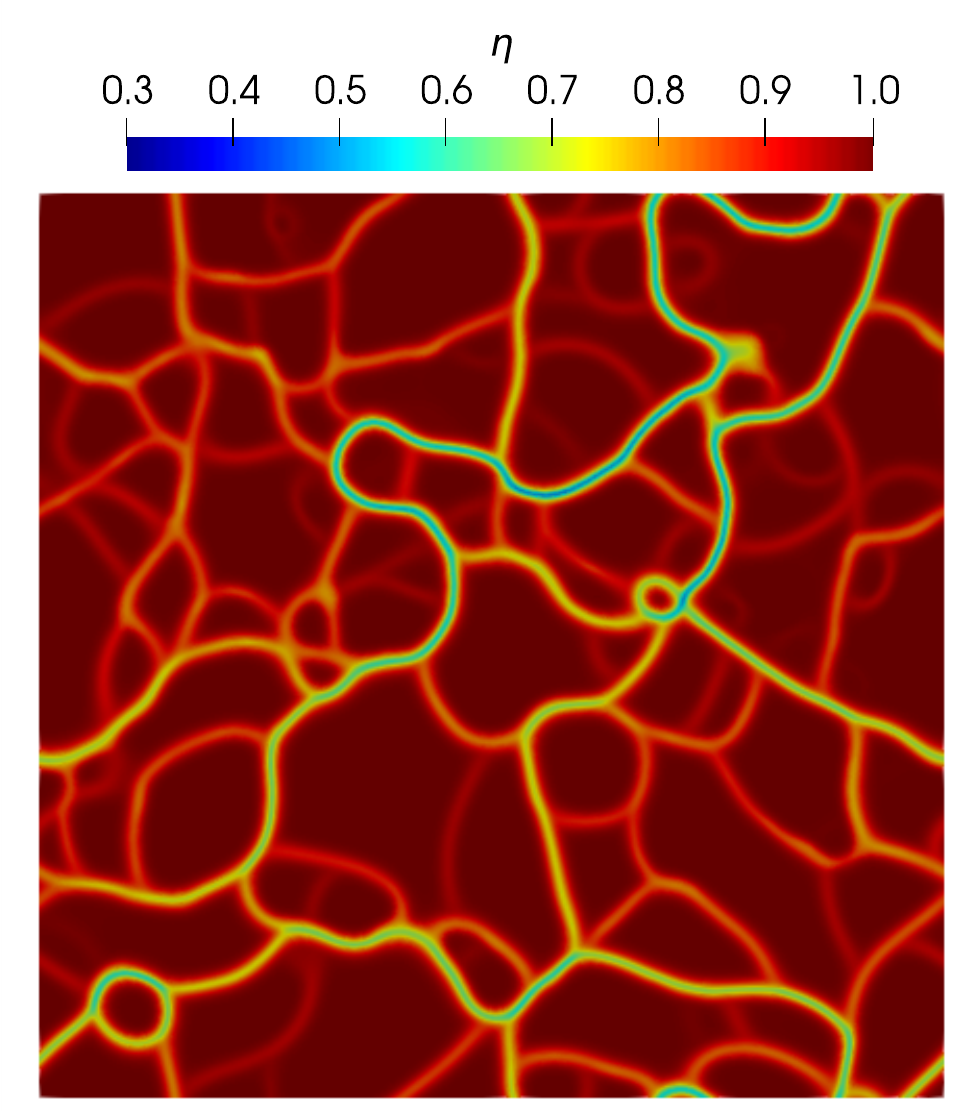}
	\includegraphics[width=0.285\textwidth,trim={5mm 0 5mm 3.2cm},clip]{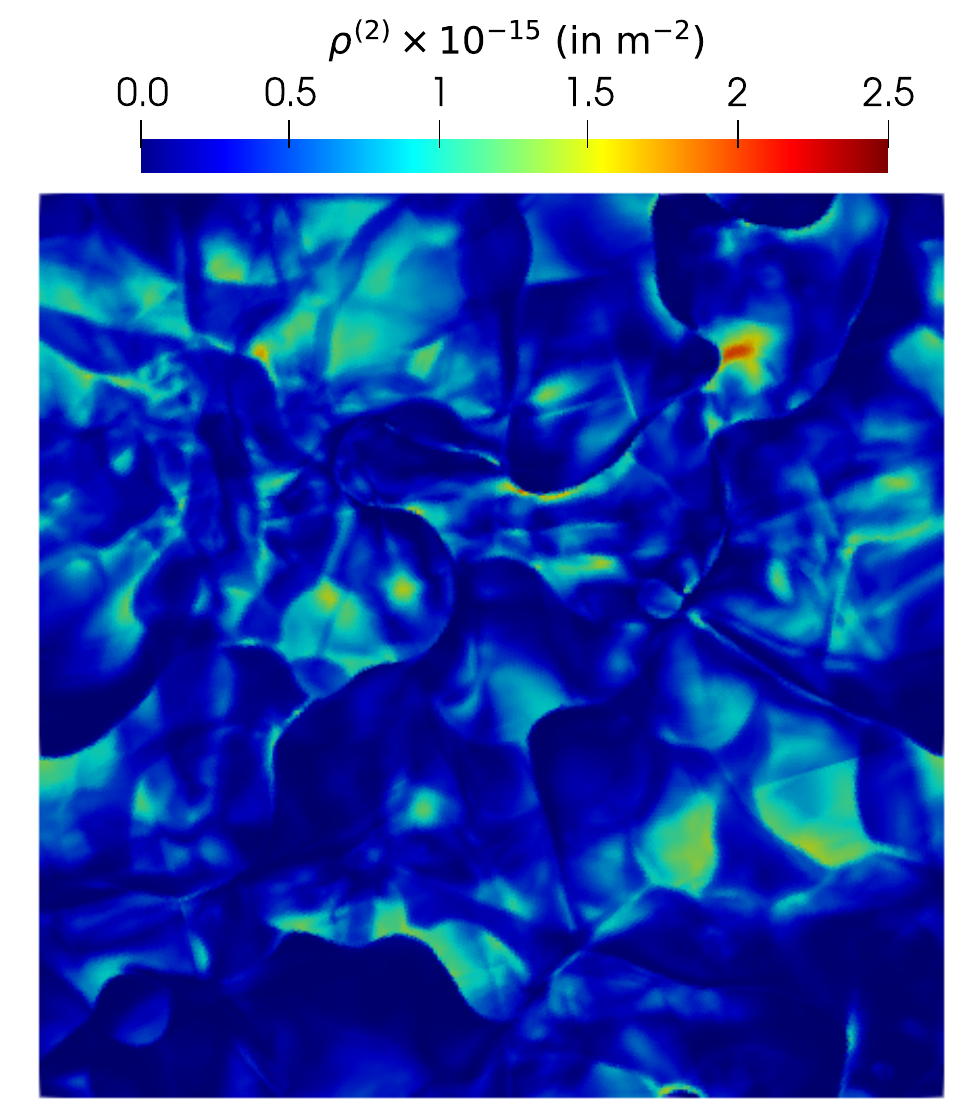}
	\caption{\MB{A granular structure with 32 grains are deformed in shear by $B_{12}=0.05$ with only slip system 2 active. Then, it is allowed to recrystallize with $c_2=0.9$ which activates grain nucleation. During the heat treatment phase, only the mean deformation gradient $B_{12}=0.05$ is held constant, which results in further local deformation. From left to right orientation $\theta$, order parameter $\eta$ and statically stored dislocation density $\rho^{(2)}$ are shown at different times.}}
	\label{fig:32grainsload5_slip2_c2_09_nodispfix}
\end{figure}

\MB{\subsubsection*{Mechanical loading and relaxation - 32 grains}}

Next, a \MB{larger polycrystal} with 32 grains is considered (Fig. \ref{fig:32grainsload5_slip2_c2_09}), where only slip system 2 is active. The structure is first \MB{sheared to} 5\% \MB{pre-strain}, then heat treated for $20\times 10^3$\;s. The top row \MB{in Fig \ref{fig:32grainsload5_slip2_c2_09}} shows the deformed state, where several sites of subgrain \MB{(marked as A$_{1/2}$, B$_{1/2}$ and C$_{1/2}$)} and slip/kink band formation are observed. At the beginning of the heat treatment phase, nucleation is triggered at several locations, and the more obvious ones are marked with white arrows in the $t=2\times 10^3$\;s contours. These locations can be \MB{distinguished} as regions of \MB{grain bulk (i.e. $\eta\approx 1$) forming inside existing GBs (i.e. $\eta<1$)}, at which $\rho$ is recovered fully. At $t=5\times 10^3$\;s, we can see that the nucleated grains expand and stabilize. At $20\times 10^3$\;s, the dislocation density is mostly recovered, and we have a \MB{recrystallized} microstructure. We can also see a clear example of subgrain nucleation and growth, where in the deformed state, the initial grain is fragmented into two sub-grains \MB{A$_1$ and A$_2$}. Then, the sub-grain \MB{A$_2$} with low energy density grows into the sub-grain \MB{A$_1$} with \MB{higher} stored dislocation density. As discussed before, some of the grains \MB{can evolve into concave shapes} due to the dominant driving force of stored dislocations. However, \MB{this is reversed} when curvature starts to take over. For example, observe the nucleated grain \MB{number 2}, which first grows into a concave shape at $5\times 10^3$\;s, then shrinks into a convex shape at $20\times 10^3$\;s. 

\MB{Fig. \ref{fig:32grainsload5_slip2_c2_09_nodispfix} shows the same case as in the previous example, however here the displacement DOFs are not constrained during the heat treatment phase. Compared to Fig. \ref{fig:32grainsload5_slip2_c2_09}, at $t=2\times 10^3$\;s the morphology is very similar, with the exception of some grains fragmenting into several sub-grains with very small misorientations (i.e. $\approx 0.5^{\degree}$). At $t=5\times 10^3$\;s, differences become more clear, mainly in the dislocation density $\rho$, which is recovered at the wake of moving GBs and at nucleation sites, but also generated throughout the domain due the deformation caused by coupled GB motion. The increased dislocation content triggers additional nucleation. At $t=20\times 10^3$\;s a significant amount of $\rho$ remains, which finally starts to recover after $t=40\times 10^3$\;s. The excessive generation of dislocation density during heat treatment stems from Eq. \eqref{eqn:rhodot}, whose recovery term, i.e. $\left(-2d\rho^\alpha/b\right)|\dot{\gamma}|$, depends strongly on the temperature and should be adjusted, which would result in reduced maximum value of $\rho$ at higher temperatures \citep{kocks2003physics}. The generation of stress and dislocations during grain growth in polycrystals have been also observed in MD simulations \citep{thomas2017reconciling}. At $t=40\times 10^3$\;s, it seems that the GB migration is accompanied with further fragmentation into sub-grains with $\Delta\theta<1^{\degree}$.} 

Finally, we would like to note that the circular shapes of the grains that form during recrystallization, especially at $t=20\times 10^3$ in Fig. \ref{fig:32grainsload5_slip2_c2_09} and \ref{fig:32grainsload5_slip2_c2_09_nodispfix}, do not fully represent the experimentally observed microstructure, \MB{though such smooth and concave shapes can occur during abnormal grain growth \citep{gruich2023effects}}. This is a consequence of the isotropic GB energy \MB{used in the simulations}, which minimizes the GB energy for a circular shape.
%, and also because of the fundamental formulation of orientation phase field, which minimizes the global energy for a single grain solution. 
When anisotropic GB energy is introduced to the HMP model, the evolved microstructure retains \MB{polygonal} shapes closer to experimental observations, as shown in \cite{staublin2022phase}. However, the focus of this paper is not to create perfect agreement with the experiments, but to investigate the new nucleation mechanism and its capabilities. Therefore, this issue is left for future work. 

\section{Conclusion}\label{sec:conc}

A thermodynamically consistent multi-physics model is presented that can simulate dislocation-induced spontaneous grain nucleation. It is an extension of our previous work \citep{tandogan2025multi}, which couples Cosserat crystal plasticity with HMP type orientation phase field. The coupled framework allows to model changes in the lattice orientation due to both mechanical deformation and grain boundary migration. \MB{The accumulated dislocation density at the grain boundaries spontaneously triggers nucleation. The proposed nucleation mechanism was investigated and the main results are:
\begin{itemize}
    \setlength\itemsep{-1mm}
    \item Nucleation mechanisms such as SIBM, sub-grain coarsening and coalescence are reproduced.
    %\item The core idea is expending stored dislocation density energy to spontaneously generate new GBs at the existing GBs.
    \item Nucleation is more likely to happen above a threshold misorientation $\Delta\theta_\mathrm{T}$, which can be controlled with the model parameters.
    \item The orientation of the nuclei is bounded by the orientation of the surrounding lattice. The exact value is determined by the distribution of dislocation density in the vicinity of GBs.
    \item The GB velocity determines whether a nucleus grows and stabilizes into a new grain or not. The nucleation intensity can be controlled by adjusting GB velocity with model parameters.
\end{itemize}}

\MB{In addition the physically motivated treatment of nucleation, the coupled Cosserat-Phase field framework has following capabilities:
\begin{itemize}
    \setlength\itemsep{-1mm}
    \item The non-local Cosserat framework allows formation of slip and kink bands during deformation.
    \item Grains can fragment into sub-grains due to heterogeneous deformation.
    \item Localized orientation gradients that form during deformation are seamlessly recognized as new GBs.
    \item GB migration is driven by curvature as well as stored dislocation density,
    \item Dislocations can be recovered partially or fully in the nucleation sites and in the wake of migrating GBs.
\end{itemize}}

The main capabilities of the model were shown in a 2D, small deformation setting with isotropic grain boundary energy. \MB{We believe the comprehensive coverage of the physical mechanisms, especially the natural treatment of nucleation, is an evidence of the potential of the proposed unified framework for grain microstructure evolution.} In the future work, inclination dependent GB energy, \MB{finite deformation} and efficient extension to 3D will be explored.
%, where the former should help with retaining granular shapes closer to experimental observations. 
%Also, the proposed nucleation mechanism has room for improvement, for example a more robust dependence on misorientation could be implemented. 
\MB{A process dependent mobility parameter that is able to cover the different time scales of loading and recrystallization should be implemented. Finally, more complex interaction mechanisms such as nucleation during loading and mechanical deformation during heat treatment could be investigated in detail.}

\section{Acknowledgment}\label{sec:acknow}
We are grateful to Dr. Anna Ask and Prof. Samuel Forest for their insight and valuable discussions.

\bibliographystyle{elsarticle-harv} 
\bibliography{mybibfile}



\end{document}